\documentclass[journal]{IEEEtran}
\usepackage{amsmath}
\usepackage{amssymb}
\usepackage{amsfonts}
\usepackage{amsthm}
\usepackage{graphicx}
\usepackage{mathtools}
\usepackage{balance}
\usepackage{bm}
\usepackage{colortbl}
\usepackage{lipsum}      
\usepackage{changepage}
\usepackage{caption}
\usepackage{color}
\usepackage{pgf,tikz}
\usepackage{balance}
\usepackage{mathtools}
\usepackage{bbm}
\usepackage{array}
\usepackage{relsize}
\usepackage{cite}
\usepackage{amsthm}
\usepackage{verbatim}
\usepackage{epstopdf}
\usepackage{array}
\usepackage{url}
\usepackage{stfloats}
\usepackage[hidelinks]{hyperref}
\usepackage{url}
\usepackage[linesnumbered,ruled]{algorithm2e}
\usepackage{algpseudocode}
\newtheorem{theorem}{Theorem}
\newtheorem{corollary}{Corollary}

\newtheorem{lemma}{Lemma}
\newtheorem{remark}{Remark}

\newtheorem{definition}{Definition}
\newtheorem{condition}{Condition}
\newtheorem{problem}{Problem}

\newcommand{\eqdef}{\mathrel{\mathop:}=}
\newcommand*{\QEDclosed}{\hfill\ensuremath{\blacksquare}}

\DeclareMathOperator*{\argmax}{arg\,max}
\DeclareMathOperator*{\argmin}{arg\,min}
\definecolor{chocolate}{rgb}{0.48, 0.25, 0.0}

\begin{document}

\title{Performance Limits of a Deep Learning-Enabled \\ Text Semantic Communication under Interference}

\author{Tilahun~M.~Getu,~\IEEEmembership{Member,~IEEE},~Walid~Saad,~\IEEEmembership{Fellow,~IEEE},\\~Georges~Kaddoum,~\IEEEmembership{Senior Member,~IEEE},~and~Mehdi~Bennis,~\IEEEmembership{Fellow,~IEEE} 
\thanks{\IEEEcompsocthanksitem T. M. Getu is with the Electrical Engineering Department, \'Ecole de Technologie Sup\'erieure (ETS), Montr\'eal, QC H3C 1K3, Canada. He was with the Communications Technology Laboratory, National Institute of Standards and Technology (NIST), Gaithersburg, MD 20899, USA and the ETS Electrical Engineering Department (e-mail: tilahun-melkamu.getu.1@ ens.etsmtl.ca).}

\thanks{\IEEEcompsocthanksitem W. Saad is with the Bradley Department of Electrical and Computer Engineering, Virginia Tech, Arlington, VA, USA, and the Cyber Security Systems and Applied AI Research Center, Lebanese American University, Beirut, Lebanon (e-mail: walids@vt.edu).}

\thanks{\IEEEcompsocthanksitem G. Kaddoum is with the Electrical Engineering Department, \'Ecole de Technologie Sup\'erieure (ETS), Montr\'eal, QC H3C 1K3, Canada, and the Cyber Security Systems and Applied AI Research Center, Lebanese American University, Beirut, Lebanon (e-mail: georges.kaddoum@etsmtl.ca).}

\thanks{\IEEEcompsocthanksitem M. Bennis is with the Centre for Wireless Communications, University of Oulu, 90570 Oulu, Finland (e-mail: mehdi.bennis@oulu.fi).}

\thanks{\IEEEcompsocthanksitem This research was supported by the U.S. Department of Commerce and its agency NIST, and in part by the U.S. National Science Foundation under Grant CNS-2225511.}

}

\markboth{IEEE Transactions on Wireless Communications}{Getu \MakeLowercase{\textit{et al.}}: Performance Limits of a Deep Learning-Enabled Text Semantic Communication under Interference}
\IEEEtitleabstractindextext{%
\vspace{-1.5cm}
\begin{abstract}
Although deep learning (DL)-enabled semantic communication (SemCom) has emerged as a 6G enabler by minimizing irrelevant information transmission -- minimizing power usage, bandwidth consumption, and transmission delay, its benefits can be limited by radio frequency interference (RFI) that causes substantial semantic noise. Such semantic noise's impact can be alleviated using an interference-resistant and robust (IR$^2$) SemCom design, though no such design exists yet. To stimulate fundamental research on IR$^2$ SemCom, the performance limits of a popular text SemCom system named \textit{DeepSC} are studied in the presence of (multi-interferer) RFI. By introducing a principled probabilistic framework for SemCom, we show that DeepSC produces semantically irrelevant sentences as the power of (multi-interferer) RFI gets very large. We also derive DeepSC’s practical limits and a lower bound on its outage probability under multi-interferer RFI, and propose a (generic) lifelong DL-based IR$^2$ SemCom system. We corroborate the derived limits with simulations and computer experiments, which also affirm the vulnerability of DeepSC to a wireless attack using RFI.   
 
\end{abstract}
}
\maketitle
\IEEEdisplaynontitleabstractindextext
\IEEEpeerreviewmaketitle
\section{Introduction}
\label{sec: introduction}
Introduced by Weaver \cite[Ch. 1]{Shannon_Weaver_Math_Theory_Commun'49} around 1949, semantic communication (SemCom) is a communications paradigm whose purpose is to convey a transmitter's intended meaning to a receiver  \cite{Chaccour_Building_NG_SemCom_Networks'22,Yang_SemCom_ComST’23}, while aiming to minimize the divergence of the receiver's interpretations -- deduced from its recovered messages -- from the meaning of the transmitted message \cite{Tong_FL_ASC'21}. SemCom transmits semantic information that is only relevant to the communication goal, significantly reducing data traffic \cite{Xie_DL-based_SemCom'21}. SemCom's ability to significantly reduce traffic means it has the potential to change the status quo viewpoint of the conventional communications systems' designers that wireless connectivity is an opaque data pipe that carries messages whose context-dependent meaning and effectiveness have been ignored \cite{Kountouris_Semantics_EmpoweredCF'21}. Contrary to conventional communication systems aiming to provide high data rates and a low symbol/bit error rate, SemCom extracts the meaning of the transmitter's message and interprets the semantic information at the receiver \cite{Luo_SemCom_Overview'22}. Accordingly, SemCom's objective is to deliver the source's intended meaning considering its dependence on not only the physical content of the message but also the users' intentions and other humanistic factors that could reflect the real quality of experience \cite{Shi_SemCom_ComMag’21}, so a SemCom system is designed to deliver the transmitted message's representative meaning \cite{Luo_SemCom_Overview'22}.

SemCom takes a meaning-centric approach to communications system design, and emphasizes conveying a transmitted message's interpretation \textit{in lieu of} reproducing the message through a symbol-by-symbol reconstruction \cite{Bao_Towards_Theory_SemCom'11}. Consequently, SemCom's meaning-centric approach to communications has made it emerge as a 6G (sixth generation) \cite{Saad_6G_Vision_20,Letaief_Edge_AI_Vision'22,Alwis_Survey_GG_Networks'21} technology enabler. As such, SemCom holds the promise of minimizing power usage, bandwidth consumption, and transmission delay by minimizing irrelevant information transmission. The transmission of irrelevant information is discarded by using efficient semantic extraction -- via a joint semantic encoding and decoding \cite{Yang_SemCom_ComST’23} -- that can be efficiently executed by leveraging state-of-the-art advancements of deep learning (DL)  \cite{Xie_DL-based_SemCom'21}. Advancements in DL \cite{YYBGH_15} and natural language processing \cite{Trands_in_DL-Based_NLP'18} have propelled the application of SemCom to text transmission \cite{Xie_DL-based_SemCom'21}, image transmission \cite{Eirina_JSCC'19}, video transmission \cite{Wang_JSAC_Semantic_Transmission’23}, audio transmission \cite{Tong_FL_ASC'21}, and visual question answering tasks \cite{Xie_MU-SemCom’22}. These applications, however, can be made ineffective by considerable semantic noise. 

Semantic noise causes semantic information misunderstanding, the manifestation of semantic decoding errors, inducing misunderstanding between a transmitter's intended meaning and a receiver's reconstructed meaning \cite{Hu_Robust_SemCom’23}. Such semantic noise can arise in semantic decoding, data transmission, and/or semantic encoding. In the semantic encoding stage, semantic noise can occur due to a mismatch between the original signal and semantically encoded signal \cite{Hu_Robust_SemCom’23} and due to adversarial examples \cite{Hu_Robust_SemCom’23}. Semantic noise can also happen during data transmission because of physical noise, signal distortion due to channel fading, or interference at a receiver \cite{Hu_Robust_SemCom’23,Luo_SemCom_Overview'22}. At a receiver, semantic noise can appear during semantic decoding in case of different message interpretations due to ambiguity in the words, sentences, or symbols used in the transmitted messages \cite{Luo_SemCom_Overview'22}; semantic ambiguity (due to \textit{dialect} and \textit{polysemy}) in the recovered symbols when a semantic symbol represents multiple sets of data with dissimilar meanings \cite{Shi_to_Semantic_Fidelity'21}; and a mismatch between the source and destination knowledge bases (even in the absence of syntactic errors) \cite{Yang_SemCom_ComST’23}. There may also be semantic noise during semantic decoding due to a malicious attacker emitting interference \cite{Hu_Robust_SemCom’23}.

Among many factors that can cause semantic noise, radio frequency interference (RFI) is a major culprit. As a major culprit, huge RFI causes substantial noise to a SemCom receiver whose channel decoder's unreliable outputs evoke significant semantic noise to the semantic decoder. Despite the semantic decoder's vulnerability to RFI, a few existing SemCom works have empirically investigated the impact of RFI/interference on the reliability of a SemCom receiver. Among these works, the authors of \cite{Hu_Robust_SemCom’23} develop an \textit{adversarial training algorithm} to combat semantic noise due to a jammer, and the authors of \cite{Sagduyu_Is_SemCom_Secure'22} empirically demonstrate that a \textit{wireless attack} using RFI can change the transmitted information's semantics. None of these works, however, quantifies the impact of RFI on the performance of a SemCom system, and the performance of any SemCom technique has not been quantified to our knowledge.

As motivated above, the main contribution of this paper is the derivation of the asymptotic performance limits of \textit{DeepSC} \cite[Fig. 2]{Xie_DL-based_SemCom'21} under RFI and multi-interferer RFI (MI RFI). Particularly, we study a semantic decoder's output in the presence of RFI to determine the performance limits of a text SemCom system experiencing RFI. RFI is generally caused by intentional or unintentional interferers and is being encountered increasingly in satellite communications, microwave radiometry, radio astronomy, ultra-wideband communication systems, radar systems, and cognitive radio communication systems \cite{TMWAR_TWC_18,TWR_WCL_2018,Getu_dissertation_19}. Consequently, SemCom systems of the near future will also be impacted by RFI from jammers, spoofers,  meaconers, and inter-cell interferers, whose RFI must be taken into account -- in a likely scenario of \textit{adversarial electronic warfare} -- to ensure the design of a fundamentally robust SemCom system, which must enable reliable SemCom regardless of any RFI. To stimulate this type of system design, we aim to quantify DeepSC's performance limits \cite{Xie_DL-based_SemCom'21}.

The asymptotic/non-asymptotic performance quantification of a DL-enabled SemCom system like DeepSC \cite[Fig. 2]{Xie_DL-based_SemCom'21} is not addressed to date and fundamentally challenging for: $1)$ The lack of \textit{interpretability} in DL models concerning \textit{optimization}, \textit{generalization}, and \textit{approximation} \cite{Poggio_Theo_Issues_Dnets_2020,arXiv_Getu_Fundamental_Limits'23_v1}; $2)$ the lack of a commonly agreed-upon definition of semantics/semantic information \cite[Ch. 10, p. 125]{Tong_Zhu__6G'21}; and $3)$ the absence of a SemCom mathematical foundation \cite{Tong_Nine_Challenges’22}. These challenges are partially overcome through the following key contributions:  
\begin{enumerate}
	
	\item We introduce a new semantic metric named \textit{the upper tail probability of a semantic similarity}.
	
	\item We introduce a principled probabilistic framework to alleviate the challenge in analyzing SemCom systems.
	
	\item We deploy our probabilistic framework to reveal the performance limits of DeepSC under RFI and MI RFI.     
	
	\item We derive DeepSC's practical limits and a lower bound on its outage probability under MI RFI.
	     
	\item Toward a fundamental 6G design for an \textit{interference-resistant and robust SemCom} (IR$^2$ SemCom), we propose a (generic) lifelong DL-based IR$^2$ SemCom system. 
	
	 \item We corroborate the derived performance limits with Monte Carlo simulations and computer experiments. 
\end{enumerate}

The rest of this paper is organized as follows. Sec. \ref{sec: systetm_setup} outlines the system description and problem formulation. Sec. \ref{sec: fundamental_limits_DeepSC} reports DeepSC's performance limits. Sec. \ref{sec: practical_limits_outage_probability_DeepSC}\linebreak reports on DeepSC's practical limits and outage probability. Sec. \ref{sec: IR2_SemCom} proposes a lifelong DL-based IR$^2$ SemCom system. Sec. \ref{sec: simulation_results} presents simulation and computer experiment results. Sec. \ref{sec: conc_summary_and_research_outlook} concludes this work.
 
\textit{Notation and Definitions}: Scalars, vectors, and matrices are denoted by italic, bold lowercase, and bold uppercase letters, respectively. $\mathbb{N}$, $\mathbb{R}$, $\mathbb{R}^{+}$, $\mathbb{C}$, and $\mathbb{C}^{1\times n}$ represent the set of natural numbers, real numbers, non-negative real numbers, complex numbers, and $n$-dimensional row vectors of complex numbers, respectively. $\eqdef$, $\sim$, $\jmath$, $(\cdot)^T$, $\| \cdot \|$, and $\bm{0}$ stand for equal by definition, distributed as, $\sqrt{-1}$, transpose, Euclidean norm, and a zero (row/column) vector, respectively. $\bm{I}_n$, $\textnormal{Re}\{\cdot\}$, $\textnormal{Im}\{\cdot\}$,  $\mathbb{E}\{\cdot\}$, $\mathbb{P}(\cdot)$, and $\mathbb{I}\{\cdot\}$ symbolize an $n\times n$ identity matrix, a real part, an imaginary part, expectation, probability, and an indicator function that returns 1 if the argument is true and 0 otherwise, respectively. For $a\in\mathbb{R}$, its absolute value is denoted by $|a|$ and $|a| \eqdef \mathbb{I}\{a\geq 0\}a-\mathbb{I}\{a < 0\}a$. For $z=a+\jmath b \in \mathbb{C}$, its magnitude $|z|$ is given by $|z| \eqdef \sqrt{a^2+b^2}$. For $n, k\in\mathbb{N}$, $[n]\eqdef \{1, 2, \ldots, n\}$ and $\mathbb{N}_{\geq k} \eqdef \{k, k+1, k+2, \ldots\}$. For $n\in\mathbb{N}_{\geq 2}$, the maximum and minimum of $a_1, a_2, \ldots, a_n \in\mathbb{R}$ are written as $\textnormal{max}(a_1, a_2, \ldots, a_n)$ and $\textnormal{min}(a_1, a_2, \ldots, a_n)$, respectively. For a row vector $\bm{a}\in\mathbb{C}^{1\times n}$, its $i$-th element is denoted by $(\bm{a})_i$, $\forall i\in[n]$.
\begin{figure*}[t!]
	\centering
	\vspace{-0.75cm}
	\includegraphics[scale=0.35]{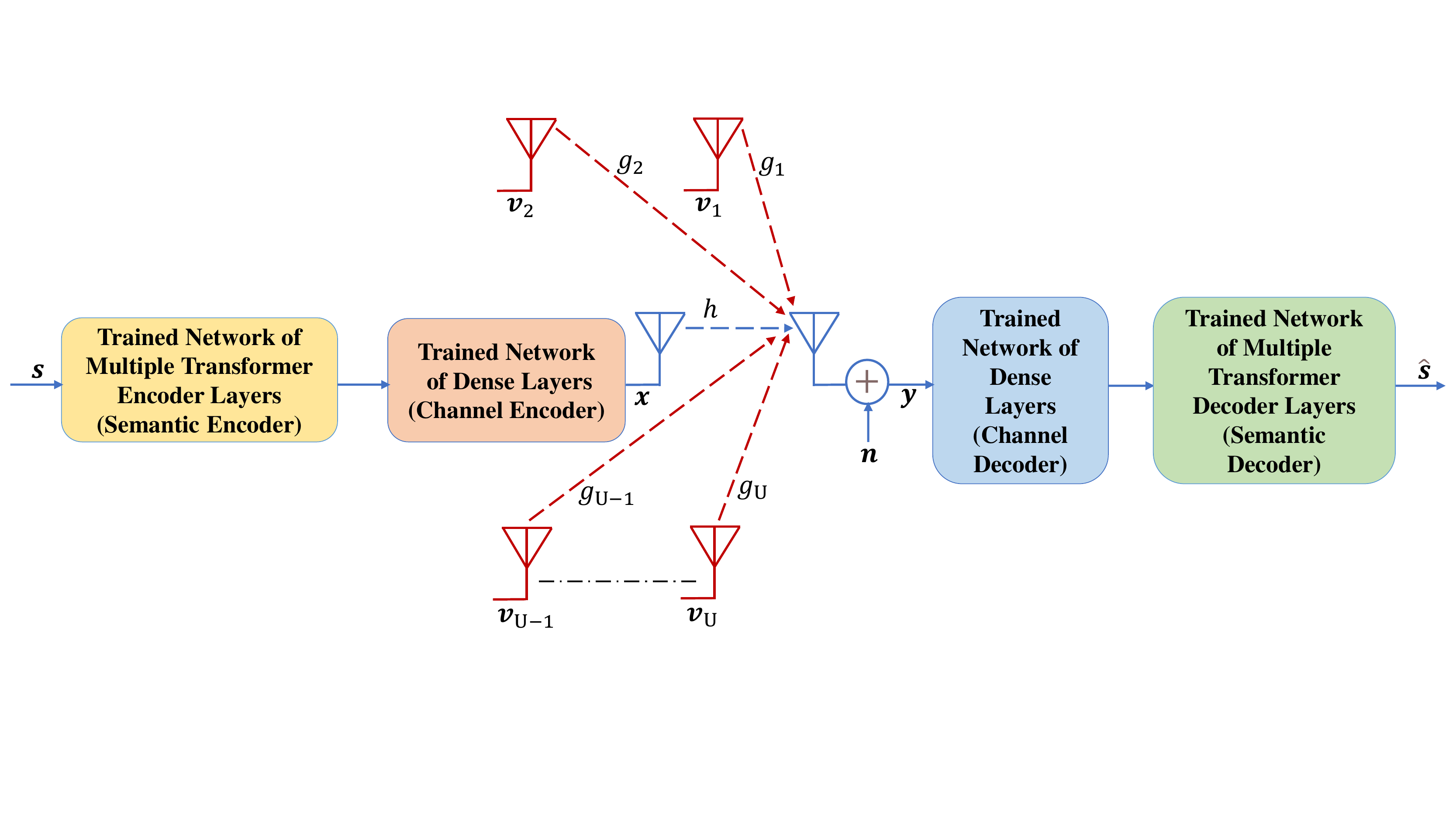}  \vspace{-1.5cm} 
	\caption{A trained DeepSC under RFI from one or more single-antenna RFI emitters.}
	\label{fig: DeepSC_System_Model}
\end{figure*}
$\mathcal{N}(0,1)$ stands for a Gaussian distribution with zero mean and unit variance. A random variable (RV) $X\sim\mathcal{N}(0,1)$ is termed a standard normal RV. A complex RV $Y\sim\mathcal{CN}(\mu, \sigma^2)$ is a complex Gaussian RV with mean $\mu$ and variance $\sigma^2$. For a complex random row vector $\bm{x}\in\mathbb{C}^{1 \times n}$, $\bm{x}=\textnormal{Re}\{\bm{x}\}+\jmath\textnormal{Im}\{\bm{x}\} \sim\mathcal{CN}(\bm{0}, \sigma^2\bm{I}_n)$ denotes a complex Gaussian random vector whose real and imaginary parts are independent Gaussian random vectors, i.e., $\textnormal{Re}\{\bm{x}\}$, $\textnormal{Im}\{ \bm{x} \} \sim \mathcal{N}(\bm{0}, \frac{1}{2}\sigma^2\bm{I}_n)$. For $Y_1, Y_2, \ldots, Y_{\nu}$ being $\nu$ independent standard normal RVs (i.e., $Y_i \in \mathcal{N}(0,1)$, $\forall i\in[\nu]$), their squared sum $X=\sum_{i=1}^{\nu} Y_i^2$ is a RV that has a chi-squared distribution ($\chi^2$-distribution) with $\nu$ degrees of freedom (DoF) \cite[Ch. 18]{NJKB_Vol_I'94}, which is denoted as $X\sim \chi_{\nu}^2$. For independent $\chi^2$-distributed RVs $X_1\sim \chi_{\nu_1}^2$ and $X_2\sim \chi_{\nu_2}^2$, the ratio $R \eqdef  \frac{X_1/\nu_1}{X_2/\nu_2}$ is a RV that has an $F$-distribution with $\nu_1,\nu_2$ DoF \cite[Ch. 27]{NJKB_Vol_II'95} -- denoted as $F_{\nu_1,\nu_2}$ and hence $R\sim F_{\nu_1,\nu_2}$, whose mean is equated as \cite[eq. (27.6a), p. 326]{NJKB_Vol_II'95}   
\begin{equation}
\label{mean_R_F_distribution}
\mathbb{E}\{R\}= \nu_2/(\nu_2-2), \hspace{2mm}  \nu_2 > 2.
\end{equation}

\section{System Description and Problem Formulation}
\label{sec: systetm_setup}
\subsection{System Model}
\label{subsec: sysetm_model}
Consider a text SemCom system dubbed DeepSC \cite{Xie_DL-based_SemCom'21}, whose transmitter consists of a semantic encoder that extracts semantic information from the source using multiple \textit{Transformer} encoder layers followed by a channel encoder made of dense layers that produce symbols to be transmitted to the DeepSC receiver \cite[Sec. IV]{Xie_DL-based_SemCom'21}. The DeepSC receiver is composed of a channel decoder made of dense layers followed by a semantic decoder built from multiple \textit{Transformer} decoder layers that are employed for symbol detection and text recovery, respectively \cite[Sec. IV]{Xie_DL-based_SemCom'21}. This DeepSC transceiver is assumed to be end-to-end pre-trained and deployed in a wireless environment experiencing narrowband\footnote{RFI can be narrowband, broadband, continuous wave, or pulsed \cite{Getu_dissertation_19}. Without loss of generality, however, we aim to study the performance limits of DeepSC subjected to narrowband RFI from one or more RFI emitters. To this end, our principled probabilistic framework (see Sec. \ref{sec: fundamental_limits_DeepSC}) can also be used to study the performance limits of DeepSC experiencing one or more RFI emitters emitting broadband, continuous wave, or pulsed RFI.} RFI from one or more single-antenna RFI emitters as shown in Fig. \ref{fig: DeepSC_System_Model}.

For $w_l$ being the $l$-th word, let $\bm{s} \eqdef [w_1, w_2, \ldots, w_L]$ be an $L$-word sentence transmitted by the DeepSC transmitter shown in Fig. \ref{fig: DeepSC_System_Model}, where the trained networks of the semantic encoder and channel encoder extract the input text's semantic features and map them into semantic symbols $\bm{x}$, hereinafter called \textit{DeepSC symbols}, given by (via \cite[eq. (1)]{Xie_DL-based_SemCom'21}) 
\begin{equation}
\label{DeepSC_symbols}
\bm{x}= C_{\hat{\bm{\alpha}}}(S_{\hat{\bm{\beta}}}(\bm{s}))\in\mathbb{C}^{1\times KL}, 
\end{equation}
where $K$ is the average number of mapped semantic symbols per word in $\bm{s}$ \cite{Mu_Heterogeneous_Commun_JSAC'23}, $S_{\hat{\bm{\beta}}}(\cdot)$ is the trained semantic encoder network with a parameter set $\hat{\bm{\beta}}$, and $C_{\hat{\bm{\alpha}}}(\cdot)$ is the trained channel encoder network with a parameter set $\hat{\bm{\alpha}}$. We use these networks to realize -- without loss of generality -- the transmission of $\bm{x}$ over an independent Rayleigh fading channel\footnote{Transmitting DeepSC symbols $\bm{x}\in\mathbb{C}^{1\times KL}$ over a Rayleigh fading channel underscores a simplifying assumption that the channel's coherence time is at least $KL$ times the duration of each DeepSC symbol $(\bm{x})_i\in\mathbb{C}$.} $h\sim \mathcal{CN}(0,1)$ in a wireless environment experiencing (MI) RFI from $U$ independent RFI emitters radiating RFI over an independent Rayleigh fading channel\footnote{In reality, an RFI emitter's (e.g., a jammer's) channel may not be precisely known and can vary in time/frequency. However, for a jammer to successfully jam futuristic SemCom systems, it must emit a huge amount of RFI power over a narrowband channel from a relatively stationary position toward the DeepSC's receiving antenna. To underscore this best-case jamming scenario (worst-case SemCom system design challenge), we move forward with Rayleigh fading RFI channels. These channels are also assumed to have a coherence time of at least $KL$ -- w.r.t. the dimension of the $u$-th RFI symbol $\bm{v}_u\in\mathbb{C}^{1 \times KL}$ for $u\in[U]$ -- times the duration of each DeepSC symbol $(\bm{x})_i\in\mathbb{C}$ for $i\in[KL]$. This assumption means that the considered RFI channels are relatively slow fading channels -- underscoring the best-case jamming scenario portrayed in our analytical study.} $g_u\sim \mathcal{CN}(0,1)$, $\forall u\in[U]$. The received DeepSC signal $\bm{y}$ thus becomes
\begin{equation}
	\label{DeepSC_received_signak_model}
	\bm{y}=h\bm{x}+\textstyle\sum_{u=1}^{U} g_u\bm{v}_u+\bm{n}\in\mathbb{C}^{1\times KL}, 
\end{equation}
where $(\bm{x})_i$ is power constrained with respect to (w.r.t.) the maximum transmission power $P_{\textnormal{max}}^s$ W as $\mathbb{E}\{[ \textnormal{Re}\{(\bm{x})_i\} ]^2 \}, \mathbb{E}\{[\textnormal{Im}\{(\bm{x})_i\}]^2 \} \leq P_{\textnormal{max}}^s$, $\forall i\in[KL]$; $\bm{v}_u \in\mathbb{C}^{1 \times KL}$ is a row vector of unknown RFI symbols transmitted by the $u$-th RFI emitter, and each $(\bm{v})_i$ is power constrained w.r.t. its minimum RFI power $P_{\textnormal{min}}^{i,u}$ W and maximum RFI power $P_{\textnormal{max}}^{i,u}$ W as $P_{\textnormal{min}}^{i,u} \leq \mathbb{E}\{[\textnormal{Re}\{(\bm{v}_u)_i\} ]^2 \}, \mathbb{E}\{[ \textnormal{Im}\{(\bm{v}_u)_i\} ]^2 \} \leq P_{\textnormal{max}}^{i,u}$, $\forall i\in[KL]$ and $\forall u\in[U]$; and $\bm{n} \sim \mathcal{CN}(\bm{0}, \sigma^2\bm{I}_{KL})$ is the additive white Gaussian noise (AWGN).\footnote{Realistically, $0 < \sigma^2, P_{\textnormal{max}}^s < \infty$ and $0 < P_{\textnormal{min}}^{i,u}, P_{\textnormal{max}}^{i,u} < \infty$, $\forall u\in [U]$.} As in Fig. \ref{fig: DeepSC_System_Model}, $\bm{y}$ is processed by the trained networks of the channel decoder and semantic decoder to yield a recovered sentence $\hat{\bm{s}}$ given by (via \cite[eq. (3)]{Xie_DL-based_SemCom'21}) 
\begin{equation}
\label{DeepSC_recovered_sentence}
\hat{\bm{s}}= S_{\hat{\bm{\theta}}}(C_{\hat{\bm{\delta}}}(\bm{y})), 
\end{equation}
where $C_{\hat{\bm{\delta}}}(\cdot)$ and $S_{\hat{\bm{\theta}}}(\cdot)$ denote the trained networks of the channel decoder and semantic decoder with parameter sets $\hat{\bm{\delta}}$ and $\hat{\bm{\theta}}$, respectively. Regarding the recovered sentence $\hat{\bm{s}}$ and the transmitted sentence $\bm{s}$, we proceed to formulate problems.    
\subsection{Problem Formulation}
\label{subsec: Analytical_motivation}
DeepSC's performance can be assessed by semantic similarity between $\bm{s}$ and $\hat{\bm{s}}$ as \cite[eq. (1)]{Mu_Heterogeneous_Commun_JSAC'23}
\begin{equation}
\label{semantic_similarity_expression_1}
\eta (\bm{s}, \hat{\bm{s}}) \eqdef \bm{B}(\bm{s})\bm{B}(\hat{\bm{s}})^T  \big( \| \bm{B}(\bm{s})\| \| \bm{B}(\hat{\bm{s}})\| \big)^{-1}  ,
\end{equation}
where $\eta (\bm{s}, \hat{\bm{s}})\in [0, 1]$ quantifies semantic similarity -- 0 and 1 indicate \textit{semantic irrelevance} and \textit{semantic consistency} \cite{Jiang_Reliable_SemCom'22}, respectively -- and $\bm{B}(\cdot)$ denotes the output of BERT (bidirectional encoder representations from Transformers \cite{Devlin’19_BERT}).\footnote{BERT is a gigantic pre-trained model comprising billions of parameters that are used to mine semantic information \cite{Xie_DL-based_SemCom'21}.} The metric $\eta (\bm{s}, \hat{\bm{s}})$ depends on $K$ and the signal-to-noise ratio (SNR) $\gamma$ \cite{Xie_DL-based_SemCom'21}. Hence, $\eta (\bm{s}, \hat{\bm{s}})$ can be expressed via the semantic similarity function $\varepsilon(K,\gamma)$ as \cite{Mu_Heterogeneous_Commun_JSAC'23}
\begin{equation}
\label{simantic_similarity_function_1}
\eta (\bm{s}, \hat{\bm{s}})=\varepsilon(K,\gamma).
\end{equation}

The authors of \cite{Mu_Heterogeneous_Commun_JSAC'23} exploited the \textit{generalized logistic function} to approximate $\varepsilon(K,\gamma)$ for any given $K$ as \cite[eq. (3)]{Mu_Heterogeneous_Commun_JSAC'23}
\begin{equation}
\label{simantic_similarity_function_approximation_1}
\varepsilon(K,\gamma) \approx\tilde{\varepsilon}_K(\gamma) = A_{K,1}+ \frac{A_{K,2}-A_{K,1}}{1+ e^{-(C_{K,1}\gamma+C_{K,2})}}, 
\end{equation}
where $A_{K,1}>0$, $A_{K,2}>0$, and $C_{K,1}>0$ are the lower asymptote, upper asymptote, and logistic growth rate, respectively; $C_{K,2}$ controls the logistic midpoint \cite{Mu_Heterogeneous_Commun_JSAC'23}. In light of (\ref{simantic_similarity_function_approximation_1}), $0\leq \varepsilon(K,\gamma) \leq 1$, $\kappa \eqdef A_{K,1}/(A_{K,1}-A_{K,2}) \geq 0$, and the generalized logistic function renders an accurate approximation for any $K$, as demonstrated in \cite[Fig. 2]{Mu_Heterogeneous_Commun_JSAC'23}. Meanwhile, employing (\ref{simantic_similarity_function_approximation_1}) in the right-hand side (RHS) of (\ref{simantic_similarity_function_1}), the following \textit{highly-accurate} approximation ensues:
\begin{equation}
	\label{simantic_similarity_function_2}
	\eta (\bm{s}, \hat{\bm{s}}) \approx  A_{K,1}+ (A_{K,2}-A_{K,1}) \big( 1+ e^{-(C_{K,1}\gamma+C_{K,2})}\big)^{-1}.
\end{equation}
DeepSC's asymptotic performance analysis is mathematically tractable via (\ref{simantic_similarity_function_2}), which paves the way for DeepSC's performance limits. To reveal these limits using \textit{probability as a lens}, we introduce a new semantic metric -- named the upper tail probability of a semantic similarity -- applicable for evaluating the performance of text SemCom systems, as defined below.
\begin{definition}
	\label{Tail_probability_definition}
	The upper tail probability of $\eta (\bm{s}, \hat{\bm{s}})$ w.r.t. $\eta_{\textnormal{min}}\in[0,1]$ is computed as
	\begin{equation}
		\label{tail_probability_defn}
		p(\eta_{\textnormal{min}}) \eqdef \mathbb{P}\big( \eta (\bm{s}, \hat{\bm{s}}) \geq \eta_{\textnormal{min}} \big),  
	\end{equation}
	where the metric $\eta (\bm{s}, \hat{\bm{s}})$ is defined in (\ref{semantic_similarity_expression_1}) and $\eta_{\textnormal{min}}$ is the minimum semantic similarity. 
\end{definition}
Regarding (\ref{tail_probability_defn}), $p(0)\eqdef\mathbb{P}\big( \eta (\bm{s}, \hat{\bm{s}}) \geq 0 \big)$ is a probabilistic metric\footnote{$p(\eta_{\textnormal{min}})$ can be used to optimize performance in SemCom networks \cite{Getu_IEEE_Access'23}.} that quantifies the probability of semantic similarity by a SemCom technique being greater than or equal to 0. Using $p(0)$ and $p(\eta_{\textnormal{min}})$, we hereunder formulate problems. 
\subsubsection{Problems on the Asymptotic Performance Analysis of DeepSC under RFI} To study the asymptotic performance of DeepSC under infinitesimally small RFI and a very low SNR, we formulate the following problem:
\begin{problem}
\label{problem_under_infinitesimally_small_RFI}
Characterize $\lim_{\sigma^2 \to \infty} p(0)$ and $\lim_{P_{\textnormal{max}}^s\to 0 } p(0)$. 
\end{problem}
Solving Problem \ref{problem_under_infinitesimally_small_RFI} will help us understand the asymptotic performance of DeepSC -- w.r.t. infinitesimally small RFI -- for very low SNR regimes. These are for a very large noise power and a very small maximum transmission power of the DeepSC symbols, as captured by Problem \ref{problem_under_infinitesimally_small_RFI}. On the other hand, to investigate the asymptotic performance of DeepSC under RFI and a very low signal-to-interference-plus-noise ratio (SINR), we formulate the underneath problem w.r.t. $P_{\textnormal{min}}^i \eqdef P_{\textnormal{min}}^{i,1}$.

\begin{problem}
\label{prob_under_single_interferer_RFI}
Derive  $\lim_{P_{\textnormal{min}}^i\to \infty } p(0)$ and    $\lim_{P_{\textnormal{max}}^s\to 0 } p(0)$.
\end{problem}
Solving Problem \ref{prob_under_single_interferer_RFI} will help us understand the asymptotic performance of DeepSC under RFI for very low SINR regimes. Such regimes can be for a very large power of the emitted RFI and a very small maximum transmission power of the DeepSC symbols, as captured by Problem \ref{prob_under_single_interferer_RFI}. Next, we proceed to our third set of formulated problems.

\subsubsection{Problems on the Asymptotic Performance Analysis of DeepSC under MI RFI}
To understand DeepSC's asymptotic performance under MI RFI and a very low SINR, we formulate the ensuing problem for $U>1$ and $\tilde{P}_{\textnormal{min}}^i \eqdef \textnormal{min} \big(P_{\textnormal{min}}^{i,1}, P_{\textnormal{min}}^{i,2}, \ldots, P_{\textnormal{min}}^{i,U}\big)$.
\begin{problem}
\label{prob_under_multi_interferer_RFI}
Derive 
$\displaystyle \lim_{P_{\textnormal{max}}^s \to 0 } p(0)$, $\displaystyle \lim_{\tilde{P}_{\textnormal{min}}^i \to \infty } p(0)$, and $\displaystyle \lim_{U\to \infty } p(0)$.
\end{problem}
Solving Problem \ref{prob_under_multi_interferer_RFI} will help us understand the asymptotic performance of DeepSC under MI RFI for very low SINR regimes. Such regimes can be for a very small maximum transmission power of the DeepSC symbols, a large power of all RFI emitters, and an enormous number of RFI emitters, as captured by Problem \ref{prob_under_multi_interferer_RFI}. Apart from Problems \ref{problem_under_infinitesimally_small_RFI}-\ref{prob_under_multi_interferer_RFI},\linebreak we also formulate the following novel problems.

\subsubsection{Problems on the Non-Asymptotic Performance Analysis of DeepSC under MI RFI}
We follow up with Problems \ref{prob_practical_limits} and \ref{prob_outage_probability} on DeepSC's non-asymptotic performance.
\begin{problem}
\label{prob_practical_limits}
For a generic $\eta_{\textnormal{min}} \in [0,1]$, derive the practical limits of DeepSC w.r.t. $p(\eta_{\textnormal{min}})$.
\end{problem}
\begin{problem}
\label{prob_outage_probability}
Derive the outage probability of DeepSC.
\end{problem}

Problems \ref{problem_under_infinitesimally_small_RFI}-\ref{prob_outage_probability} are novel problems considering the fact that the performance analysis of any SemCom system is fundamentally challenging, as noted in Sec. \ref{sec: introduction}. Overcoming this limitation, we present the performance limits of DeepSC. 
\section{Performance Limits of DeepSC}
\label{sec: fundamental_limits_DeepSC} 
\subsection{Performance Limits of DeepSC Subjected to RFI}
\label{subsec: fundamental_limits_under_RFI}
We use (\ref{simantic_similarity_function_2}) and (\ref{tail_probability_defn}) w.r.t. the system model in Sec. \ref{subsec: sysetm_model} to derive the following theorem.
\begin{theorem}
\label{thm: fund_limit_under_no_interference}
Per the approximation in (\ref{simantic_similarity_function_2}) and Sec. \ref{subsec: sysetm_model}'s settings, DeepSC manifests the following performance limits -- for a given $K$ -- under infinitesimally small RFI: $i) \lim_{\sigma^2 \to \infty} p(0)   = 0$; $ii) \lim_{P_{\textnormal{max}}^s\to 0 } p(0) = 0 $, where these results are valid for $\alpha \leq \kappa \leq 1$ given $\alpha \eqdef e^{C_{K,2}}/(1+e^{C_{K,2}})$ and $\kappa \eqdef A_{K,1}/ (A_{K,1}-A_{K,2})$.
  
\proof The proof is in Appendix \ref{sec: proof_fund_limit_under_no_interference}. 
\end{theorem}

Per Theorem \ref{thm: fund_limit_under_no_interference}, $\lim_{\sigma^2 \to \infty} p(0)=0$ and $\lim_{P_{\textnormal{max}}^s\to 0 } p(0)=0$\linebreak corroborate -- whenever $\sigma^2 \to \infty$ and $P_{\textnormal{max}}^s\to 0$ -- that $\eta (\bm{s}, \hat{\bm{s}}) = 0$, which affirms \textit{maximum semantic dissimilarity} (or semantic irrelevance) \cite{Jiang_Reliable_SemCom'22}. Consequently, Theorem \ref{thm: fund_limit_under_no_interference} translates to the following remarks.
\begin{remark}
\label{remark_insight_with_no_RFI}
DeepSC -- per Sec. \ref{subsec: sysetm_model}'s settings  -- exhibits the following performance limits when it is subjected to infinitesimally small RFI: $i)$ DeepSC will generate semantically irrelevant sentences as the noise power gets large; $ii)$ DeepSC will generate semantically irrelevant sentences as the DeepSC symbols' maximum transmission power tends to zero Watt (W). 
\end{remark}
\begin{remark}
\label{remark_insight_with_no_RFI_2}
Contrary to the analytically unsubstantiated sentiment that SemCom techniques (such as DeepSC) work well in very low SNR regimes while outperforming the techniques of conventional communication systems, Theorem \ref{thm: fund_limit_under_no_interference} asserts that DeepSC will generate semantically irrelevant sentences as the noise power gets large and the DeepSC symbols' maximum transmission power approaches zero W.
\end{remark}

Similarly, the impact of moderate/strong RFI on DeepSC's performance is quantified below.  
\begin{theorem}
\label{thm: fund_limit_with_interference}
According to Sec. \ref{subsec: sysetm_model}'s settings and the approximation in (\ref{simantic_similarity_function_2}), DeepSC manifests the following performance limits -- for a given $K$ -- under RFI: $i) \lim_{P_{\textnormal{min}}^i\to \infty } p(0) = 0$; $ii) \lim_{P_{\textnormal{max}}^s\to 0 } p(0) = 0$, where these results are true for $\alpha \leq \kappa \leq 1$. 
  
\proof The proof is in Appendix \ref{sec: proof_fund_limit_with_interference}. 
\end{theorem}

Theorem \ref{thm: fund_limit_with_interference} asserts the following insight. 
\begin{remark}
\label{remark_insight_with_RFI}
DeepSC -- per Sec. \ref{subsec: sysetm_model}'s settings  -- displays the following performance limits when it is subjected to RFI: $i)$ DeepSC will recover semantically irrelevant sentences as the DeepSC symbols' maximum transmission power nears zero W; $ii)$ DeepSC will produce semantically irrelevant sentences as the power emitted by RFI becomes large, which attests to the fact that strong RFI can destroy the faithfulness of SemCom by producing a huge amount of semantic noise.
\end{remark}

Strong RFI emitter radiates significant semantic noise, which can affect the reliability of SemCom that can be decimated by MI RFI. Therefore, we henceforth expose the performance limits of DeepSC subjected to MI RFI.
\subsection{Performance Limits of DeepSC Subjected to MI RFI}
\label{subsec: fundamental_limits_with_MU-RFI}
Considering MI RFI modeled as in Sec. \ref{subsec: sysetm_model}, the approximation in (\ref{simantic_similarity_function_2}), and (\ref{tail_probability_defn}), we derive the performance limits of DeepSC, as formalized in the following theorem.

\begin{theorem}
	\label{thm: fund_limit_with_MU-interference}
	Pursuant to Sec. \ref{subsec: sysetm_model}'s settings and the approximation in (\ref{simantic_similarity_function_2}), DeepSC exhibits the following performance limits -- for a given $K$ -- under MI RFI: $i) \lim_{P_{\textnormal{max}}^s \to 0 } p(0) = 0$; $ii) \lim_{\tilde{P}_{\textnormal{min}}^i \to \infty } p(0)  = 0$; $iii) \lim_{U\to \infty } p(0)  = 0$, where these results are valid for $\alpha \leq \kappa \leq 1$, $U>1$, and $P_{\textnormal{max}}^s \leq \beta (U-1)\tilde{P}_{\textnormal{min}}^i$ such that $\beta =  \ln\big[ \kappa/(1-\kappa)\big]/C_{K,1} -C_{K,2}/C_{K,1}$.
	\proof The proof is in Appendix \ref{sec: proof_fund_limit_with_MU-interference}. 
\end{theorem}

The following intuitive remark stems from Theorem \ref{thm: fund_limit_with_MU-interference}.
\begin{remark}
\label{remark_concerning_multi-user_RFI}
DeepSC -- per Sec. \ref{subsec: sysetm_model}'s settings  -- manifests the following performance limits when it is subjected to MI RFI (i.e., $U\geq 2$): $i)$ DeepSC will generate semantically irrelevant sentences as the DeepSC symbols' maximum transmission power tends to zero W; $ii)$ DeepSC will produce semantically irrelevant sentences as all RFI emitters get strong; $iii)$ DeepSC will produce semantically irrelevant sentences as the number of RFI emitters becomes enormous (i.e., $U\to\infty$).
\end{remark}

Summarizing, the following remarks ensue.
\begin{remark}
\label{summary_remark_1}
Although the performance analyses (Appendices \ref{sec: proof_fund_limit_under_no_interference}-\ref{sec: proof_fund_limit_with_MU-interference}) that led to the performance limits per Theorems \ref{thm: fund_limit_under_no_interference}-\ref{thm: fund_limit_with_MU-interference} are regarding DeepSC, the introduced probabilistic framework -- with the proposed semantic metric -- is particularly applicable to many text SemCom techniques and generally relevant in speech SemCom, image SemCom, and video SemCom.  
\end{remark}
\begin{remark}
\label{summary_remark_2}
In view of a \textit{wireless attack} with RFI that changes the semantics of information transmitted using SemCom \cite{Sagduyu_Is_SemCom_Secure'22}, Theorems \ref{thm: fund_limit_with_interference} and \ref{thm: fund_limit_with_MU-interference} underscore DeepSC's security vulnerability.
\end{remark}
Aside from Theorems \ref{thm: fund_limit_under_no_interference}-\ref{thm: fund_limit_with_MU-interference}, DeepSC's practical limits and outage probability provide useful insights, as detailed below.

\section{The Practical Limits and Outage Probability of DeepSC}
\label{sec: practical_limits_outage_probability_DeepSC}
\subsection{The Practical Limits of DeepSC under MI RFI}
\label{subsec: practical_limits}
In the practical design of DeepSC, $\eta_{\textnormal{min}} \geq 0.8$ can be a desirable range. Hence, understanding the practical limits of DeepSC for a given $\eta_{\textnormal{min}} \in [0,1]$ and $K$ is crucial -- leading to DeepSC's practical limits, as formalized below.
\begin{theorem}
\label{thm: practical_limits_with_MU-interference}
Per Sec. \ref{subsec: sysetm_model}'s settings and the approximation in (\ref{simantic_similarity_function_2}), DeepSC manifests the following performance limits for a given $\eta_{\textnormal{min}}$ and $K$: $i)$ $p(\eta_{\textnormal{min}}) \leq P_{\textnormal{max}}^s/[\beta_K(\eta_{\textnormal{min}}) (U-1)\tilde{P}_{\textnormal{min}}^i]$; $ii)$ $\lim_{P_{\textnormal{max}}^s \to 0} p(\eta_{\textnormal{min}})=0$; $iii)$ $\lim_{\tilde{P}_{\textnormal{min}}^i \to \infty} p(\eta_{\textnormal{min}})=0$; $iv)$ $\lim_{U \to \infty} p(\eta_{\textnormal{min}})=0$, where these results are justified for $U>1$, $\alpha \leq \kappa_K(\eta_{\textnormal{min}}) \leq 1$, and $P_{\textnormal{max}}^s \leq \beta_K(\eta_{\textnormal{min}}) (U-1) \tilde{P}_{\textnormal{min}}^i$ such that $\kappa_K(\eta_{\textnormal{min}}) \eqdef (\eta_{\textnormal{min}}-A_{K,1})/(A_{K,2}-A_{K,1})$ and $\beta_K(\eta_{\textnormal{min}})  \eqdef \ln[ \kappa_K(\eta_{\textnormal{min}})/(1-\kappa_K(\eta_{\textnormal{min}}))]/C_{K,1} - C_{K,2}/C_{K,1}$.
	\proof The proof is deferred to Appendix \ref{sec: proof_practical_limits_with_MU-interference}. 
\end{theorem}
\begin{remark}
Under the satisfaction of Theorem \ref{thm: practical_limits_with_MU-interference}'s conditions and regardless of $\eta_{\textnormal{min}} \in [0,1]$, DeepSC under MI RFI manifests the performance limits $\lim_{P_{\textnormal{max}}^s \to 0} p(\eta_{\textnormal{min}})=0$, $\lim_{\tilde{P}_{\textnormal{min}}^i \to \infty} p(\eta_{\textnormal{min}})=0$, and $\lim_{U \to \infty} p(\eta_{\textnormal{min}})=0$.
\end{remark}
\begin{remark}
Quantifying the practical limits of DeepSC under MI RFI, Theorem \ref{thm: practical_limits_with_MU-interference} subsumes Theorem \ref{thm: fund_limit_with_MU-interference} -- a generalization of Theorem \ref{thm: fund_limit_with_interference}.
\end{remark}
Via Theorem \ref{thm: practical_limits_with_MU-interference}, DeepSC's outage probability is bounded.

\subsection{The Outage Probability of DeepSC under MI RFI}
\label{subsec: DeepSC_outage_probability}
Adopting a popular performance metric of wireless communication \cite{Simon_Alouni_DC_over_Fading_Channels'05}, DeepSC's outage probability is defined as
\begin{equation}
\label{outage_prob_defn}
P_{\textnormal{out}}(\eta_{\textnormal{min}}) \eqdef  \mathbb{P}\big( \eta (\bm{s}, \hat{\bm{s}}) < \eta_{\textnormal{min}} \big) \stackrel{(a)}{=} 1- p(\eta_{\textnormal{min}}),
\end{equation}
where $(a)$ is due to (\ref{tail_probability_defn}) and (\ref{outage_prob_defn}) leads to the following lemma.
\begin{lemma}
\label{DeepSC_outage_probability_lower_bound}
Per Sec. \ref{subsec: sysetm_model}'s settings and the approximation in (\ref{simantic_similarity_function_2}), the outage probability of DeepSC -- for a given $\eta_{\textnormal{min}}$ and $K$ -- under MI RFI satisfies the bound: $P_{\textnormal{out}}(\eta_{\textnormal{min}}) \geq 1-P_{\textnormal{max}}^s/[\beta_K(\eta_{\textnormal{min}}) (U-1)\tilde{P}_{\textnormal{min}}^i]$, where this is valid for\linebreak $\alpha < \kappa_K(\eta_{\textnormal{min}}) \leq 1$ and $P_{\textnormal{max}}^s \leq \beta_K(\eta_{\textnormal{min}}) (U-1) \tilde{P}_{\textnormal{min}}^i$.

\proof The bound follows from (\ref{outage_prob_defn}) and Theorem \ref{thm: practical_limits_with_MU-interference}.   \QEDclosed 
\end{lemma}
Deploying Lemma \ref{DeepSC_outage_probability_lower_bound}, the underneath optimization ensues.
\subsection{Optimization Based On the Outage Probability of DeepSC}
\label{subsec: DeepSC_outage_probability_opt}
According to (\ref{outage_prob_defn}), the $K$ that minimizes $P_{\textnormal{out}}(\eta_{\textnormal{min}})$ is the one that maximizes $p(\eta_{\textnormal{min}})$ given $K \in  \{1, 2, \ldots, K_{\textnormal{max}} \}$. Thus, the following optimization problem ensues:
\begin{equation}
\label{opt_K_outage_prob_1}
\argmax_{K}  \quad    p(\eta_{\textnormal{min}})  \hspace{3mm} \textnormal{s.t.}  \hspace{3mm}   0 \leq \eta_{\textnormal{min}} \leq 1; \hspace{2mm} K \leq K_{\textnormal{max}}.
\end{equation}
Consequently, the underneath lemma follows.
\begin{lemma}
\label{lemma: K_opt}
The solution of (\ref{opt_K_outage_prob_1}) is $K^{\star}= K_{\textnormal{max}}$.

\proof Using Theorem \ref{thm: practical_limits_with_MU-interference}, (\ref{opt_K_outage_prob_1}) can also be cast as 
\begin{subequations}
	\begin{align}
		\label{opt_K_outage_prob_1_1}
		\argmax_{K}  \quad &   P_{\textnormal{max}}^s/[\beta_K(\eta_{\textnormal{min}}) (U-1)\tilde{P}_{\textnormal{min}}^i]    \\
		\label{opt_K_outage_prob_2_1}
		\textnormal{s.t.}  \quad &  0 \leq \eta_{\textnormal{min}} \leq 1; \hspace{2mm} K \leq K_{\textnormal{max}}.
	\end{align}
\end{subequations}
Discarding constants from the optimization objective in (\ref{opt_K_outage_prob_1_1}) produces an equivalent optimization problem given by
\begin{equation}
\label{opt_K_outage_prob_1_2}
\argmin_{K}  \quad    \beta_K(\eta_{\textnormal{min}}) \hspace{3mm} \textnormal{s.t.}  \hspace{3mm} 0 \leq \eta_{\textnormal{min}} \leq 1; \hspace{2mm} K \leq K_{\textnormal{max}}.
\end{equation}
As it can be seen in \cite[Fig. 2]{Mu_Heterogeneous_Commun_JSAC'23}, $\tilde{\varepsilon}_K(\gamma) \geq \tilde{\varepsilon}_{\tilde{K}}(\gamma)$ for any $\gamma$ and $K > \tilde{K}$. For any $K > \tilde{K}$ and $\gamma$, thus, $\mathbb{P}( \tilde{\varepsilon}_K(\gamma) \geq \eta_{\textnormal{min}} ) \geq  \mathbb{P}( \tilde{\varepsilon}_{\tilde{K}}(\gamma) \geq \eta_{\textnormal{min}} )$, and it follows from (\ref{practical_limits_with_RFI_1})-(\ref{practical_limits_with_RFI_7}) that  
\begin{equation}
	\label{prob_comparison_1}
	\mathbb{P}\big( \gamma \geq \beta_K(\eta_{\textnormal{min}})  \big) \geq   \mathbb{P}\big( \gamma \geq \beta_{\tilde{K}}(\eta_{\textnormal{min}})  \big), 
\end{equation}
where $\beta_{\tilde{K}}(\eta_{\textnormal{min}}) = \beta_K(\eta_{\textnormal{min}}) \big|_{K=\tilde{K}}$. Hence, it follows from (\ref{prob_comparison_1}) that $\beta_K(\eta_{\textnormal{min}}) \leq \beta_{\tilde{K}}(\eta_{\textnormal{min}})$ w.r.t. all $K > \tilde{K}$ and hence $\beta_{K_{\textnormal{max}}}(\eta_{\textnormal{min}}) \leq \beta_{K_{\textnormal{max}}-1}(\eta_{\textnormal{min}}) \leq \ldots \leq \beta_K(\eta_{\textnormal{min}}) \leq \ldots \leq \beta_1(\eta_{\textnormal{min}})$. Using this relation in (\ref{opt_K_outage_prob_1_2}), the optimal $K$ that minimizes the outage probability of DeepSC is $K^{\star}= K_{\textnormal{max}}$. This ends the proof of Lemma \ref{lemma: K_opt}.\QEDclosed  
\end{lemma}
\begin{remark}
\label{rem: K_opt}
The optimal $K$ that minimizes the outage probability of DeepSC subjected to MI RFI is $K_{\textnormal{max}}$.
\end{remark}
\begin{figure}[htb!]
	\centering
	\vspace{-0.68cm}
	\includegraphics[scale=0.25]{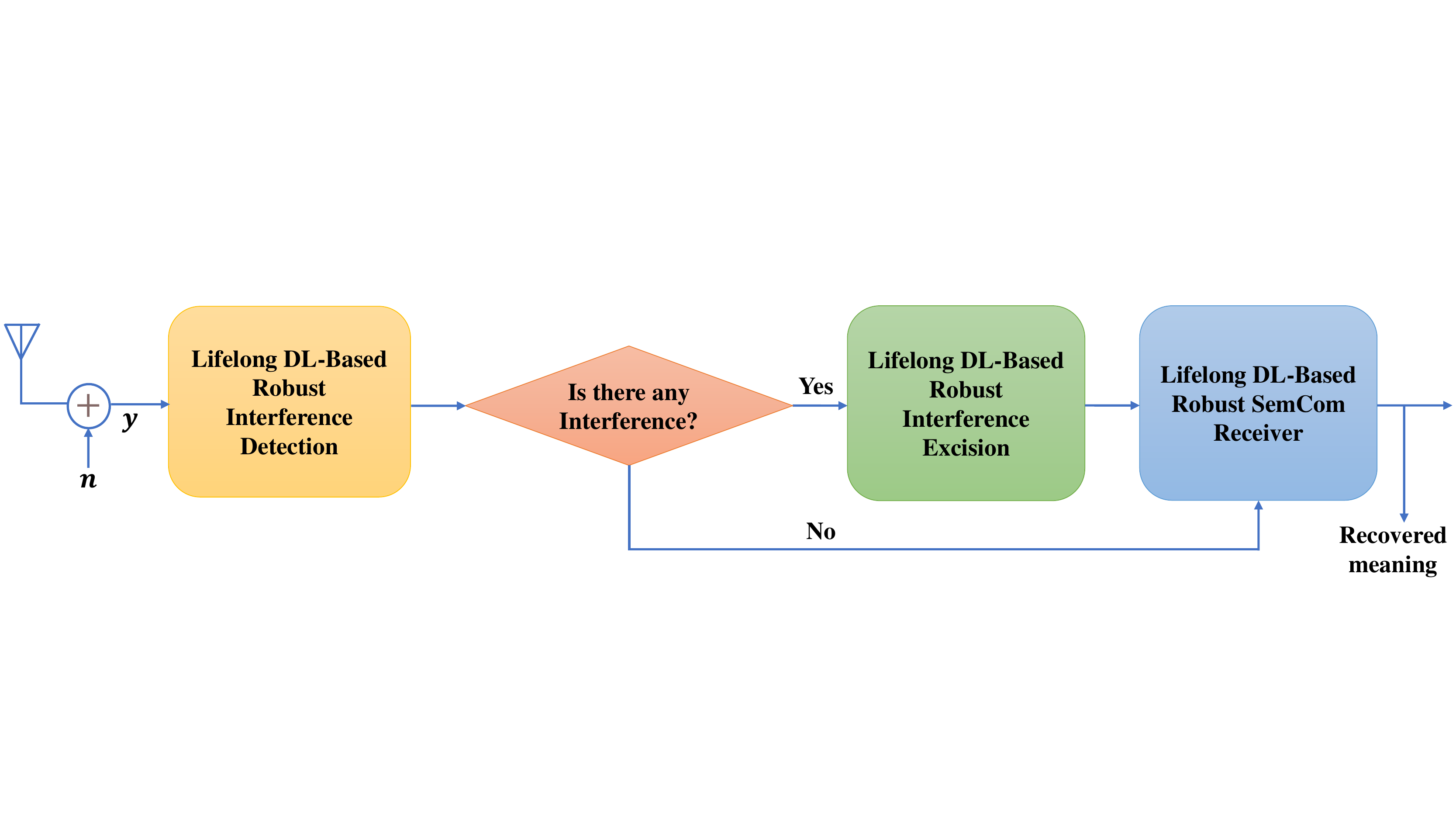}  \vspace{-1.0cm} 
	\caption{A (generic) lifelong DL-based IR$^2$ SemCom system.}
	\label{fig: LL_DL_Based_IR2_SemCom_20230812}
\end{figure}  
\section{Toward IR$^2$ 6G Wireless Systems}
\label{sec: IR2_SemCom}
In radio frequency (RF) operating systems such as radio astronomy, microwave radiometry, and satellite communication, RFI causes performance loss for orbital and terrestrial interferers, malicious interferers, spoofing attacks, and RF jammers \cite{Getu_dissertation_19}. RFI also impacts wireless systems based on ultra-wideband communication, radars, and cognitive radios due to wideband interferers, wideband jammers, and imperfect spectrum sensing, respectively \cite{Getu_dissertation_19}. By the same token, interference is also a big concern for medical devices such as medical implants \cite{Gollakota_dissertation_13}, public safety networks \cite{Tech_Report_NIST-1885_2015}, wireless body area networks \cite{Tech_Report_NIST-1885_2015}, and critical infrastructure such as Smart Grid, smart manufacturing, and the healthcare industry \cite{Tech_Report_NIST-1885_2015}.\linebreak These use cases of interference and RFI in the aforementioned contemporary wireless systems underscore the need for IR$^2$ wireless systems in 6G and beyond. For 6G and beyond, on the other hand, SemCom holds promise in minimizing bandwidth consumption, power usage, and transmission delay. Accordingly, there is a need for IR$^2$ SemCom systems as such systems are also impacted by RFI\footnote{RFI impacts bit-based communication (BitCom) and SemCom systems differently. In BitCom, huge RFI introduces huge noise to the channel encoder which then impacts the detector's (mainly) linear mapping from symbols to bits. Whereas in SemCom, large RFI causes large noise to the channel encoder which then introduces large semantic noise to the semantic decoder's highly non-linear mapping from semantic symbols to semantic meaning.}, in particular, and interference, in general.

Inspired by the traditional interference-resistant wireless communication advocated in \cite{Getu_dissertation_19} and state-of-the-art advancements in lifelong DL \cite{Synthe_lect_DLLR_16,DeLange_CL_Survey'22,Masana_Class_IL'20}, we propose a \textit{generic lifelong DL-based IR$^2$ SemCom system} schematized in Fig. \ref{fig: LL_DL_Based_IR2_SemCom_20230812}. As seen in Fig. \ref{fig: LL_DL_Based_IR2_SemCom_20230812}, the lifelong DL-based robust interference detection module learns the presence of any interference in real-time by learning multiple distributions of interference on a continual basis. If this module flags the presence of interference, the lifelong DL-based robust interference excision module excises the received interference in real-time -- irrespective of its angle of arrival (AoA) -- while learning the AoA of interference in a lifelong manner. The output of this module is then fed to the lifelong DL-based robust SemCom receiver, which is designed to accurately recover the meaning of the transmitted message. 

To gain insight into the performance of a lifelong DL-based IR$^2$ DeepSC per Fig. \ref{fig: LL_DL_Based_IR2_SemCom_20230812}, let $\breve{p}(\eta_{\textnormal{min}}) \eqdef \mathbb{P} ( \varepsilon(K,\breve{\gamma}) \geq \eta_{\textnormal{min}} )$ be its exhibited SemCom performance given that $\breve{\gamma}$ is the SINR after applying a lifelong DL-based robust interference excision. Suppose the optimal lifelong DL network at the $t$-th deployment instant for a robust interference excision is $\bm{\Phi}_t^{\star}(\cdot)$. Hence, the signal output after a robust interference excision is $\bm{\Phi}_t^{\star}(\bm{y})$. For $\bm{y}$ defined in (\ref{DeepSC_received_signak_model}), let us presume that $\bm{\Phi}_t^{\star}(h\bm{x}+\textstyle\sum_{u=1}^{U} g_u\bm{v}_u+\bm{n}) \approx \bm{\Phi}_{t,1}^{\star}(h\bm{x}) + \bm{\Phi}_{t,2}^{\star}(\textstyle\sum_{u=1}^{U} g_u\bm{v}_u) + \bm{\Phi}_{t,3}^{\star}(\bm{n})$ for $  \big\{ \bm{\Phi}_{t,j}^{\star}(\cdot) \big\}_{j=1}^3 $ being the respective lifelong DL networks for the received signal, impinging RFI, and noise during the $t$-th deployment instant. W.r.t. this approximation, the underneath corollary asserts the performance gain of the lifelong DL-based\linebreak IR$^2$ DeepSC over DeepSC under large $U$.
\begin{corollary}
\label{lemma: IR2_SemCom}
Assume a lifelong DL-based IR$^2$ DeepSC per Fig. \ref{fig: LL_DL_Based_IR2_SemCom_20230812} that is equipped with a perfect interference detector. If $\bm{\Phi}_{t,2}^{\star}(\textstyle\sum_{u=1}^{U} g_u\bm{v}_u) \approx \bm{0}$, $\forall t \in \mathbb{N}$, and $ 0 < (\bm{\Phi}_{t,1}^{\star}(h\bm{x}))_i, (\bm{\Phi}_{t,3}^{\star}(\bm{n}))_i  < \infty$, $\forall t \in \mathbb{N}$ and $i \in [KL]$, the following are true with a high probability:    
\begin{itemize}
	\item $\displaystyle\lim_{U\to \infty}\breve{p}(\eta_{\textnormal{min}}) >> \displaystyle\lim_{U\to \infty} p(\eta_{\textnormal{min}})$ for high SNR regimes;  
	\item $\displaystyle\lim_{U\to \infty}\breve{p}(\eta_{\textnormal{min}}) \geq \displaystyle\lim_{U\to \infty} p(\eta_{\textnormal{min}})$ for low SNR regimes.
\end{itemize}

\proof It is provided in our preprint in \cite[Appendix E]{arXiv_Getu_DeepSC_Performance_Limits'23_v2}.  
\end{corollary} 
\begin{figure*}[t!]
	\begin{minipage}[htb]{0.46\linewidth}
		\centering
		\includegraphics[width=\textwidth]{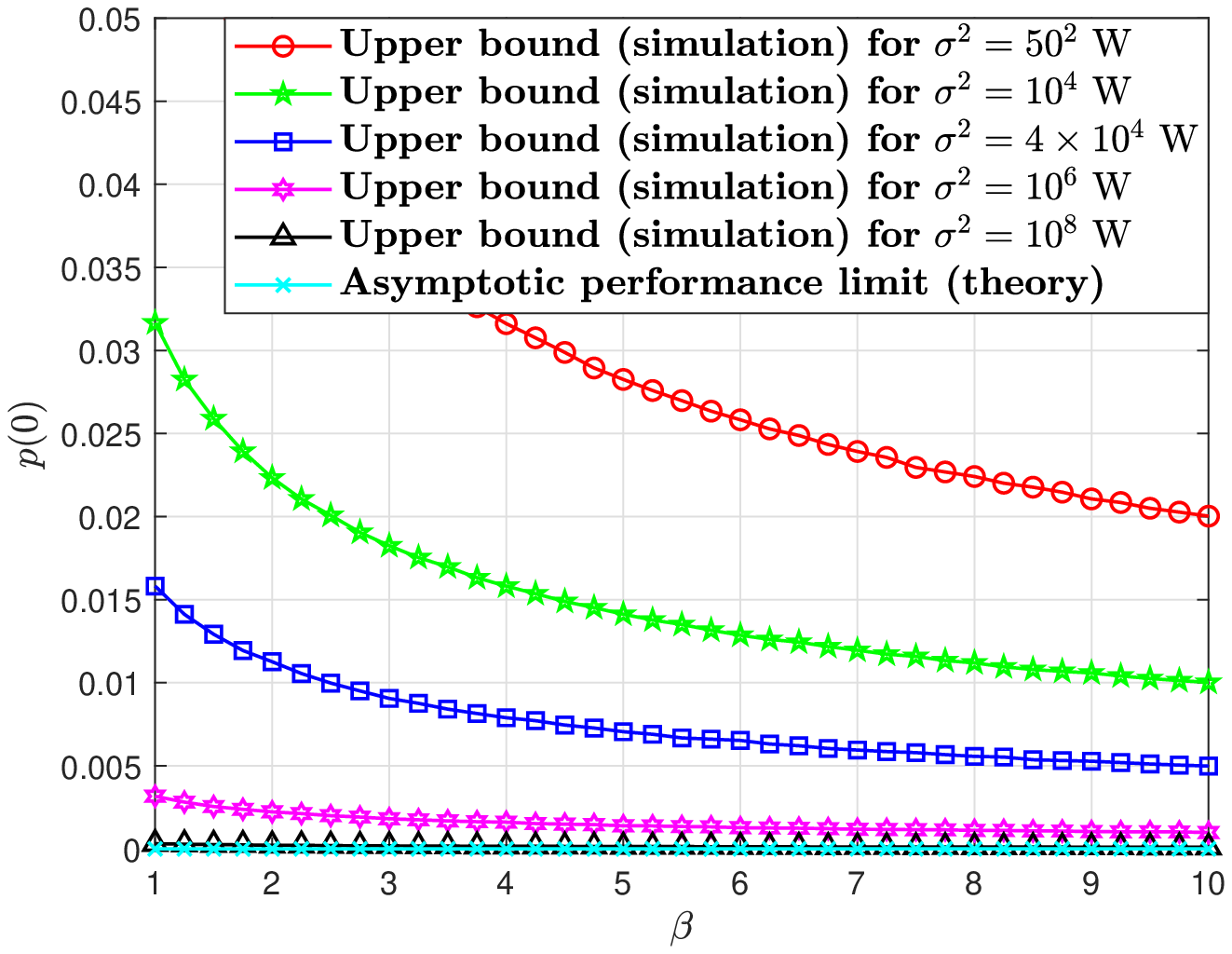}
		\caption{$p(0)$ versus $\beta$ under infinitesimally small RFI, fixed $P_{\textnormal{max}}^s=5$ W, and varying $\sigma^2$: $N=10^7$.    }
		\label{fig: Perf_comparison_plot_wrt_no_RFI_fixed_P_max_s_varying_sigma}
	\end{minipage}
	\hspace{0.5cm}
	\begin{minipage}[htb]{0.46\linewidth}
		\centering
		\includegraphics[width=\textwidth]{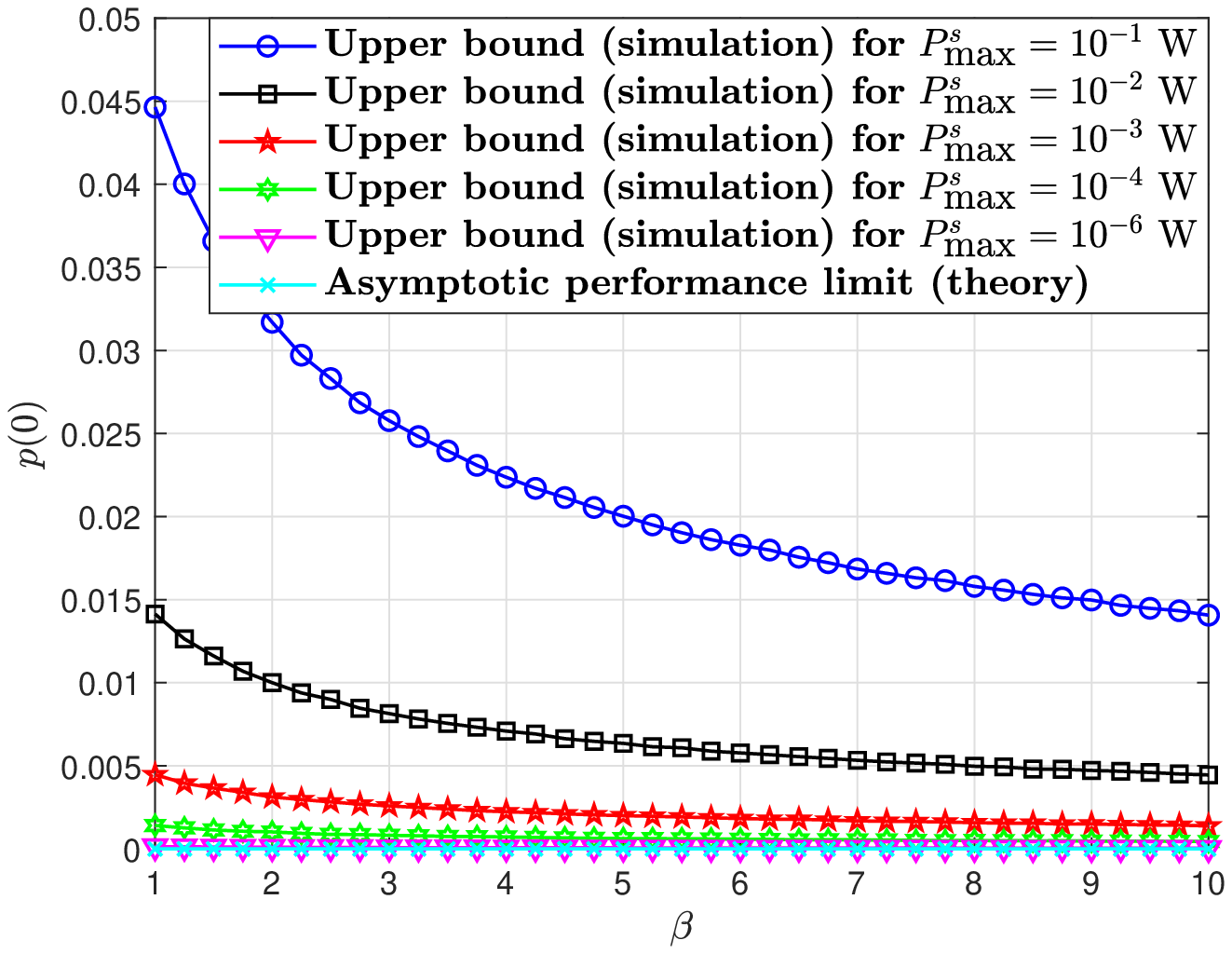}
		\caption{$p(0)$ versus $\beta$ under infinitesimally small RFI, fixed $\sigma^2=100$ W, and varying $P_{\textnormal{max}}^s$: $N=10^7$.  }
		\label{fig: Perf_comparison_plot_wrt_no_RFI_fixed_sigma_varying_P_max_s}
	\end{minipage}
\end{figure*}
\section{Simulation and Computer Experiment Results}
\label{sec: simulation_results}
Simulation results validating Theorems \ref{thm: fund_limit_under_no_interference}-\ref{thm: fund_limit_with_MU-interference} are presented in this section, which also presents computer experiment results that substantiate our theory. 
\subsection{Simulation Results with Infinitesimally Small RFI}
\label{subsec: results_under_an_infinitesimally_small_RFI}
W.r.t. the infinitesimally small RFI examined in Theorem \ref{thm: fund_limit_under_no_interference},\linebreak it follows from Appendix \ref{sec: proof_fund_limit_under_no_interference}, (\ref{tail_probability_with_no_RFI_6_1}), and (\ref{SNR_definition_with_no_RFI_4}) that $p(0) \leq \mathbb{P}\Big( \frac{2P_{\textnormal{max}}^s}{\sigma^2}\frac{\big([\sqrt{2}\textnormal{Re}\{h\}]^2+[\sqrt{2}\textnormal{Im}\{h\}]^2\big) }{[\sqrt{2}/\sigma\textnormal{Re}\{(\bm{n})_i\}]^2 } \geq \beta \Big)$, where $\beta \in \mathbb{R}^+$ is a constant per (\ref{beta_defn}) and $\sqrt{2}\textnormal{Re}\{h\}, \sqrt{2}\textnormal{Im}\{h\}, \sqrt{2}/\sigma\textnormal{Re}\{(\bm{n})_i\}\sim \mathcal{N}(0,1)$ are independent standard normal RVs. Considering $N$ independent realizations $h$ and $(\bm{n})_i$, the upper bound of $p(0)$ can be numerically computed as $p(0) \leq \frac{1}{N} \sum_{k=1}^N \mathbb{I}\Big\{  \frac{2P_{\textnormal{max}}^s}{\sigma^2}\frac{\big(X_k^2+Y_k^2\big) }{Z_k^2 } \geq \beta \Big\}$, where $X_k\eqdef\sqrt{2}\textnormal{Re}\{h_k\}, Y_k\eqdef\sqrt{2}\textnormal{Im}\{h_k\} \sim \mathcal{N}(0,1)$ are independent standard normal RVs regarding the $k$-th realization of $h$, and $Z_k\eqdef\sqrt{2}/\sigma\textnormal{Re}\{(\bm{n})_{i,k}\} \sim \mathcal{N}(0,1)$ is an independent standard normal RV for the $k$-th realization of $(\bm{n})_i$. Hence, we carry out Monte Carlo simulations in MATLAB$\textsuperscript{\textregistered}$  by generating the $3N$ independent standard normal RVs $\big\{X_k, Y_k, Z_k\big\}_{k=1}^N$ deployed to compute the upper bound in the RHS of the above numerical expression by $1)$ fixing $P_{\textnormal{max}}^s$ and varying $\sigma^2$ and $2)$ fixing $\sigma^2$ and varying $P_{\textnormal{max}}^s$. For these settings and the asymptotic performance limits of Theorem \ref{thm: fund_limit_under_no_interference}, the $p(0)$ versus $\beta$ plots are shown in Figs. \ref{fig: Perf_comparison_plot_wrt_no_RFI_fixed_P_max_s_varying_sigma} and \ref{fig: Perf_comparison_plot_wrt_no_RFI_fixed_sigma_varying_P_max_s}.

Fig. \ref{fig: Perf_comparison_plot_wrt_no_RFI_fixed_P_max_s_varying_sigma} demonstrates that $p(0)$ approaches zero as $\sigma^2$ gets large -- regardless of $\beta$ -- and hence $\lim_{\sigma^2 \to \infty} p(0) = 0$. This validates the first part of Theorem \ref{thm: fund_limit_under_no_interference}. Similarly, Fig. \ref{fig: Perf_comparison_plot_wrt_no_RFI_fixed_sigma_varying_P_max_s} shows that $p(0)$ tends to zero as $P_{\textnormal{max}}^s$ gets infinitesimally small, irrespective of $\beta$, and hence $\lim_{P_{\textnormal{max}}^s\to 0 } p(0) = 0$. This verifies the second part of Theorem \ref{thm: fund_limit_under_no_interference}, which is thus substantiated.
\begin{figure*}[t!]
	\begin{minipage}[htb]{0.46\linewidth}
		\centering
		\includegraphics[width=\textwidth]{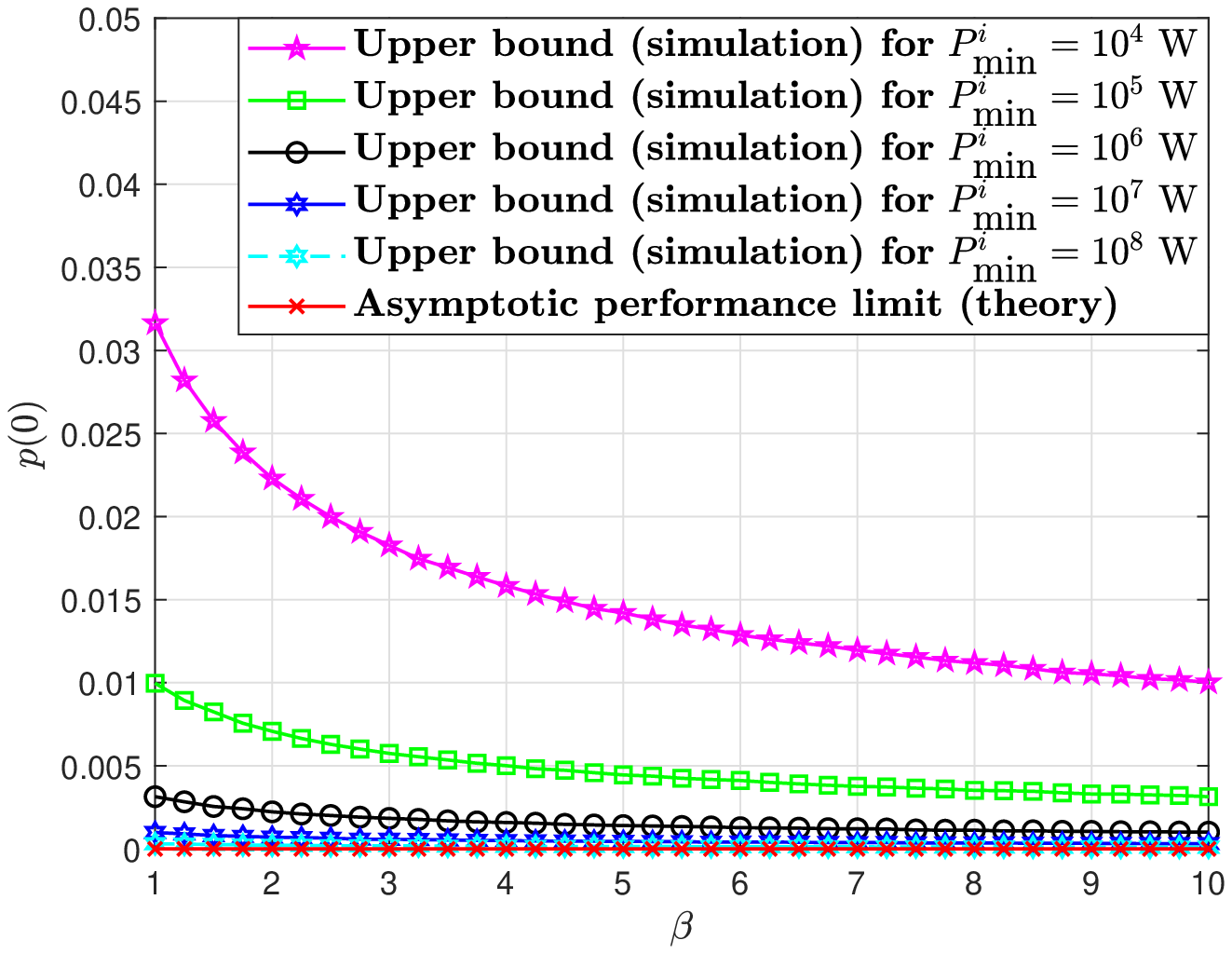}
		\caption{$p(0)$ versus $\beta$ under an RFI, fixed $P_{\textnormal{max}}^s=10$ W, and varying $P_{\textnormal{min}}^i$: $N=10^7$.}
		\label{fig: Perf_comparison_plot_wrt_RFI_fixed_P_max_s_varying_P_min_i}
	\end{minipage}
	\hspace{0.5cm}
	\begin{minipage}[htb]{0.46\linewidth}
		\centering
		\includegraphics[width=\textwidth]{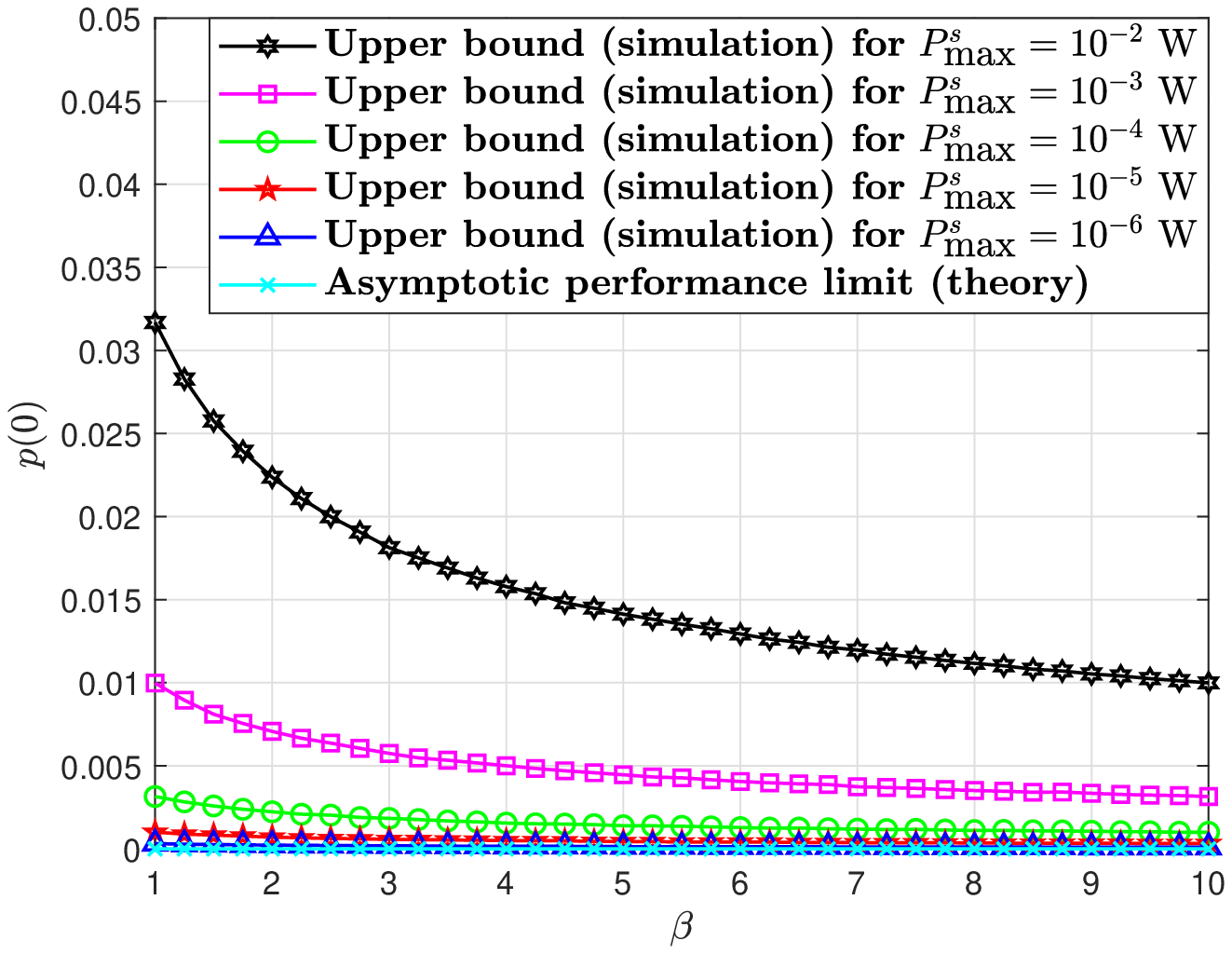}
		\caption{$p(0)$ versus $\beta$ under an RFI, fixed $P_{\textnormal{min}}^i=10$ W, and varying $P_{\textnormal{max}}^s$: $N=10^7$.}
		\label{fig: Perf_comparison_plot_wrt_RFI_fixed_P_min_i_varying_P_max_s}
	\end{minipage}
\end{figure*}

\subsection{Simulation Results with RFI}
\label{subsec: results_under_a_single-user_RFI}
In reference to the RFI addressed in Theorem \ref{thm: fund_limit_with_interference}, it follows from Appendix \ref{sec: proof_fund_limit_with_interference}, (\ref{tail_probability_with_RFI_1}), and (\ref{SNR_definition_with_RFI_7}) that $p(0) \leq \mathbb{P}\Big(\frac{P_{\textnormal{max}}^s\big([\sqrt{2}\textnormal{Re}\{h\}]^2+[\sqrt{2}\textnormal{Im}\{h\}]^2\big)}{P_{\textnormal{min}}^i[\sqrt{2}\textnormal{Re}\{g\}]^2} \geq \beta \Big)$, where $\beta \geq 0$ is a constant defined in (\ref{beta_defn}) and $\sqrt{2}\textnormal{Re}\{h\}, \sqrt{2}\textnormal{Im}\{h\}, \sqrt{2}\textnormal{Re}\{g\} \sim \mathcal{N}(0,1)$ are independent standard normal RVs. If we consider $N$ independent realizations of $\{h, g\}$, the upper bound of $p(0)$ can be numerically obtained as $p(0) \leq \frac{1}{N} \sum_{k=1}^N \mathbb{I}\Big\{\frac{P_{\textnormal{max}}^s}{P_{\textnormal{min}}^i} \frac{\big(A_k^2+B_k^2\big)}{C_k^2}  \geq \beta \Big\}$, where $A_k\eqdef \sqrt{2}\textnormal{Re}\{h_k\}, B_k\eqdef\sqrt{2}\textnormal{Im}\{h_k\}, C_k\eqdef \sqrt{2}\textnormal{Re}\{g_k\} \sim \mathcal{N}(0,1)$ are independent standard normal RVs for the $k$-th realization of the independent Rayleigh fading channels $\{h,g\}$. Thus, we conduct Monte Carlo simulations in MATLAB$\textsuperscript{\textregistered}$ by generating the $3N$ independent standard normal RVs $\big\{A_k, B_k, C_k\big\}_{k=1}^N$ employed to calculate the upper bound in the RHS of the above numerical expression by $1)$ fixing $P_{\textnormal{max}}^s$ and varying $P_{\textnormal{min}}^i$ and $2)$ fixing $P_{\textnormal{min}}^i$ and varying $P_{\textnormal{max}}^s$. For these settings and the asymptotic performance limits of Theorem \ref{thm: fund_limit_with_interference}, the $p(0)$ versus $\beta$ plots are depicted in Figs. \ref{fig: Perf_comparison_plot_wrt_RFI_fixed_P_max_s_varying_P_min_i} and \ref{fig: Perf_comparison_plot_wrt_RFI_fixed_P_min_i_varying_P_max_s}.

Fig. \ref{fig: Perf_comparison_plot_wrt_RFI_fixed_P_max_s_varying_P_min_i} shows that $p(0)$ approaches zero as $P_{\textnormal{min}}^i$ gets huge, irrespective of $\beta$, and thus $\lim_{P_{\textnormal{min}}^i\to \infty } p(0) = 0$. This validates the first part of Theorem \ref{thm: fund_limit_with_interference}. Similarly, Fig. \ref{fig: Perf_comparison_plot_wrt_RFI_fixed_P_min_i_varying_P_max_s} demonstrates that $p(0)$ approaches zero as $P_{\textnormal{max}}^s$ becomes infinitesimally small, regardless of $\beta$, and thus $\lim_{P_{\textnormal{max}}^s\to 0 } p(0) = 0$. This verifies the second part of Theorem \ref{thm: fund_limit_with_interference}, which is now verified. 
\begin{figure*}[t!]
	\begin{minipage}[htb]{0.46\linewidth}
		\centering
		\includegraphics[width=\textwidth]{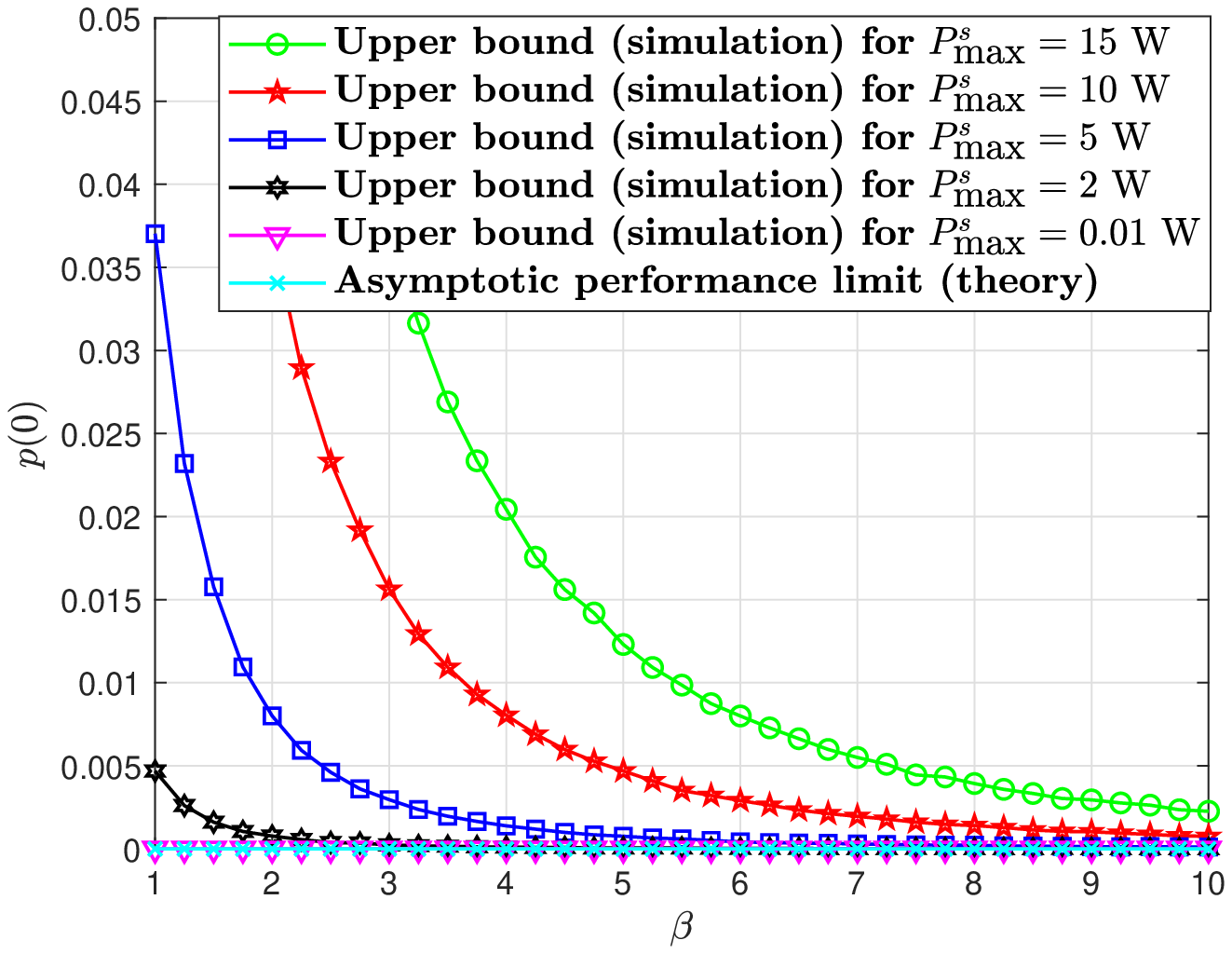}  
		\caption{$p(0)$ versus $\beta$ under MI RFI, fixed $\tilde{P}_{\textnormal{min}}^i=10$ W, fixed $U=3$, and varying $P_{\textnormal{max}}^s$: $N=10^6$.}
		\label{fig: Perf_comparison_plot_MU_RFI_fixed_t_P_min_i_U_varying_P_max}
	\end{minipage}
	\hspace{0.5cm}
	\begin{minipage}[htb]{0.46\linewidth}
		\centering
		\includegraphics[width=\textwidth]{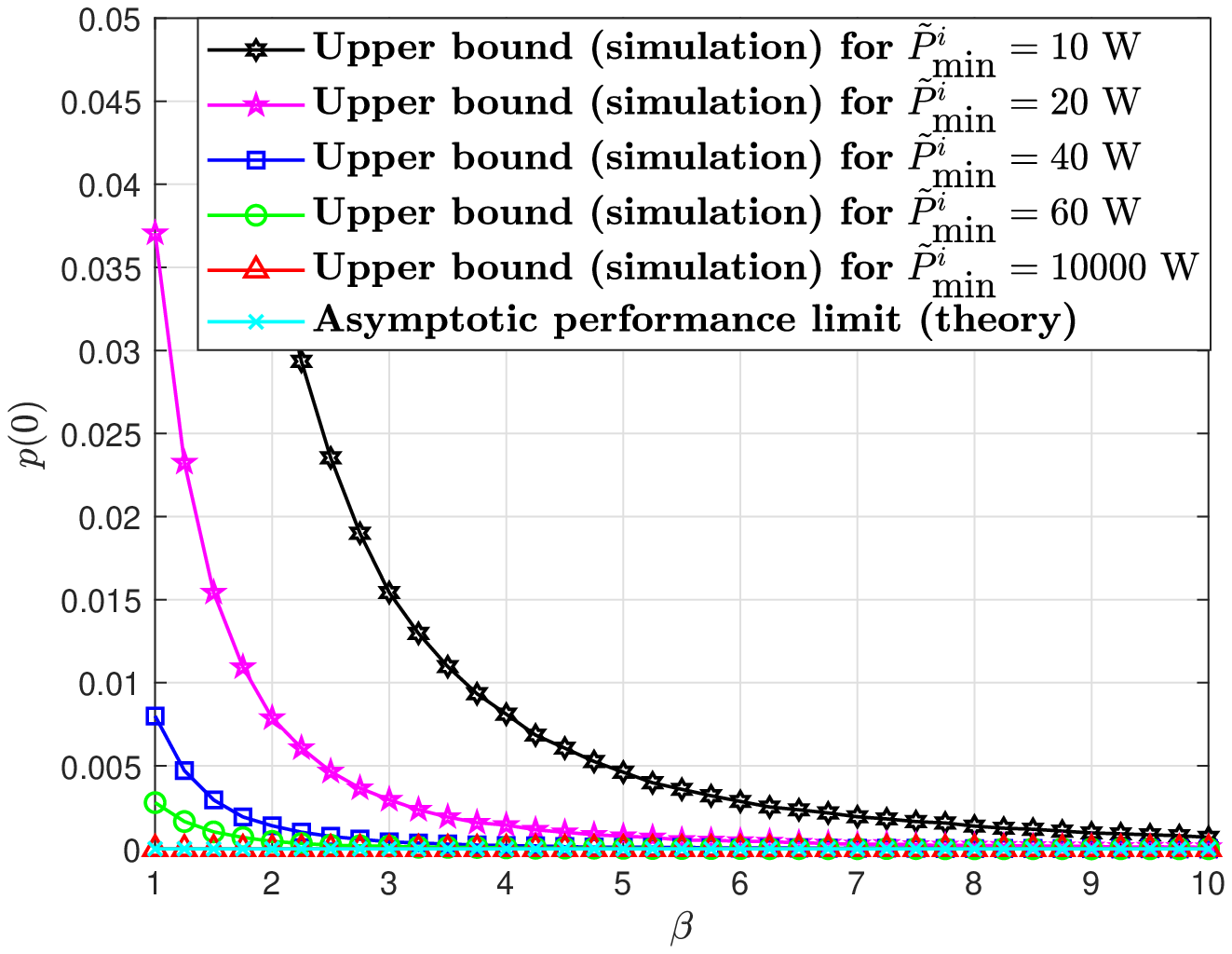}
		\caption{$p(0)$ versus $\beta$ under MI RFI, fixed $P_{\textnormal{max}}^s=10$ W, fixed $U=3$, and varying $\tilde{P}_{\textnormal{min}}^i$: $N=10^6$.}
		\label{fig: Perf_comparison_plot_MU_RFI_fixed_P_max_s_U_varying_t_P_min_i}
	\end{minipage}
\end{figure*}
\begin{figure*}[t!]
	\begin{minipage}[htb]{0.46\linewidth}
		\centering
		\includegraphics[width=\textwidth]{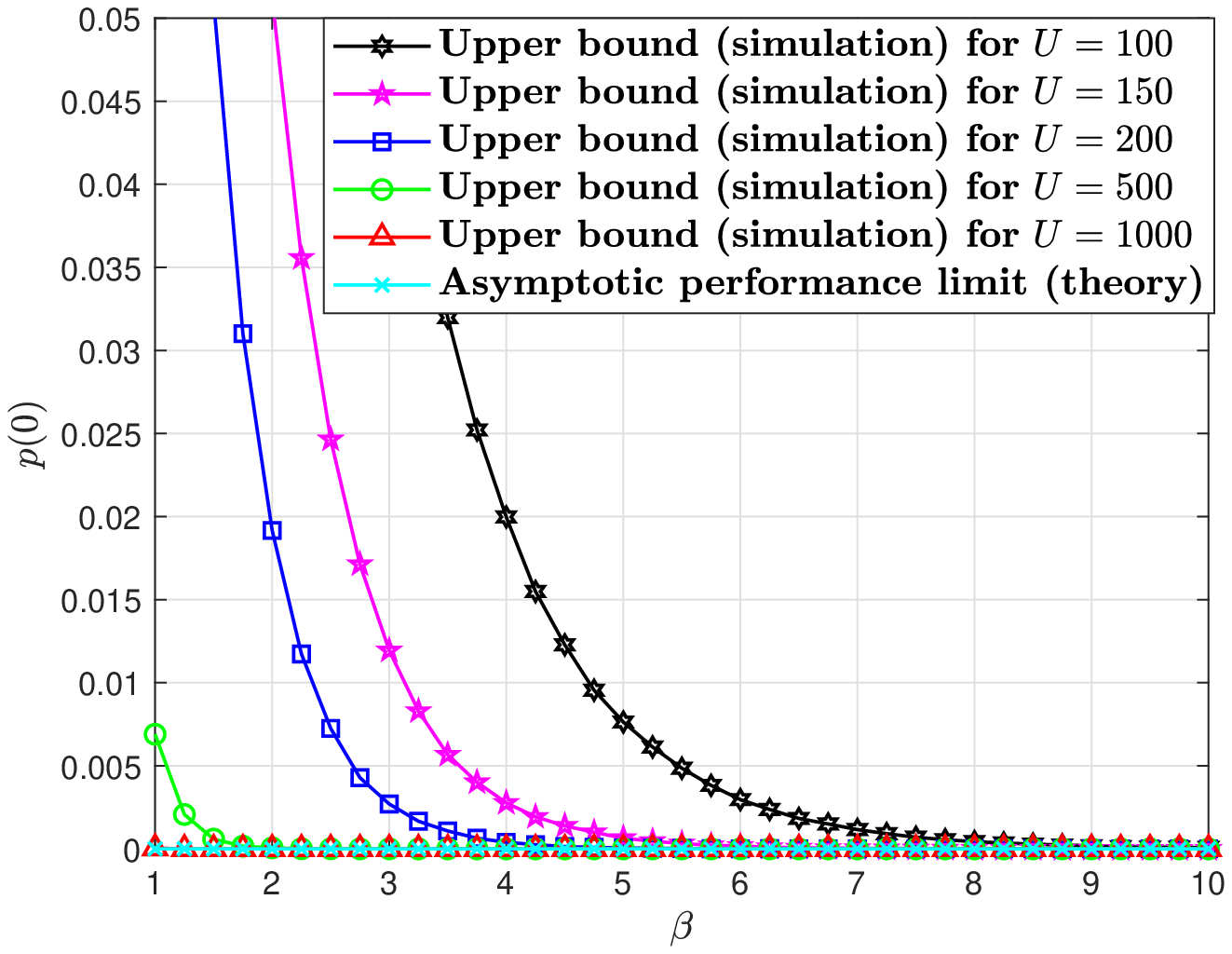}
		\caption{$p(0)$ versus $\beta$ under MI RFI, fixed $\tilde{P}_{\textnormal{min}}^i=0.1$ W, fixed $P_{\textnormal{max}}^s =10$ W, and varying $U$: $N=10^6$.}
		\label{fig: Perf_comparison_plot_MU_RFI_fixed_P_max_s_t_P_min_i_varying_U_1}
	\end{minipage}
	\hspace{0.5cm}
	\begin{minipage}[htb]{0.46\linewidth}
		\centering
		\includegraphics[width=\textwidth]{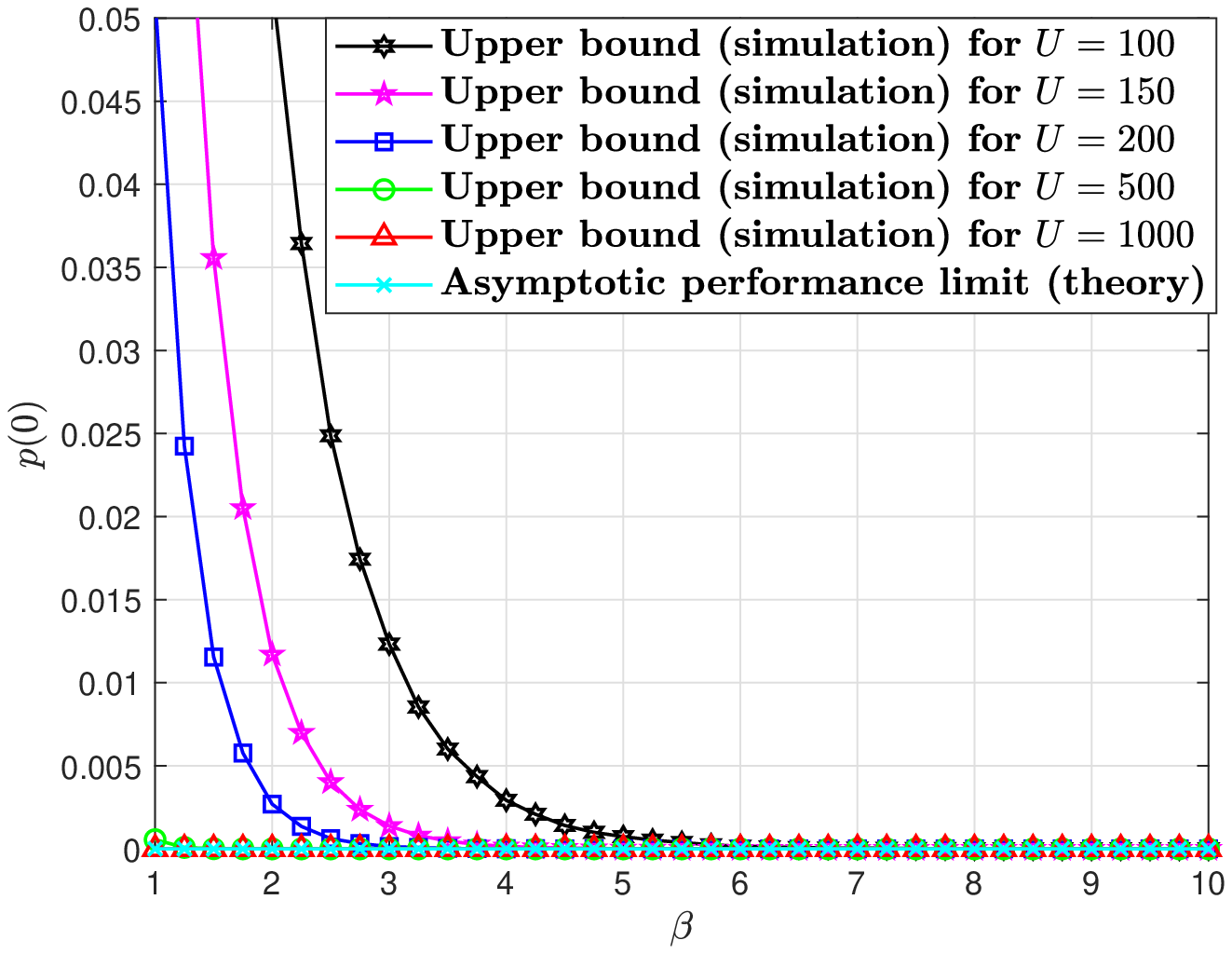}
		\caption{$p(0)$ versus $\beta$ under MI RFI, fixed $\tilde{P}_{\textnormal{min}}^i=0.15$ W, fixed $P_{\textnormal{max}}^s =10$ W, and varying $U$: $N=10^6$.}
		\label{fig: Perf_comparison_plot_MU_RFI_fixed_P_max_s_t_P_min_i_varying_U_2}
	\end{minipage}
\end{figure*}
\subsection{Simulation Results with MI RFI}
\label{subsec: results_under_a_multi-user_RFI}
Regarding the MI RFI considered in Theorem \ref{thm: fund_limit_with_MU-interference}, it follows from Appendix \ref{sec: proof_fund_limit_with_MU-interference}, (\ref{tail_probability_with_MU-RFI_1}), and (\ref{SNR_definition_with_MU-RFI_5}) that $p(0) \leq \mathbb{P}\Big( \frac{P_{\textnormal{max}}^s\big([\sqrt{2}\textnormal{Re}\{h\}]^2+[\sqrt{2}\textnormal{Im}\{h\}]^2\big)}{\tilde{P}_{\textnormal{min}}^i \big(\sum_{u=1}^U [\sqrt{2}\textnormal{Re}\{g_u\}]^2+[\sqrt{2}\textnormal{Im}\{g_u\}]^2\big)} \geq \beta\Big)$, where $\beta \in \mathbb{R}^{+}$\linebreak is a constant defined in (\ref{beta_defn}); $\sqrt{2}\textnormal{Re}\{h\}, \sqrt{2}\textnormal{Im}\{h\} \sim \mathcal{N}(0,1)$ are mutually independent standard normal RVs; and $\sqrt{2}\textnormal{Re}\{g_u\}, \sqrt{2}\textnormal{Im}\{g_u\} \sim \mathcal{N}(0,1)$ are mutually independent standard normal RVs, $\forall u\in[U]$. Considering $N$ independent realizations of $\{h, g_1, g_2, \ldots, g_U\}$, the upper bound of $p(0)$ can be computed numerically as $p(0) \leq \frac{1}{N} \sum_{k=1}^N \mathbb{I}\Big\{ \frac{P_{\textnormal{max}}^s}{\tilde{P}_{\textnormal{min}}^i} \frac{\big(D_k^2+E_k^2\big)}{ \sum_{u=1}^U \big( F_{u,k}^2+G_{u,k}^2 \big) }  \geq \beta \Big\}$, where the RVs $D_k\eqdef\sqrt{2}\textnormal{Re}\{h_k\}, E_k\eqdef\sqrt{2}\textnormal{Im}\{h_k\}, F_{u,k}\eqdef\sqrt{2}\textnormal{Re}\{g_{u,k}\}, G_{u,k}\eqdef\sqrt{2}\textnormal{Im}\{g_{u,k}\} \sim \mathcal{N}(0,1)$ are independent standard normal RVs for the $k$-th realization of the independent Rayleigh fading channels $\{h, g_1, \ldots, g_U\}$. Accordingly, we execute Monte Carlo simulations in MATLAB$\textsuperscript{\textregistered}$ by generating the $2N$ independent standard normal RVs $\big\{D_k, E_k \big\}_{k=1}^N$ and the $2NU$ independent standard normal RVs $\big\{\big\{F_{u,k}, G_{u,k} \big\}_{u=1}^U\big\}_{k=1}^N$ deployed to determine the upper bound in the RHS of the above numerical expression by $1)$ fixing $\big\{\tilde{P}_{\textnormal{min}}^i,U\big\} $ and varying $P_{\textnormal{max}}^s$; $2)$ fixing $\big\{ P_{\textnormal{max}}^s, U \big\}$ and varying $\tilde{P}_{\textnormal{min}}^i$; and $3)$ fixing $\big\{ \tilde{P}_{\textnormal{min}}^i,P_{\textnormal{max}}^s \big\}$ and varying $U$. For these settings and the asymptotic performance limits of Theorem \ref{thm: fund_limit_with_MU-interference}, the $p(0)$ versus $\beta$ plots are shown in Figs. \ref{fig: Perf_comparison_plot_MU_RFI_fixed_t_P_min_i_U_varying_P_max}-\ref{fig: Perf_comparison_plot_MU_RFI_fixed_P_max_s_t_P_min_i_varying_U_2}.

Fig. \ref{fig: Perf_comparison_plot_MU_RFI_fixed_t_P_min_i_U_varying_P_max} shows that $p(0)$ approaches zero as $P_{\textnormal{max}}^s$ gets infinitesimally small, regardless of $\beta$, and hence $\lim_{P_{\textnormal{max}}^s \to 0 } p(0)= 0$. This validates the first part of Theorem \ref{thm: fund_limit_with_MU-interference}. Fig. \ref{fig: Perf_comparison_plot_MU_RFI_fixed_P_max_s_U_varying_t_P_min_i} shows that $p(0)$ tends to zero as $\tilde{P}_{\textnormal{min}}^i$ gets gigantic, irrespective of $\beta$, and thus $\lim_{\tilde{P}_{\textnormal{min}}^i \to \infty } p(0) = 0$. This validates the second part of Theorem \ref{thm: fund_limit_with_MU-interference}. Figs. \ref{fig: Perf_comparison_plot_MU_RFI_fixed_P_max_s_t_P_min_i_varying_U_1} and \ref{fig: Perf_comparison_plot_MU_RFI_fixed_P_max_s_t_P_min_i_varying_U_2} corroborate\footnote{Figs. \ref{fig: Perf_comparison_plot_MU_RFI_fixed_P_max_s_t_P_min_i_varying_U_1} and \ref{fig: Perf_comparison_plot_MU_RFI_fixed_P_max_s_t_P_min_i_varying_U_2} also reveal that a slight increase in the minimum MI RFI power constraint -- from $\tilde{P}_{\textnormal{min}}^i=0.1$ W to $\tilde{P}_{\textnormal{min}}^i=0.15$ W -- results in worse performance by shifting the $p(0)$ versus $\beta$ curve to the left.} that $p(0)$ gets close to zero as $U$ gets enormous -- regardless of the minimum MI RFI power constraint and $\beta$ -- and hence $\lim_{U\to \infty } p(0)  = 0$. This verifies the third part of Theorem \ref{thm: fund_limit_with_MU-interference},\linebreak which is now corroborated. 
\subsection{Simulation Results Regarding DeepSC's Practical Limits}
\begin{figure*}[t!]
	\begin{minipage}[htb]{0.46\linewidth}
		\centering
		\includegraphics[width=\textwidth]{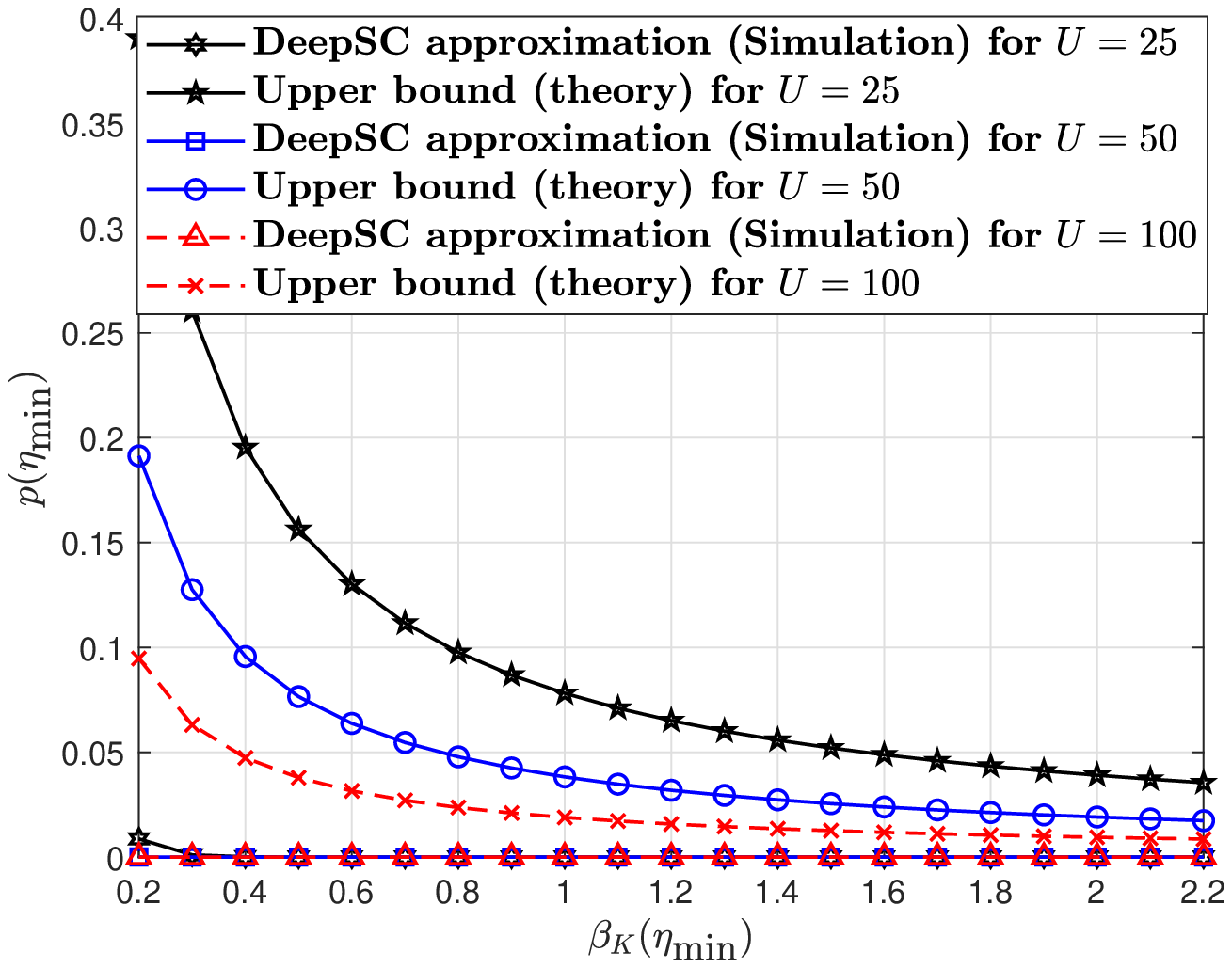}
		\caption{$p(\eta_{\textnormal{min}})$ versus $\beta_K(\eta_{\textnormal{min}})$ under $(\tilde{P}_{\textnormal{min}}^i, P_{\textnormal{max}}^s)=(0.8, 1.5)$ W and $U \in \{25, 50, 100\}$: $N=10^6$.}
		\label{fig: Practical_limits_of_DeepSC_P_min_0.8W}
	\end{minipage}
	\hspace{0.5cm}
	\begin{minipage}[htb]{0.46\linewidth}
		\centering
		\includegraphics[width=\textwidth]{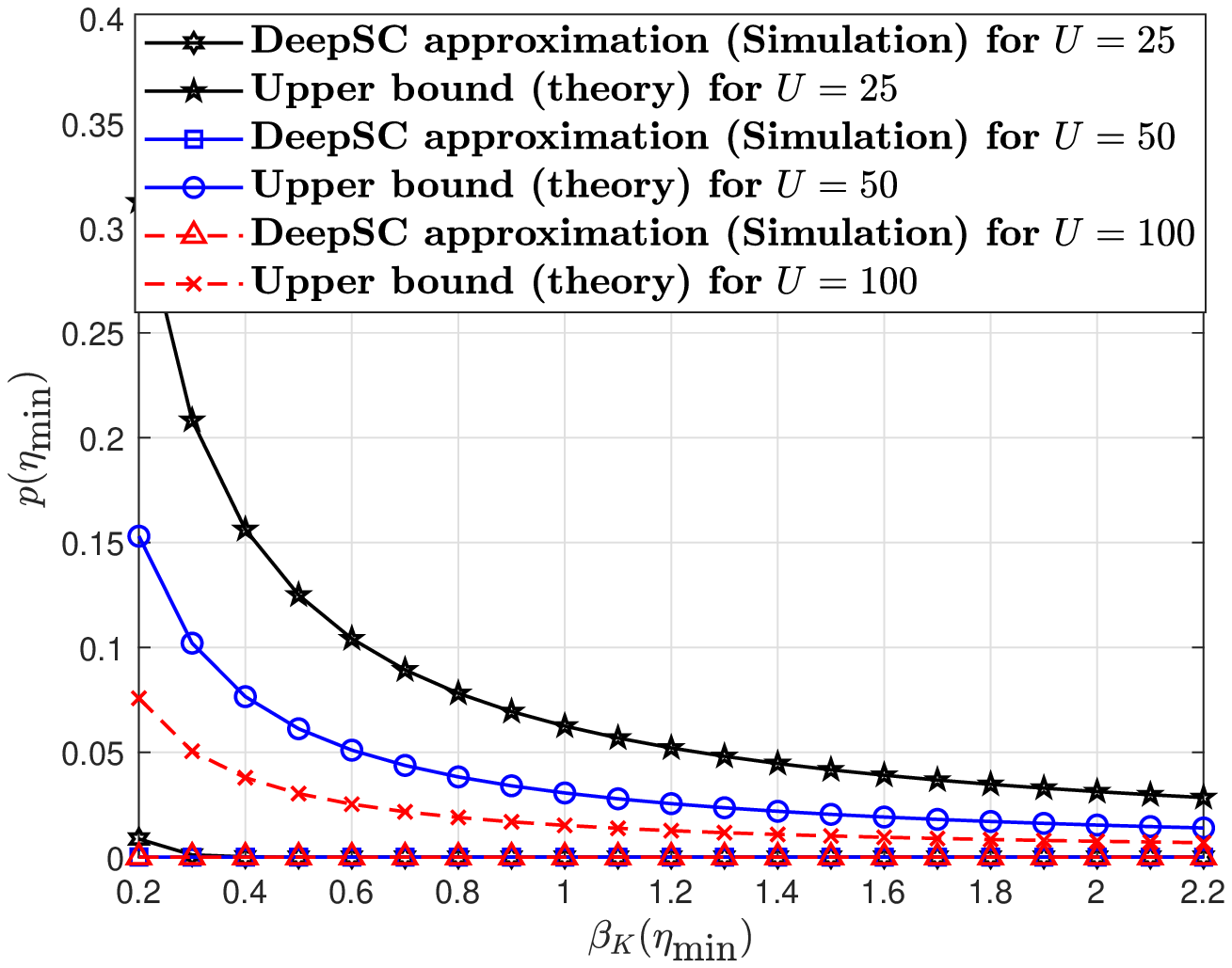}
		\caption{$p(\eta_{\textnormal{min}})$ versus $\beta_K(\eta_{\textnormal{min}})$ under $(\tilde{P}_{\textnormal{min}}^i, P_{\textnormal{max}}^s)=(1, 1.5)$ W and $U \in \{25, 50, 100\}$: $N=10^6$.}
		\label{fig: Practical_limits_of_DeepSC_P_min_1W}
	\end{minipage}
\end{figure*}

Per Appendix \ref{sec: proof_fund_limit_with_MU-interference}, the SINR under MI RFI at the $m$-th realization  is $\gamma_m \eqdef  \frac{|h_m|^2|(\bm{x})_i|^2}{\sum_{u=1}^U |g_{u,m}|^2|(\bm{v}_u)_i|^2+|(\bm{n})_m|^2}$, where $h_m, g_m \sim \mathcal{CN}(0,1)$. Without loss of generality and per the constraints of Theorem \ref{thm: practical_limits_with_MU-interference}, we consider $|(\bm{x})_i|^2=|(\bm{v}_u)_i|^2=1$ and $(\bm{n})_m \sim \mathcal{CN}(0,1)$, $P_{\textnormal{max}}^s=1.5$ W, $\tilde{P}_{\textnormal{min}}^i \in \{0.8, 1\}$ W, and $U \in \{25, 50, 100\}$. Per this setting, the Monte Carlo simulation of DeepSC's (approximated) performance is empirically evaluated as $p(\eta_{\textnormal{min}}) = \frac{1}{N} \sum_{m=1}^N \mathbb{I}\big\{ \gamma_m  \geq \beta_K(\eta_{\textnormal{min}}) \big\}$, considering $N=10^6$. This produces the $p(\eta_{\textnormal{min}})$ versus $\beta_K(\eta_{\textnormal{min}})$ plots depicted in Figs. \ref{fig: Practical_limits_of_DeepSC_P_min_0.8W} and \ref{fig: Practical_limits_of_DeepSC_P_min_1W}, where the upper bound -- formalized by Theorem \ref{thm: practical_limits_with_MU-interference} -- gets very close to the simulation results as $\beta_K(\eta_{\textnormal{min}})$ gets bigger and $U$ gets larger.
\begin{figure}[htb!]
	\raggedright
	\includegraphics[scale=.55]{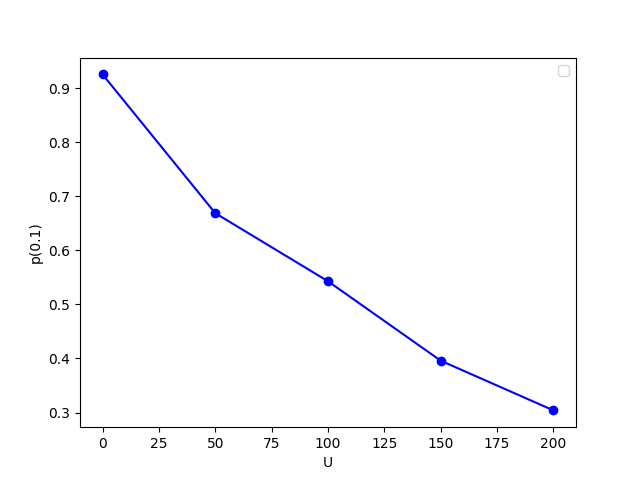}
	\caption{$p(0.1)$ versus $U$ exhibited by our trained DeepSC.}
	\label{fig: Prob_SSM-0828-I1}
\end{figure}

\subsection{Computer Experiment Result}
We carry out extensive computer experiments on the training of DeepSC and its testing without and with MI RFI using a standard dataset named the \textit{proceedings of the European Parliament} (Europarl). Employing 1.5 million training sentences and 300,000 validation sentences \textit{tokenized and vectorized} from Europarl, we train the DeepSC model with \texttt{Adam} in Keras 2.9 with TensorFlow 2.9\linebreak as a backend. The trained model is then tested without and with time-varying Gaussian MI RFI \cite{TMGWA17,GeAR_conf_16} by deploying 10,000 testing sentences that were also tokenized and vectorized from Europarl. This led to the $p(0.1)$ versus $U$ curve shown in Fig. \ref{fig: Prob_SSM-0828-I1}, where DeepSC is demonstrated producing semantically-irrelevant sentences -- regarding a nearly-semantically-irrelevant $\eta_{\textnormal{min}}$ of 0.1 -- as the number of Gaussian RFI emitters gets large. This is what is predicted by Theorem \ref{thm: practical_limits_with_MU-interference}.\linebreak Our concluding summary and research outlook thus follow.
\section{Concluding Summary and Research Outlook}
\label{sec: conc_summary_and_research_outlook}
As a 6G enabler that only transmits semantically-relevant information, SemCom promises to minimize power usage, bandwidth consumption, and transmission delay. However, the fidelity of text, image, audio, and video SemCom techniques can be destroyed by a considerable semantic noise caused by RFI. Quantifying RFI's impact while introducing a principled probabilistic framework applicable primarily for SemCom, we characterized the asymptotic performance limits, practical limits, and outage probability of DeepSC subjected to RFI and MI RFI. These performance limits of DeepSC were validated by extensive Monte Carlo simulations and computer experiments -- asserting that strong RFI can destroy the faithfulness of SemCom. Toward 6G and beyond, this paper affirms the need for an \textit{adversarial electronic warfare-resistant} SemCom system by proposing a (generic) lifelong DL-based IR$^2$ SemCom system.

As promising research outlook, this paper stimulates multiple lines of research on the design, analysis, and optimization of multi-input multi-output (MIMO) SemCom, ultra-massive MIMO SemCom, and interference mitigation in multi-cell multi-interferer MIMO SemCom networks. Finally, the probabilistic framework introduced by this paper paves the way for the performance analysis of speech SemCom, image SemCom, and video SemCom systems; and for the fundamental non-asymptotic performance analysis of these SemCom systems.

\appendices

\section{Proof of Theorem \ref{thm: fund_limit_under_no_interference}}
\label{sec: proof_fund_limit_under_no_interference}
In this appendix, we set out to analyze the fundamental performance limits of DeepSC subjected to infinitesimally small RFI and hence $U=1$. We will thus drop the subscript $u$ and let $\bm{v} \eqdef \bm{v}_u = \bm{v}_1$ and $g \eqdef g_u=g_1 \sim\mathcal{CN}(0,1)$.

To begin, if we use (\ref{simantic_similarity_function_1}) in the RHS of (\ref{tail_probability_defn}) for any given $K$,
\begin{equation}
\label{tail_probability_with_no_RFI_1}
p(0) = \mathbb{P}\big( \varepsilon(K,\gamma) \geq 0 \big) \stackrel{(a)}{\approx} \mathbb{P}\big( \tilde{\varepsilon}_K(\gamma) \geq 0 \big), 
\end{equation}
where $(a)$ is due to (\ref{simantic_similarity_function_approximation_1}). If we then substitute (\ref{simantic_similarity_function_approximation_1}) into the RHS of (\ref{tail_probability_with_no_RFI_1}) and consider $\kappa \eqdef A_{K,1}/(A_{K,1}-A_{K,2}) \geq 0$,
\begin{subequations}
\begin{align}
\label{tail_probability_with_no_RFI_2}
p(0) & \approx \mathbb{P}\Big( \frac{A_{K,2}-A_{K,1}}{1+ e^{-(C_{K,1}\gamma+C_{K,2})}} \geq -A_{K,1} \Big)  \\
\label{tail_probability_with_no_RFI_2_1}
&= \mathbb{P}\Big( \frac{1}{1+ e^{-(C_{K,1}\gamma+C_{K,2})}} \geq \kappa \Big)  \\
\label{tail_probability_with_no_RFI_3}
&= \mathbb{P}\big( e^{-(C_{K,1}\gamma+C_{K,2})} \leq 1/\kappa-1 \big)  \\
\label{tail_probability_with_no_RFI_4}
&=\mathbb{P}\big( -(C_{K,1}\gamma+C_{K,2}) \leq \ln\big[( 1-\kappa)/\kappa \big]  \big)    \\
\label{tail_probability_with_no_RFI_5}
&\stackrel{(a)}{=}\mathbb{P}\big( (C_{K,1}\gamma+C_{K,2}) \geq \ln\big[ \kappa/(1-\kappa) \big] \big)   \\
\label{tail_probability_with_no_RFI_6}
&\stackrel{(b)}{=}\mathbb{P}\big( \gamma \geq \ln\big[ \kappa/(1-\kappa)\big]/C_{K,1} -C_{K,2}/C_{K,1}  \big)    \\ 
\label{tail_probability_with_no_RFI_6_1}
& =\mathbb{P}\big( \gamma \geq \beta  \big),
\end{align}
\end{subequations}
where $(a)$ is due to multiplying both sides of (\ref{tail_probability_with_no_RFI_4}) by -1, $\gamma$ is the SNR upon the reception of the DeepSC symbol, $(b)$ is due to the logistic growth rate constraint $C_{K,1}> 0$, and
\begin{equation}
\label{beta_defn}
\beta \eqdef \ln\big[ \kappa/(1-\kappa)\big]/C_{K,1} -C_{K,2}/C_{K,1},
\end{equation}
where $\beta\in\mathbb{R}$ and constant for a given $K$. For the non-negative argument constraint of the $\ln(\cdot)$ function, it follows from the RHS of (\ref{tail_probability_with_no_RFI_4}) and (\ref{tail_probability_with_no_RFI_5}) that $\kappa\leq 1$ and $\kappa \geq 0$, respectively. Consequently, the following condition ensues.
\begin{condition}
\label{condition_kappa_1}
Regarding $\kappa \eqdef A_{K,1}/(A_{K,1}-A_{K,2})$, $\kappa \in [0,1]$.
\end{condition}

Under the satisfaction of Condition \ref{condition_kappa_1}, we proceed to bound the RHS of (\ref{tail_probability_with_no_RFI_6_1}) w.r.t. the statistics of the SNR $\gamma$. To this end, if we consider infinitesimally small interference ($g\bm{v}\to \bm{0}$) and the model in (\ref{DeepSC_received_signak_model}), the SNR is given by
\begin{equation}
\label{SNR_definition_with_no_RFI}
\gamma \eqdef  \frac{|h|^2|(\bm{x})_i|^2}{|(\bm{n})_i|^2} =\frac{\big([\textnormal{Re}\{h\}]^2+[\textnormal{Im}\{h\}]^2\big)|(\bm{x})_i|^2}{[\textnormal{Re}\{(\bm{n})_i\}]^2+[\textnormal{Im}\{(\bm{n})_i\}]^2}, 
\end{equation}
where $i\in[KN]$; $\textnormal{Re}\{h\}, \textnormal{Im}\{h\} \sim \mathcal{N}(0,1/2)$ are independent Gaussian RVs; and $\textnormal{Re}\{(\bm{n})_i\}, \textnormal{Im}\{(\bm{n})_i\} \sim \mathcal{N}(0,\sigma^2/2)$\linebreak are other independent Gaussian RVs. In light of the DeepSC symbols' power constraint of Sec. \ref{subsec: sysetm_model}, $\mathbb{E}\{[\textnormal{Re}\{(\bm{x})_i\}]^2  \}, \mathbb{E}\{[\textnormal{Im}\{(\bm{x})_i\}]^2 \} \leq P_{\textnormal{max}}^s$. Thus, $\forall i\in[KL]$, $[\textnormal{Re}\{(\bm{x})_i\}]^2 \leq P_{\textnormal{max}}^s$, $[\textnormal{Im}\{(\bm{x})_i\}]^2 \leq P_{\textnormal{max}}^s$, and hence $[\textnormal{Re}\{(\bm{x})_i\}]^2  + [\textnormal{Im}\{(\bm{x})_i\}]^2   =|(\bm{x})_i|^2 \leq 2P_{\textnormal{max}}^s$. If we substitute this constraint into the RHS of (\ref{SNR_definition_with_no_RFI}), we will obtain
\begin{subequations}
\begin{align}
\label{SNR_definition_with_no_RFI_2_1}
\gamma & \leq \frac{2P_{\textnormal{max}}^s\big([\textnormal{Re}\{h\}]^2+[\textnormal{Im}\{h\}]^2\big)}{[\textnormal{Re}\{(\bm{n})_i\}]^2+[\textnormal{Im}\{(\bm{n})_i\}]^2}    \\
\label{SNR_definition_with_no_RFI_2}
&\stackrel{(a)}{\leq} \frac{2P_{\textnormal{max}}^s\big([\textnormal{Re}\{h\}]^2+[\textnormal{Im}\{h\}]^2\big)}{[\textnormal{Re}\{(\bm{n})_i\}]^2}, 
\end{align}
\end{subequations}
where $(a)$ is because $[\textnormal{Im}\{(\bm{n})_i\}]^2 \geq 0$. Multiplying the numerator and denominator of the RHS of (\ref{SNR_definition_with_no_RFI_2}) by $2/\sigma^2$ gives
\begin{subequations}
\begin{align}
\label{SNR_definition_with_no_RFI_4_1}
\gamma  & \leq \frac{2P_{\textnormal{max}}^s\big([\textnormal{Re}\{h\}]^2+[\textnormal{Im}\{h\}]^2\big) \times 2/\sigma^2}{[\textnormal{Re}\{(\bm{n})_i\}]^2 \times 2/\sigma^2 }    \\
\label{SNR_definition_with_no_RFI_4}
&=\frac{2P_{\textnormal{max}}^s}{\sigma^2}\frac{\big([\sqrt{2}\textnormal{Re}\{h\}]^2+[\sqrt{2}\textnormal{Im}\{h\}]^2\big) }{[\sqrt{2}/\sigma\textnormal{Re}\{(\bm{n})_i\}]^2 }   \\
\label{SNR_definition_with_no_RFI_5}
 &\stackrel{(a)}{=} 2P_{\textnormal{max}}^s/\sigma^2 \times \big[ ( A/C )^2 + ( B/C)^2 \big],
\end{align}
\end{subequations}
where $(a)$ is for $A \eqdef \sqrt{2}\textnormal{Re}\{h\}$, $B \eqdef \sqrt{2}\textnormal{Im}\{h\}$, and $C \eqdef \sqrt{2}/\sigma\textnormal{Re}\{(\bm{n})_i\}$.

Note that $A, B, C \sim \mathcal{N}(0,1)$ are independent standard normal RVs since $\textnormal{Re}\{h\}, \textnormal{Im}\{h\} \sim \mathcal{N}(0,1/2)$ and  $\textnormal{Re}\{(\bm{n})_i\} \sim \mathcal{N}(0,\sigma^2/2)$ are independent Gaussian RVs. If we let $X \eqdef A/C$ and $Y \eqdef  B/C$, then (\ref{SNR_definition_with_no_RFI_5}) can be expressed as
\begin{equation}
\label{SNR_definition_with_no_RFI_6}
\gamma \leq 2P_{\textnormal{max}}^s/\sigma^2 \times ( X^2 +Y^2 ).
\end{equation}
As $X $ and $Y $ are the ratios of independent standard normal RVs, they would have the following probability density function (PDF) \cite[eq. (7.1), p. 61]{Marvin_K_SImon_Handbook'06}; \cite[eq. (10.6), p. 101]{Marvin_K_SImon_Handbook'06}:
\begin{equation}
\label{PDF_X_PDF_Y}
p_X(x)=\big[ \pi (x^2+1) \big]^{-1}  \hspace{2mm} \textnormal{and} \hspace{2mm} p_Y(y)=\big[ \pi (y^2+1) \big]^{-1}.
\end{equation}
For the inequality in (\ref{SNR_definition_with_no_RFI_6}) and $\beta$ defined in (\ref{beta_defn}),
\begin{equation}
\label{prob_inequality_with_no_RFI_1}
\mathbb{P}(\gamma \geq \beta) \leq \mathbb{P}\big( 2P_{\textnormal{max}}^s/\sigma^2 \times ( X^2 +Y^2 ) \geq \beta \big).  
\end{equation}
If we use the inequality in (\ref{prob_inequality_with_no_RFI_1}) in the RHS of (\ref{tail_probability_with_no_RFI_6_1}) under the satisfaction of Condition \ref{condition_kappa_1},
\begin{subequations}
\begin{align}
\label{tail_probability_with_no_RFI_7_3_1}
p(0) & \leq \mathbb{P}\big( 2P_{\textnormal{max}}^s( X^2 +Y^2 )/\sigma^2 \geq \beta \big)    \\
\label{tail_probability_with_no_RFI_7_3}
&\stackrel{(a)}{=} \mathbb{P}\Big(  \sqrt{ X^2 +Y^2 } \geq \sigma \sqrt{\beta/ (2P_{\textnormal{max}}^s) } \Big),
\end{align}
\end{subequations}
where $(a)$ follows from applying the square root function because it is a monotonically increasing function. Accordingly, the outermost RHS of (\ref{tail_probability_with_no_RFI_7_3}) enforces the constraint $\beta \geq 0$. $\beta \geq 0$ w.r.t. (\ref{beta_defn}), thus, translates to the following condition:
\begin{condition}
\label{beta_condition}
For $\alpha=e^{C_{K,2}}/(1+e^{C_{K,2}})$, $\kappa\in [\alpha, \infty)$. 
\end{condition}

Therefore, under the fulfillment of Conditions \ref{condition_kappa_1} and \ref{beta_condition}, it follows from (\ref{tail_probability_with_no_RFI_7_3}) that
\begin{equation}
\label{tail_probability_with_no_RFI_7_4}
p(0) \leq \mathbb{P}\big(  \sqrt{ X^2 +Y^2 } \geq t \big),
\end{equation}
where $t\in\mathbb{R}^{+}$ and $t\eqdef \sigma \sqrt{ \beta/(2P_{\textnormal{max}}^s) }$. Since $X^2, Y^2 \geq 0$, it follows from \cite[Lemma 11, p. 71]{Getu_DSFC_Estimation'22} that
\begin{equation}
\label{squareroot_inequality_identity}
\sqrt{ X^2 +Y^2 } \leq  \sqrt{ X^2 }+\sqrt{ Y^2 } \stackrel{(a)}{=} | X| + |Y|,
\end{equation}
where $(a)$ is since $\sqrt{x^2}=|x|=\mathbb{I}\{x\geq 0\}x-\mathbb{I}\{x < 0\}x$. W.r.t. (\ref{squareroot_inequality_identity}) and $t\in\mathbb{R}^{+}$, $\mathbb{P}\big( \sqrt{ X^2 +Y^2 } \geq t \big)  \leq  \mathbb{P}\big( | X| + |Y| \geq t \big)$. Using this inequality in the RHS of (\ref{tail_probability_with_no_RFI_7_4}),
\begin{equation}
\label{tail_probability_with_no_RFI_7_5}
p(0) \leq \mathbb{P}\big( | X| + |Y| \geq t \big)=\mathbb{P}\big( | X|  \geq t -|Y|\big).
\end{equation}
We then resume our analysis from (\ref{tail_probability_with_no_RFI_7_5}) provided that Conditions \ref{condition_kappa_1} and \ref{beta_condition} are satisfied.

Since $B, C\sim\mathcal{N}(0,1)$ are mutually independent standard normal RVs, $\mathbb{P}(B,C\geq 0)=\mathbb{P}(B \leq 0)\mathbb{P}(C \leq 0)+\mathbb{P}(B\geq 0)\mathbb{P}(C\geq 0)=1/2$. Similarly, $\mathbb{P}(B,C < 0)=\mathbb{P}(B < 0)\mathbb{P}(C \geq 0)+\mathbb{P}(B \geq 0)\mathbb{P}(C < 0)=1/2$. Conditioning on these probabilities, thus, $|Y|=|B|/|C|=B/C=Y$ with probability 1/2 and $|Y|=|B|/|C|=-B/C=-Y$, also with probability 1/2. Using these probabilities, applying the \textit{total probability theorem} \cite[p. 28]{DPJN08} to the RHS of (\ref{tail_probability_with_no_RFI_7_5}) gives
\begin{equation}
\label{tail_probability_with_no_RFI_7_6}
p(0) \leq  1/2 \big[ \mathbb{P}\big( | X|  \geq t + Y\big)+\mathbb{P}\big( | X|  \geq t -Y\big) \big],
\end{equation}
where $X$ and $Y$ are ratio RVs with PDFs given by (\ref{PDF_X_PDF_Y}). Since the values $t + Y$ and $t -Y$ vary for a given $t$ of the different random values assumed by the RV $Y$, we must average w.r.t. $P_Y$ -- the PDF of $Y$ per (\ref{PDF_X_PDF_Y}) -- to simplify $\mathbb{P}\big( | X|  \geq t + Y\big)$ and $\mathbb{P}\big( | X|  \geq t -Y\big)$. To this end, it follows from the \textit{law of total probability} (w.r.t. a continuous RV) \cite{Prob_Stat_Random_Process'14} that 
\begin{subequations}
\begin{align}
\label{probability_geq_t+Y_1}
\mathbb{P}\big( | X|  \geq t + Y\big)&=\int_{-\infty}^{\infty} \mathbb{P}\big( | X|  \geq t + y \big) P_Y(y) dy    \\
\label{probability_geq_t-Y_1}
\mathbb{P}\big( | X|  \geq t -Y\big)&= \int_{-\infty}^{\infty} \mathbb{P}\big( | X|  \geq t - y \big) P_Y(y) dy.
\end{align}
\end{subequations}
 
We are therefore going to compute $\mathbb{P}\big( | X|  \geq t + y \big)$ and $\mathbb{P}\big( | X|  \geq t - y \big)$ w.r.t. a given $y$ and $t$, as defined in above. Meanwhile, for $X=A/C$, where $A, C \sim \mathcal{N}(0,1)$, $\mathbb{P}(A,C\geq 0)=\mathbb{P}(A \leq 0)\mathbb{P}(C \leq 0)+\mathbb{P}(A\geq 0)\mathbb{P}(C\geq 0)=1/2$. Similarly, $\mathbb{P}(A,C < 0)=\mathbb{P}(A < 0)\mathbb{P}(C \geq 0)+\mathbb{P}(A \geq 0)\mathbb{P}(C < 0)=1/2$. Hence, $|X|=|A|/|C|=A/C=X$ with probability 1/2 and $|X|=|A|/|C|=-A/C=-X$, also with probability 1/2. Using these probabilities, exploiting the total probability theorem \cite[p. 28]{DPJN08} leads to
\begin{multline}
\label{probability_abs_X_geq_t+Y_2}
\hspace{-2mm}\mathbb{P}\big( | X|  \geq t + y\big) = \frac{\mathbb{P}\big( X  \geq t + y\big) + \mathbb{P}\big( X  \leq  -(t + y)\big)}{2}  \stackrel{(a)}{=} \\  \mathbb{P}\big( X  \geq t + y\big) \stackrel{(b)}{=}  \frac{1}{\pi}\int_{t + y}^{\infty}    \frac{dx}{x^2+1}, 
\end{multline}
where $(a)$ is due to the symmetric PDF of X (i.e., $P_X$) -- which is given by (\ref{PDF_X_PDF_Y}) -- that leads to the equality in $(b)$. Similarly,
\begin{multline}
\label{probability_abs_X_geq_t-Y_2}
\hspace{-2mm} \mathbb{P}\big( | X|  \geq t - y\big)  =\frac{\mathbb{P}\big( X  \geq t - y\big) + \mathbb{P}\big( X  \leq -(t - y)\big) }{2}  =  \\ \mathbb{P}\big( X  \geq t - y\big)  \stackrel{(a)}{=}  \frac{1}{\pi}\int_{t - y}^{\infty}   \frac{dx}{x^2+1} , 
\end{multline}
where $(a)$ is because of the PDF $P_X$ equated in (\ref{PDF_X_PDF_Y}). 

To proceed from (\ref{probability_abs_X_geq_t+Y_2}) and (\ref{probability_abs_X_geq_t-Y_2}), we require the following identity \cite[eq. (2.141.2), p. 74]{ISGI07}:  
\begin{equation}
\label{arctanx_identity}
\int \frac{dx}{x^2+1}=\arctan x.
\end{equation}
If we deploy (\ref{arctanx_identity}) and the identity $\arctan \infty=\pi/2$, it follows directly from (\ref{probability_abs_X_geq_t+Y_2}) and (\ref{probability_abs_X_geq_t-Y_2}) that
\begin{multline}
\label{probability_abs_X_geq_t+Y_3}
\mathbb{P}\big( | X|  \geq t + y\big)  = 1/\pi \big[ \pi/2 -\arctan (t+y) \big] \hspace{2mm}  \textnormal{and}  \\  \mathbb{P}\big( | X|  \geq t - y\big) = 1/\pi \big[ \pi/2-\arctan (t-y) \big].
\end{multline}
Substituting (\ref{probability_abs_X_geq_t+Y_3}) and (\ref{PDF_X_PDF_Y}) into the RHS of (\ref{probability_geq_t+Y_1}) and (\ref{probability_geq_t-Y_1}):
\begin{multline}
\label{probability_geq_t+Y_2}
\mathbb{P}\big( | X|  \geq t + Y\big)  = \int_{-\infty}^{\infty} f_1(t,y)  \frac{dy}{\pi(y^2+1)} 	\hspace{2mm} \textnormal{and}  \\  \mathbb{P}\big( | X|  \geq t -Y\big) = \int_{-\infty}^{\infty} f_2(t,y)  \frac{dy}{\pi(y^2+1)},
\end{multline}
where $f_1(t,y) = 1/\pi \big[ \pi/2 -\arctan (t+y) \big]$ and $f_2(t,y) = 1/\pi \big[ \pi/2 -\arctan (t-y) \big]$. Consequently, plugging (\ref{probability_geq_t+Y_2}) into the RHS of (\ref{tail_probability_with_no_RFI_7_6}) results in the expression
\begin{equation}
\label{tail_probability_with_no_RFI_7_7}
p(0) \leq \frac{1}{2} \int_{-\infty}^{\infty} \big[ f_1(t,y) + f_2(t,y) \big]  \frac{dy}{\pi(y^2+1)},
\end{equation}
where $f_1(t,y)$ and $f_2(t,y)$ are defined in above and can be simplified using $t$ -- as defined in above -- to  
\begin{multline}
\label{f_1_t_and_y_def_2}
f_1(t,y) =  1/2-1/\pi\times\arctan (\sigma\sqrt{\beta/(2P_{\textnormal{max}}^s)}+y)  \hspace{2mm} \textnormal{and} \hspace{2mm}  \\ f_2(t,y) =  1/2-1/\pi\times\arctan (\sigma\sqrt{\beta/(2P_{\textnormal{max}}^s)}-y).
\end{multline}

Hence, applying limit and its properties to (\ref{tail_probability_with_no_RFI_7_7}) produces  
\begin{equation}
\label{tail_probability_with_no_RFI_7_8}
\lim_{\sigma^2 \to \infty} p(0) \leq \frac{1}{2\pi} \int_{-\infty}^{\infty} S_{1,2}(t, y, \sigma^2)  \frac{dy}{y^2+1}, 
\end{equation}
where $S_{1,2}(t, y, \sigma^2) =  \lim_{\sigma^2 \to \infty} f_1(t,y) + \lim_{\sigma^2 \to \infty} f_2(t,y) $. For $y\in\mathbb{R}$ and $P_{\textnormal{max}}^s \in (0, \infty)$, it follows from (\ref{f_1_t_and_y_def_2}) and the identity $\arctan \infty=\pi/2$ that $\lim_{\sigma^2 \to \infty} f_1(t,y)  =  \lim_{\sigma^2 \to \infty} f_2(t,y) =1/2-\arctan \infty/\pi=0$, and hence $S_{1,2}(t, y, \sigma^2)=0$. Plugging this value into the RHS of (\ref{tail_probability_with_no_RFI_7_8}), 
\begin{equation}
	\label{tail_probability_with_no_RFI_7_9}
	\lim_{\sigma^2 \to \infty} p(0) \leq 0.
\end{equation}
From the axioms of probability \cite{DPJN08}, $0 \leq p(0) \leq 1$ and hence $0 \leq \lim_{\sigma^2 \to \infty} p(0) \leq 1$. If we intersect this inequality and (\ref{tail_probability_with_no_RFI_7_9}), $\lim_{\sigma^2 \to \infty} p(0) = 0$. This is true $\forall\kappa \in [\alpha,1]$ -- of the fulfillment of Conditions \ref{condition_kappa_1} and \ref{beta_condition} -- and the first part of Theorem \ref{thm: fund_limit_under_no_interference} is corroborated.

Applying limit and its properties, it follows from (\ref{tail_probability_with_no_RFI_7_7}) that 
\begin{equation}
	\label{tail_probability_with_no_RFI_7_8_2}
	\lim_{P_{\textnormal{max}}^s \to 0} p(0) \leq \frac{1}{2\pi} \int_{-\infty}^{\infty} S_{1,2}(t, y, P_{\textnormal{max}}^s) \frac{dy}{y^2+1}, 
\end{equation} 
where $S_{1,2}(t, y, P_{\textnormal{max}}^s) = \lim_{P_{\textnormal{max}}^s \to 0} f_1(t,y) + \lim_{P_{\textnormal{max}}^s \to 0} f_2(t,y)$. It follows from (\ref{f_1_t_and_y_def_2}) and the identity $\arctan \infty=\pi/2$, w.r.t. $\sigma^2 \in (0, \infty)$ and $y\in\mathbb{R}$, that $\lim_{P_{\textnormal{max}}^s \to 0} f_1(t,y)  =  \lim_{P_{\textnormal{max}}^s \to 0} f_2(t,y) =1/2-\arctan \infty/\pi =0$, and thus $S_{1,2}(t, y, P_{\textnormal{max}}^s)=0$. Substituting this value into the RHS of (\ref{tail_probability_with_no_RFI_7_8_2}) leads to: 
\begin{equation}
	\label{tail_probability_with_no_RFI_7_9_2}
	\lim_{P_{\textnormal{max}}^s \to 0} p(0) \leq 0.
\end{equation}
From the axioms of probability \cite{DPJN08} and the properties of limit, $0 \leq \lim_{P_{\textnormal{max}}^s \to 0} p(0) \leq 1$. If we intersect this inequality and (\ref{tail_probability_with_no_RFI_7_9_2}), $\lim_{P_{\textnormal{max}}^s \to 0} p(0) = 0$. This is valid $\forall\kappa \in [\alpha,1]$ -- of the satisfaction of Conditions \ref{condition_kappa_1} and \ref{beta_condition} -- and the second part of Theorem \ref{thm: fund_limit_under_no_interference} is verified. This ends Theorem \ref{thm: fund_limit_under_no_interference}'s proof.   \QEDclosed

\section{Proof of Theorem \ref{thm: fund_limit_with_interference}}
\label{sec: proof_fund_limit_with_interference}
In this appendix, we set forth to analyze the performance limits of DeepSC subjected to RFI (i.e., $U=1$). Hence, we discard the subscript/superscript $u$ and let $\bm{v} \eqdef \bm{v}_u=\bm{v}_1$, $g \eqdef g_u=g_1\sim\mathcal{CN}(0,1)$, $P_{\textnormal{max}}^i \eqdef P_{\textnormal{max}}^{i,u}$, and $P_{\textnormal{min}}^i \eqdef P_{\textnormal{min}}^{i,u}$.    

To start, it follows from Appendix \ref{sec: proof_fund_limit_under_no_interference} and (\ref{tail_probability_with_no_RFI_1})-(\ref{tail_probability_with_no_RFI_6_1}) that  
\begin{equation}
\label{tail_probability_with_RFI_1}
p(0) \approx \mathbb{P}\big( \gamma \geq \beta  \big),
\end{equation}
where (\ref{tail_probability_with_RFI_1}) is valid under the fulfillment of Condition \ref{condition_kappa_1}, $\beta$ is defined in (\ref{beta_defn}) and constant for a given $K$, and $\gamma$ is the effective SNR in the presence of narrowband RFI -- modeled as in Sec. \ref{subsec: sysetm_model} -- $g\bm{v}\in\mathbb{R}^{1 \times KL}$. In the presence of narrowband RFI $g\bm{v}$, where $g\sim\mathcal{N}(0,1)$, the RFI is treated as noise by the channel decoder -- which readily introduces semantic noise to the semantic decoder -- and the SNR is identical to the SINR $\gamma   \eqdef \frac{\big([\textnormal{Re}\{h\}]^2+[\textnormal{Im}\{h\}]^2\big)|(\bm{x})_i|^2}{\big([\textnormal{Re}\{g\}]^2+[\textnormal{Im}\{g\}]^2\big)|(\bm{v})_i|^2+[\textnormal{Re}\{(\bm{n})_i\}]^2+[\textnormal{Im}\{(\bm{n})_i\}]^2}$. As a result, $\gamma  \stackrel{(a)}{\leq} \frac{\big([\textnormal{Re}\{h\}]^2+[\textnormal{Im}\{h\}]^2\big)|(\bm{x})_i|^2}{\big([\textnormal{Re}\{g\}]^2+[\textnormal{Im}\{g\}]^2\big)|(\bm{v})_i|^2 }$: 
\begin{subequations}
\begin{align}
\label{SNR_definition_with_RFI_4}
 \gamma &\stackrel{(b)}{\leq}  \frac{2P_{\textnormal{max}}^s\big([\textnormal{Re}\{h\}]^2+[\textnormal{Im}\{h\}]^2\big)}{\big([\textnormal{Re}\{g\}]^2+[\textnormal{Im}\{g\}]^2\big)|(\bm{v})_i|^2 }  \\
\label{SNR_definition_with_RFI_5}
&\stackrel{(c)}{\leq} 
\frac{2P_{\textnormal{max}}^s\big([\textnormal{Re}\{h\}]^2+[\textnormal{Im}\{h\}]^2\big)}{[\textnormal{Re}\{g\}]^2|(\bm{v})_i|^2 }, 
\end{align}
\end{subequations}
where $i\in[KL]$; $(a)$ is because of the constraint $[\textnormal{Re}\{(\bm{n})_i\}]^2+[\textnormal{Im}\{(\bm{n})_i\}]^2 \geq 0$; $(b)$ follows from the power constraint $[\textnormal{Re}\{(\bm{x})_i\}]^2, [\textnormal{Im}\{(\bm{x})_i\}]^2 \leq P_{\textnormal{max}}^s$ and hence $[\textnormal{Re}\{(\bm{x})_i\}]^2  + [\textnormal{Im}\{(\bm{x})_i\}]^2   =|(\bm{x})_i|^2 \leq 2P_{\textnormal{max}}^s$; and $(c)$ is due to the constraint $[\textnormal{Im}\{g\}]^2|(\bm{v})_i|^2 \geq 0$. 
	
To bound the RHS of (\ref{SNR_definition_with_RFI_5}), we bound $[\textnormal{Re}\{g\}]^2|(\bm{v})_i|^2$ by employing the $u$-th RFI power constraint of Sec. \ref{subsec: sysetm_model}. Per the model in Sec. \ref{subsec: sysetm_model}, the unknown RFI symbols satisfy the power constraint $P_{\textnormal{min}}^i \leq \mathbb{E}\{[\textnormal{Re}\{(\bm{v})_i\} ]^2 \}, \mathbb{E}\{[ \textnormal{Im}\{(\bm{v})_i\} ]^2 \} \leq P_{\textnormal{max}}^i$. This constraint can thus be expressed in terms of $|(\bm{v})_i|^2=[\textnormal{Re}\{(\bm{v})_i\} ]^2  + [ \textnormal{Im}\{(\bm{v})_i\} ]^2$ as $2P_{\textnormal{min}}^i \leq |(\bm{v})_i|^2 \leq 2P_{\textnormal{max}}^i$ or $2P_{\textnormal{max}}^i \geq |(\bm{v})_i|^2 \geq 2P_{\textnormal{min}}^i$. From these inequalities, the following relations follow: $1/|(\bm{v})_i|^2    \leq 1/2P_{\textnormal{min}}^i  $ and   
\begin{equation}
\label{RFI_power_constraint_2}
\frac{2P_{\textnormal{max}}^s\big([\textnormal{Re}\{h\}]^2+[\textnormal{Im}\{h\}]^2\big)}{|(\bm{v})_i|^2[\textnormal{Re}\{g\}]^2}  \stackrel{(a)}{\leq}	 \frac{2P_{\textnormal{max}}^s\big([\textnormal{Re}\{h\}]^2+[\textnormal{Im}\{h\}]^2\big)}{2P_{\textnormal{min}}^i[\textnormal{Re}\{g\}]^2},
\end{equation}
where $(a)$ follows from multiplying both sides by $\frac{2P_{\textnormal{max}}^s\big([\textnormal{Re}\{h\}]^2+[\textnormal{Im}\{h\}]^2\big)}{[\textnormal{Re}\{g\}]^2} \geq 0$. As a result, using (\ref{RFI_power_constraint_2}) in the RHS of (\ref{SNR_definition_with_RFI_5}) leads to the relationship
\begin{subequations}
\begin{align}
\label{SNR_definition_with_RFI_7_1}
\gamma  & \leq  \frac{2P_{\textnormal{max}}^s\big([\textnormal{Re}\{h\}]^2+[\textnormal{Im}\{h\}]^2\big)}{2P_{\textnormal{min}}^i[\textnormal{Re}\{g\}]^2}    \\
\label{SNR_definition_with_RFI_7}
&=\frac{P_{\textnormal{max}}^s\big([\sqrt{2}\textnormal{Re}\{h\}]^2+[\sqrt{2}\textnormal{Im}\{h\}]^2\big)}{P_{\textnormal{min}}^i[\sqrt{2}\textnormal{Re}\{g\}]^2} \\
\label{SNR_definition_with_RFI_8}
& \leq P_{\textnormal{max}}^s/P_{\textnormal{min}}^i \big[ \big(\tilde{A}/\tilde{C} \big)^2 + \big( \tilde{B}/\tilde{C} \big)^2 \big],
\end{align}
\end{subequations}
where $\textnormal{Re}\{h\}, \textnormal{Im}\{h\}, \textnormal{Re}\{g\} \sim \mathcal{N}(0,1/2)$ and\linebreak $\tilde{A}\eqdef \sqrt{2}\textnormal{Re}\{h\}, \tilde{B} \eqdef \sqrt{2}\textnormal{Im}\{h\}, \tilde{C} \eqdef \sqrt{2}\textnormal{Re}\{g\} \sim \mathcal{N}(0,1)$.

Thus, w.r.t. $\beta$, it follows from (\ref{SNR_definition_with_RFI_8}) that 
\begin{subequations}
\begin{align}
\label{prob_relation_with_RFI_2_1}
\mathbb{P}\big( \gamma \geq \beta  \big)  & \leq \mathbb{P}\big( P_{\textnormal{max}}^s/P_{\textnormal{min}}^i\big[ \big( \tilde{A}/\tilde{C} \big)^2 + \big( \tilde{B}/\tilde{C} \big)^2 \big] \geq \beta  \big)     \\
\label{prob_relation_with_RFI_2}
&=\mathbb{P}\big(  \big( \tilde{A}/\tilde{C} \big)^2 + \big( \tilde{B}/ \tilde{C} \big)^2 \geq  \beta P_{\textnormal{min}}^i/P_{\textnormal{max}}^s \big).
\end{align}
\end{subequations}
If we now let $\tilde{X} \eqdef \tilde{A}/\tilde{C}$ and $\tilde{Y}\eqdef \tilde{B}/\tilde{C}$, and use these RVs in the RHS of (\ref{prob_relation_with_RFI_2}), $\mathbb{P}\big( \gamma \geq \beta  \big)   \leq \mathbb{P}\big(  \tilde{X}^2 + \tilde{Y}^2  \geq \beta P_{\textnormal{min}}^i/P_{\textnormal{max}}^s  \big)$:
\begin{equation}
\label{prob_relation_with_RFI_5}
\mathbb{P}\big( \gamma \geq \beta  \big)   \stackrel{(a)}{\leq}  \mathbb{P}\big(  \sqrt{\tilde{X}^2 + \tilde{Y}^2 }  \geq  \tilde{t}  \big)    
\end{equation}
where $(a)$ is true under the satisfaction of Conditions \ref{condition_kappa_1} and \ref{beta_condition} as well as $\tilde{t} \eqdef  \sqrt{\beta P_{\textnormal{min}}^i/P_{\textnormal{max}}^s}$. Plugging (\ref{prob_relation_with_RFI_5}) into (\ref{tail_probability_with_RFI_1}),  
\begin{equation}
\label{tail_probability_with_RFI_2_1}
p(0) \leq \mathbb{P}\big(  \sqrt{\tilde{X}^2 + \tilde{Y}^2 }  \geq  \tilde{t}  \big).
\end{equation}
Since $\tilde{X}^2, \tilde{Y}^2 \geq 0$, it follows via \cite[Lemma 11, p. 71]{Getu_DSFC_Estimation'22} that
\begin{equation}
	\label{squareroot_inequality_identity_addition}
	\sqrt{ \tilde{X}^2 +\tilde{Y}^2 } \leq  \sqrt{ \tilde{X}^2 }+\sqrt{ \tilde{Y}^2 } \stackrel{(a)}{=} | \tilde{X}| + |\tilde{Y}|,
\end{equation}
where $(a)$ is for $\sqrt{x^2}=|x|=\mathbb{I}\{x\geq 0\}x-\mathbb{I}\{x < 0\}x$. Using the inequality in (\ref{squareroot_inequality_identity_addition}) in the RHS of (\ref{tail_probability_with_RFI_2_1}), it follows that
\begin{subequations}
\begin{align}
\label{tail_probability_with_RFI_2_3_1}
p(0)  \leq \mathbb{P}\big(  \sqrt{\tilde{X}^2 + \tilde{Y}^2 }   \geq  \tilde{t}  \big) &\leq \mathbb{P}\big(  | \tilde{X}| + |\tilde{Y}|  \geq  \tilde{t}  \big)   \\
\label{tail_probability_with_RFI_2_3}
&=\mathbb{P}\big(  | \tilde{X}|  \geq  \tilde{t}-|\tilde{Y}|  \big).
\end{align}
\end{subequations}
The RV $\tilde{Y}$ is defined in above as the ratio of two standard normal RVs. Thus, $|\tilde{Y}|=\tilde{Y}$ with probability of $1/2$ and\linebreak $|\tilde{Y}|=-\tilde{Y}$, also with a probability $1/2$. Using these probabilities, implementing the total probability theorem \cite[p. 28]{DPJN08}\linebreak to the RHS of (\ref{tail_probability_with_RFI_2_3}) leads to the inequality  
\begin{equation}
\label{tail_probability_with_RFI_2_4}
p(0) \leq  \big[ \mathbb{P}\big(  | \tilde{X}|  \geq  \tilde{t}+\tilde{Y}  \big)+ \mathbb{P}\big(  | \tilde{X}|  \geq  \tilde{t}-\tilde{Y}  \big) \big]/2.
\end{equation}
We can now use Appendix \ref{sec: proof_fund_limit_under_no_interference}'s results to simplify (\ref{tail_probability_with_RFI_2_4}). 

The RVs in the RHS of (\ref{tail_probability_with_RFI_2_4}) are the ratios between the independent standard normal RVs $\tilde{A}, \tilde{C}\sim\mathcal{N}(0,1)$ and $\tilde{B}, \tilde{C}\sim\mathcal{N}(0,1)$, respectively. Similarly, the RVs $X$ and $Y$ -- defined in Appendix \ref{sec: proof_fund_limit_under_no_interference} -- are the ratios between the independent standard normal RVs $A, C \sim\mathcal{N}(0,1)$ and $B, C \sim\mathcal{N}(0,1)$, respectively. Consequently, $X$ and $\tilde{X}$ as well as $Y$ and $\tilde{Y}$ have the same distributions. To this end,  
\begin{multline}
\label{prob_equality_1}
\mathbb{P}(  | \tilde{X}|  \geq  \tilde{t}+\tilde{Y}  )  =  \mathbb{P}(  | X|  \geq  t+Y  ) \big|_{t=\tilde{t}}  \hspace{2mm}\textnormal{and}  \\  \mathbb{P}(  | \tilde{X}|  \geq  \tilde{t}-\tilde{Y}  ) = \mathbb{P}(  | X|  \geq  t-Y  ) \big|_{t=\tilde{t}}.
\end{multline}
We can now exploit Appendix \ref{sec: proof_fund_limit_under_no_interference}'s results for $\mathbb{P}\big(  | X|  \geq  t+Y  \big)$ and $\mathbb{P}\big(  | X|  \geq  t-Y  \big)$ to simplify the RHSs of (\ref{prob_equality_1}). To this end, plugging (\ref{probability_geq_t+Y_2}) into the RHS of (\ref{prob_equality_1}) results in
\begin{multline}
\label{probability_geq_tilde_t+Y_2}
\mathbb{P}\big(  | \tilde{X}|  \geq  \tilde{t}+\tilde{Y}  \big)  = \int_{-\infty}^{\infty} f_1(\tilde{t},y)  \frac{dy}{\pi(y^2+1)}  \hspace{2mm} \textnormal{and} \hspace{2mm}  \\ \mathbb{P}\big(  | \tilde{X}|  \geq  \tilde{t}-\tilde{Y}  \big) = \int_{-\infty}^{\infty} f_2(\tilde{t},y)  \frac{dy}{\pi(y^2+1)}, 
\end{multline}
where $f_1(\tilde{t},y)  =  1/2-1/\pi\times\arctan \big(\sqrt{\beta P_{\textnormal{min}}^i/P_{\textnormal{max}}^s}+y\big)$ and $f_2(\tilde{t},y) =  1/2-1/\pi\times\arctan \big(\sqrt{\beta P_{\textnormal{min}}^i/P_{\textnormal{max}}^s }-y\big)$. Meanwhile, deploying (\ref{probability_geq_tilde_t+Y_2}) in the RHS of (\ref{tail_probability_with_RFI_2_4}) gives      
\begin{equation}
\label{tail_probability_with_RFI_2_5}
p(0)  \leq \frac{1}{2} \int_{-\infty}^{\infty} \big[ f_1(\tilde{t},y) + f_2(\tilde{t},y) \big]  \frac{dy}{\pi(y^2+1)}.
\end{equation} 

Implementing limit and its properties in (\ref{tail_probability_with_RFI_2_5}) gives us 
\begin{equation}
	\label{tail_probability_with_RFI_2_6}
	\lim_{P_{\textnormal{min}}^i \to \infty} p(0) \leq \frac{1}{2\pi} \int_{-\infty}^{\infty}  \tilde{S}_{1,2} (\tilde{t}, y, P_{\textnormal{min}}^i)  \frac{dy}{y^2+1}, 
\end{equation}
where $\tilde{S}_{1,2} (\tilde{t}, y, P_{\textnormal{min}}^i) = \lim_{P_{\textnormal{min}}^i \to \infty} f_1(\tilde{t},y) + \lim_{P_{\textnormal{min}}^i \to \infty} f_2(\tilde{t},y)$. For a given $y\in\mathbb{R}$, it follows from the identity $\arctan \infty=\pi/2$ and the above definition of $f_1(\tilde{t},y)$ and $f_2(\tilde{t},y)$ that $\lim_{P_{\textnormal{min}}^i \to \infty} f_1(\tilde{t},y)  =\lim_{P_{\textnormal{min}}^i \to \infty} f_2(\tilde{t},y)=  
1/2-\arctan\infty/\pi =0$, and hence $\tilde{S}_{1,2} (\tilde{t}, y, P_{\textnormal{min}}^i)=0$. Consequently, plugging this value into the RHS of (\ref{tail_probability_with_RFI_2_6}) gives 
\begin{equation}
	\label{tail_probability_with_RFI_2_7}
	\lim_{P_{\textnormal{min}}^i \to \infty} p(0) \leq 0.
\end{equation}
From the axioms of probability \cite{DPJN08} and the properties of limit, $0 \leq \lim_{P_{\textnormal{min}}^i \to \infty} p(0) \leq 1$. Intersecting this inequality and (\ref{tail_probability_with_RFI_2_7}) leads to the result $\lim_{P_{\textnormal{min}}^i \to \infty } p(0) = 0$. This is true $\forall\kappa \in [\alpha,1]$ -- under the satisfaction of Conditions \ref{condition_kappa_1} and \ref{beta_condition} -- and the first part of Theorem \ref{thm: fund_limit_with_interference} is validated.

Furthermore, applying limit and its properties to (\ref{tail_probability_with_RFI_2_5}) gives    
\begin{equation}
	\label{tail_probability_with_RFI_2_6_2}
	\lim_{P_{\textnormal{max}}^s \to 0} p(0) \leq \frac{1}{2\pi} \int_{-\infty}^{\infty} \tilde{S}_{1,2} (\tilde{t}, y, P_{\textnormal{max}}^s)   \frac{dy}{y^2+1},
\end{equation}
where $\tilde{S}_{1,2} (\tilde{t}, y, P_{\textnormal{max}}^s) =\lim_{P_{\textnormal{max}}^s \to 0} f_1(\tilde{t},y) + \lim_{P_{\textnormal{max}}^s \to 0} f_2(\tilde{t},y)$. It follows from the identity $\arctan \infty=\pi/2$ and the aforementioned definition of $f_1(\tilde{t},y)$ and $f_2(\tilde{t},y)$ -- w.r.t. any $y\in\mathbb{R}$ and $P_{\textnormal{min}}^i\in(0, \infty)$ -- that $\lim_{P_{\textnormal{max}}^s \to 0} f_1(\tilde{t},y)  =  \lim_{P_{\textnormal{max}}^s \to 0} f_2(\tilde{t},y)=
1/2-\arctan\infty/\pi=0$, and thus $\tilde{S}_{1,2} (\tilde{t}, y, P_{\textnormal{max}}^s) =0$. Replacing this value into the RHS of (\ref{tail_probability_with_RFI_2_6_2}) leads to   
\begin{equation}
	\label{tail_probability_with_RFI_2_7_2}
	\lim_{P_{\textnormal{max}}^s \to 0} p(0) \leq 0.
\end{equation}
From the axioms of probability \cite{DPJN08} and the properties of limit, $0 \leq \lim_{P_{\textnormal{max}}^s \to 0} p(0) \leq 1$. Intersecting this inequality and (\ref{tail_probability_with_RFI_2_7_2}) leads to $\lim_{ P_{\textnormal{max}}^s \to 0 } p(0) = 0$. This is applicable $\forall\kappa \in [\alpha,1]$ -- under Conditions \ref{condition_kappa_1} and \ref{beta_condition} -- and the second part of Theorem \ref{thm: fund_limit_with_interference} is validated. This ends Theorem \ref{thm: fund_limit_with_interference}'s proof.   \QEDclosed

\section{Proof of Theorem \ref{thm: fund_limit_with_MU-interference}}
\label{sec: proof_fund_limit_with_MU-interference}
To begin, it follows from Appendix \ref{sec: proof_fund_limit_under_no_interference} and (\ref{tail_probability_with_no_RFI_1})-(\ref{tail_probability_with_no_RFI_6_1}) that   
\begin{equation}
	\label{tail_probability_with_MU-RFI_1}
	p(0) \approx \mathbb{P}\big( \gamma \geq \beta  \big),
\end{equation}
where (\ref{tail_probability_with_MU-RFI_1}) is subject to Condition \ref{condition_kappa_1} being met, $\beta$ is given in (\ref{beta_defn}) and constant for a given $K$, and $\gamma$ is the effective SNR under MI RFI -- per Sec. \ref{subsec: sysetm_model} -- $\sum_{u=1}^U g_u\bm{v}_u\in\mathbb{R}^{1 \times KL}$. In the presence of narrowband MI RFI $\sum_{u=1}^U g_u\bm{v}_u$, where $g_u\sim\mathcal{CN}(0,1)$ $\forall u\in[U]$, the RFI is treated as noise by the channel decoder -- which introduces semantic noise to the semantic decoder -- and the effective SNR is the same as the SINR $\gamma   \eqdef  \frac{|h|^2|(\bm{x})_i|^2}{\sum_{u=1}^U |g_u|^2|(\bm{v}_u)_i|^2+|(\bm{n})_i|^2}$:  
\begin{equation}
\label{SNR_definition_with_MU-RFI_2}
\gamma \stackrel{(a)}{\leq}\frac{\big([\textnormal{Re}\{h\}]^2+[\textnormal{Im}\{h\}]^2\big)|(\bm{x})_i|^2}{\sum_{u=1}^U \big([\textnormal{Re}\{g_u\}]^2+[\textnormal{Im}\{g_u\}]^2\big)|(\bm{v}_u)_i|^2},
\end{equation}
where $i\in[KL]$ and $(a)$ is for $|(\bm{n})_i|^2 \geq 0$. Per Sec. \ref{subsec: sysetm_model}'s settings, it follows that $[\textnormal{Re}\{(\bm{x})_i\}]^2, [\textnormal{Im}\{(\bm{x})_i\}]^2 \leq P_{\textnormal{max}}^s$ and hence $[\textnormal{Re}\{(\bm{x})_i\}]^2  + [\textnormal{Im}\{(\bm{x})_i\}]^2   =|(\bm{x})_i|^2 \leq 2P_{\textnormal{max}}^s$. Deploying this bound in the RHS of (\ref{SNR_definition_with_MU-RFI_2}) leads to
\begin{subequations}
\begin{align}
\label{SNR_definition_with_MU-RFI_4_1}
\gamma  & \leq \frac{2P_{\textnormal{max}}^s\big([\textnormal{Re}\{h\}]^2+[\textnormal{Im}\{h\}]^2\big)}{\sum_{u=1}^U \big([\textnormal{Re}\{g_u\}]^2+[\textnormal{Im}\{g_u\}]^2\big)|(\bm{v}_u)_i|^2}   \\
\label{SNR_definition_with_MU-RFI_4}
&=\frac{P_{\textnormal{max}}^s\big([\sqrt{2}\textnormal{Re}\{h\}]^2+[\sqrt{2}\textnormal{Im}\{h\}]^2\big)}{\sum_{u=1}^U \big([\textnormal{Re}\{g_u\}]^2+[\textnormal{Im}\{g_u\}]^2\big)|(\bm{v}_u)_i|^2},
\end{align}
\end{subequations} 
where $\textnormal{Re}\{g_u\}, \textnormal{Im}\{g_u\} \sim \mathcal{N}(0, 1/2)$ $\forall u\in[U]$ and $\textnormal{Re}\{h\}, \textnormal{Im}\{h\} \sim \mathcal{N}(0, 1/2)$ are all independent Gaussian RVs. To bound the RHS of (\ref{SNR_definition_with_MU-RFI_4}), we will bound $|(\bm{v}_u)_i|^2$.

From the system model in Sec. \ref{subsec: sysetm_model}, the unknown symbols of the $u$-th RFI emitter satisfy the power-constraint $P_{\textnormal{min}}^{i,u} \leq \mathbb{E}\{[\textnormal{Re}\{(\bm{v}_u)_i\} ]^2 \}, \mathbb{E}\{[ \textnormal{Im}\{(\bm{v}_u)_i\} ]^2 \} \leq P_{\textnormal{max}}^{i,u}$ $\forall i\in[KL]$ and all $u\in[U]$. This constraint can hence be expressed as $2P_{\textnormal{min}}^{i,u} \leq |(\bm{v}_u)_i|^2 \leq 2P_{\textnormal{max}}^{i,u}$ or $2P_{\textnormal{max}}^{i,u} \geq |(\bm{v}_u)_i|^2 \geq 2P_{\textnormal{min}}^{i,u}$ for $|(\bm{v}_u)_i|^2=[\textnormal{Re}\{(\bm{v}_u)_i\} ]^2  + [ \textnormal{Im}\{(\bm{v}_u)_i \} ]^2$. Thus, it is true $\forall u\in[U]$ that $ 2\tilde{P}_{\textnormal{max}}^i  \geq |(\bm{v}_u)_i|^2 \geq 2\tilde{P}_{\textnormal{min}}^i$ and
\begin{equation}
\label{inequality_constraint_with_MU-RFI_1}
1/2\tilde{P}_{\textnormal{max}}^i  \leq  1/|(\bm{v}_u)_i|^2  \leq   1/2\tilde{P}_{\textnormal{min}}^i ,
\end{equation}
where $\tilde{P}_{\textnormal{max}}^i \eqdef \textnormal{max} \big(P_{\textnormal{max}}^{i,1}, P_{\textnormal{max}}^{i,2}, \ldots, P_{\textnormal{max}}^{i,U}\big)$ and $\tilde{P}_{\textnormal{min}}^i \eqdef \textnormal{min} \big(P_{\textnormal{min}}^{i,1}, P_{\textnormal{min}}^{i,2}, \ldots, P_{\textnormal{min}}^{i,U}\big)$. The inequality in the RHS of (\ref{inequality_constraint_with_MU-RFI_1}) leads to the inequality $\frac{P_{\textnormal{max}}^s([\sqrt{2}\textnormal{Re}\{h\}]^2+[\sqrt{2}\textnormal{Im}\{h\}]^2)}{\sum_{u=1}^U ([\textnormal{Re}\{g_u\}]^2+[\textnormal{Im}\{g_u\}]^2)|(\bm{v}_u)_i|^2} \leq     \frac{P_{\textnormal{max}}^s([\sqrt{2}\textnormal{Re}\{h\}]^2+[\sqrt{2}\textnormal{Im}\{h\}]^2)}{2\tilde{P}_{\textnormal{min}}^i (\sum_{u=1}^U [\textnormal{Re}\{g_u\}]^2+[\textnormal{Im}\{g_u\}]^2)}$, where this follows from the RHS of (\ref{inequality_constraint_with_MU-RFI_1}) via multiplication by $\frac{P_{\textnormal{max}}^s\big([\sqrt{2}\textnormal{Re}\{h\}]^2+[\sqrt{2}\textnormal{Im}\{h\}]^2\big)}{\sum_{u=1}^U \big([\textnormal{Re}\{g_u\}]^2+[\textnormal{Im}\{g_u\}]^2\big)}$. Using the above inequality in the RHS of (\ref{SNR_definition_with_MU-RFI_4}) gives us     
\begin{subequations}
\begin{align}
\label{SNR_definition_with_MU-RFI_5}
\gamma     &   \leq \frac{P_{\textnormal{max}}^s\big([\sqrt{2}\textnormal{Re}\{h\}]^2+[\sqrt{2}\textnormal{Im}\{h\}]^2\big)}{\tilde{P}_{\textnormal{min}}^i \big(\sum_{u=1}^U [\sqrt{2}\textnormal{Re}\{g_u\}]^2+[\sqrt{2}\textnormal{Im}\{g_u\}]^2\big)}    \\
\label{SNR_definition_with_MU-RFI_6}
& = P_{\textnormal{max}}^s \big(\sum_{i=1}^2 X_i^2 \big) / \big[\tilde{P}_{\textnormal{min}}^i \big(\sum_{u=1}^{2U} Y_u^2 \big)\big] ,
\end{align}  
\end{subequations}
where $X_1 \eqdef \sqrt{2}\textnormal{Re}\{h\},  X_2 \eqdef \sqrt{2}\textnormal{Im}\{h\} \sim \mathcal{N}(0,1)$ and $Y_u \eqdef \sqrt{2}\textnormal{Re}\{g_u\},  Y_{u+U} \eqdef \sqrt{2}\textnormal{Im}\{g_u\} \sim \mathcal{N}(0,1)$, $\forall u\in[U]$,\linebreak are mutually independent standard normal RVs. Thus, it follows from the inequality in (\ref{SNR_definition_with_MU-RFI_6}) that
\begin{equation}
\label{Prob_inequality_wrt_SNR_MU-RFI_2}
\mathbb{P}\big( \gamma \geq \beta  \big) \leq \mathbb{P}\big( \sum_{i=1}^2 X_i^2/\big[\sum_{u=1}^{2U} Y_u^2 \big]  \geq \beta\tilde{P}_{\textnormal{min}}^i/P_{\textnormal{max}}^s \big).
\end{equation}

Deploying (\ref{Prob_inequality_wrt_SNR_MU-RFI_2}) in the RHS of (\ref{tail_probability_with_MU-RFI_1}), meanwhile, gives us   
\begin{equation}
\label{tail_probability_with_MU-RFI_3}
p(0)  \leq  \mathbb{P}\big(  Z_1/Z_2 \geq \beta\tilde{P}_{\textnormal{min}}^i/P_{\textnormal{max}}^s \big), 
\end{equation}
where $\big(Z_1, Z_2\big)\eqdef \big(\sum_{i=1}^2 X_i^2, \sum_{u=1}^{2U} Y_u^2\big)$. For these sums of mutually independent squared standard normal RVs, $Z_1$ and $Z_2$ are mutually independent $\chi^2$-distributed RVs with a DoF of $2$ and $2U$, respectively \cite[Ch. 18]{NJKB_Vol_I'94}. Consequently, $Z_1 \sim \chi_2^2$ and  $Z_2 \sim \chi_{2U}^2$. W.r.t. these independent $\chi^2$-distributed RVs, it follows from from (\ref{tail_probability_with_MU-RFI_3}) that $p(0)  \leq \mathbb{P}\big(  \frac{Z_1/4U}{Z_2/4U}   \geq  \frac{\beta\tilde{P}_{\textnormal{min}}^i}{P_{\textnormal{max}}^s} \big)=\mathbb{P}\big(  \frac{Z_1/2 \times 1/2U}{Z_2/2U \times 1/2}   \geq  \frac{\beta\tilde{P}_{\textnormal{min}}^i}{P_{\textnormal{max}}^s} \big)$. Accordingly,     
\begin{equation}
\label{tail_probability_with_MU-RFI_4}
 p(0)         
\stackrel{(a)}{\leq}\mathbb{P}\big(  R    \geq  \beta U\tilde{P}_{\textnormal{min}}^i/P_{\textnormal{max}}^s  \big), 
\end{equation}
where $(a)$ follows from letting $R \eqdef \frac{Z_1/2}{Z_2/2U}$. As $R$ is the ratio of two normalized independent $\chi^2$-distributed RVs that are normalized by their respective DoF, $R$ is an $F$-distributed RV with $2$ and $2U$ DoFs \cite[Ch. 27]{NJKB_Vol_II'95}. Hence, $R \sim F_{2,2U}$ and   
\begin{equation}
\label{tail_probability_with_MU-RFI_9}
p(0) \leq \mathbb{P}\big(  R    \geq  \beta U\tilde{P}_{\textnormal{min}}^i/P_{\textnormal{max}}^s\big).
\end{equation}
Subject to Condition \ref{beta_condition} being met, $\beta> 0$, which paves the way for the simplification of (\ref{tail_probability_with_MU-RFI_9}) using \textit{Markov’s inequality} \cite{vershynin_2018}.\linebreak Applying Markov’s inequality \cite[Proposition 1.2.4]{vershynin_2018} -- subject to Condition \ref{beta_condition} being met -- to the RHS of (\ref{tail_probability_with_MU-RFI_9}) gives
\begin{equation}
\label{tail_probability_with_MU-RFI_10-13}
 p(0)  \leq \mathbb{E}\{R\}/c      \stackrel{(a)}{=}2UP_{\textnormal{max}}^s/ [2(U-1)\beta U\tilde{P}_{\textnormal{min}}^i],  
\end{equation}
where $c \eqdef \beta U\tilde{P}_{\textnormal{min}}^i/P_{\textnormal{max}}^s$ and $(a)$ follows from (\ref{mean_R_F_distribution}) that $\mathbb{E}\{R\}= 2U/(2U-2)$ given that $\nu_2=2U > 2$. Thus, subject to Conditions \ref{condition_kappa_1} and \ref{beta_condition} being met, it follows for $U>1$ that
\begin{equation}
\label{tail_probability_with_MU-RFI_13}
p(0)  \leq  P_{\textnormal{max}}^s/[\beta (U-1) \tilde{P}_{\textnormal{min}}^i] .
\end{equation}
For (\ref{tail_probability_with_MU-RFI_13}) to be a valid probability bound, $\frac{P_{\textnormal{max}}^s}{\beta (U-1) \tilde{P}_{\textnormal{min}}^i} \leq 1$ and hence the following condition:  
\begin{condition}
\label{upper_bound_consstraint}
$P_{\textnormal{max}}^s \leq \beta (U-1) \tilde{P}_{\textnormal{min}}^i$.
\end{condition}

Therefore, under the satisfaction of Conditions \ref{condition_kappa_1}, \ref{beta_condition}, and \ref{upper_bound_consstraint},
\begin{subequations}
\begin{align}
\label{asymptotic_result_1}
\lim_{P_{\textnormal{max}}^s \to 0} p(0)  & \leq  0,  \hspace{1mm} \textnormal{for $U>1$ and $\tilde{P}_{\textnormal{min}}^i \in (0, \infty)$};    \\
\label{asymptotic_result_2}
\lim_{\tilde{P}_{\textnormal{min}}^i \to \infty} p(0) & \leq  0,  \hspace{1mm} \textnormal{for $U>1$ and $ P_{\textnormal{max}}^s\in (0, \infty)$};   \\
\label{asymptotic_result_3}
\lim_{U \to \infty} p(0) &\leq  0, \hspace{1mm} \textnormal{for $P_{\textnormal{max}}^s, \tilde{P}_{\textnormal{min}}^i \in (0, \infty)$.}
\end{align}
\end{subequations}
Moreover, from the axiomatic constraint of probability \cite{DPJN08}, it is evident that $0 \leq \lim_{P_{\textnormal{max}}^s \to 0} p(0) \leq 1$. Intersecting this inequality and the inequality in (\ref{asymptotic_result_1}), $\lim_{P_{\textnormal{max}}^s \to 0} p(0) =  0$. This proves the first part of Theorem \ref{thm: fund_limit_with_MU-interference}. Similarly, $0 \leq \lim_{\tilde{P}_{\textnormal{min}}^i \to \infty} p(0) \leq 1$ and $0 \leq \lim_{U \to \infty} p(0) \leq 1$. Intersecting these inequalities and the inequalities in (\ref{asymptotic_result_2}) and (\ref{asymptotic_result_3}) results in $\lim_{\tilde{P}_{\textnormal{min}}^i \to \infty} p(0) =0$ and $\lim_{U \to \infty} p(0) =0$, respectively. These results validate the second and third part of Theorem \ref{thm: fund_limit_with_MU-interference}, respectively. This also ends the proof of Theorem \ref{thm: fund_limit_with_MU-interference}.   \QEDclosed

\section{Proof of Theorem \ref{thm: practical_limits_with_MU-interference}}
\label{sec: proof_practical_limits_with_MU-interference}
Using (\ref{simantic_similarity_function_1}) in the RHS of (\ref{tail_probability_defn}) for any given $K$ and $\eta_{\textnormal{min}}$,
\begin{equation}
\label{practical_limits_with_RFI_1}
p(\eta_{\textnormal{min}}) = \mathbb{P}\big( \varepsilon(K,\gamma) \geq \eta_{\textnormal{min}} \big) \stackrel{(a)}{\approx} \mathbb{P}\big( \tilde{\varepsilon}_K(\gamma) \geq \eta_{\textnormal{min}} \big), 
\end{equation}
where $(a)$ is due to (\ref{simantic_similarity_function_approximation_1}). Plugging (\ref{simantic_similarity_function_approximation_1}) into the RHS of (\ref{practical_limits_with_RFI_1}), 
\begin{subequations}
	\begin{align}
		\label{practical_limits_with_RFI_2}
		\hspace{-2mm} p(\eta_{\textnormal{min}}) & \approx \mathbb{P}\Big( \frac{A_{K,2}-A_{K,1}}{1+ e^{-(C_{K,1}\gamma+C_{K,2})}} \geq \eta_{\textnormal{min}}-A_{K,1} \Big)  \\
		\label{practical_limits_with_RFI_3}
		&= \mathbb{P}\Big( \frac{1}{1+ e^{-(C_{K,1}\gamma+C_{K,2})}} \geq \kappa_K(\eta_{\textnormal{min}}) \Big)  \\
		\label{practical_limits_with_RFI_3_1}
		&= \mathbb{P}\big( e^{-(C_{K,1}\gamma+C_{K,2})} \leq 1/\kappa_K(\eta_{\textnormal{min}})-1 \big)  \\
		\label{practical_limits_with_RFI_4}
		&=\mathbb{P}\Big( -(C_{K,1}\gamma+C_{K,2}) \leq \ln \frac{1-\kappa_K(\eta_{\textnormal{min}})}{\kappa_K(\eta_{\textnormal{min}})}   \Big)    \\
		\label{practical_limits_with_RFI_5}
		&=\mathbb{P}\Big( (C_{K,1}\gamma+C_{K,2}) \geq \ln \frac{\kappa_K(\eta_{\textnormal{min}})}{(1-\kappa_K(\eta_{\textnormal{min}}))} \Big)   \\
		\label{practical_limits_with_RFI_6}
		&\stackrel{(a)}{=}\mathbb{P}\Big( \gamma \geq \frac{\ln[ \kappa_K(\eta_{\textnormal{min}})/(1-\kappa_K(\eta_{\textnormal{min}}))]}{C_{K,1}} - \frac{C_{K,2}}{C_{K,1}}  \Big)    \\ 
		\label{practical_limits_with_RFI_7}
		& =\mathbb{P}\big( \gamma \geq \beta_K(\eta_{\textnormal{min}})  \big),
	\end{align}
\end{subequations}  
where $\kappa_K(\eta_{\textnormal{min}}) \eqdef (\eta_{\textnormal{min}}-A_{K,1})/(A_{K,2}-A_{K,1}) \geq 0$, $(a)$ is due to the logistic growth rate constraint $C_{K,1}> 0$, $\gamma$ is the SINR defined in Appendix \ref{sec: proof_fund_limit_with_MU-interference}, and
\begin{equation}
\label{beta_eta_min_defn}
\beta_K(\eta_{\textnormal{min}})  \eqdef  \frac{\ln\big[ \kappa_K(\eta_{\textnormal{min}})/(1-\kappa_K(\eta_{\textnormal{min}}))\big]}{C_{K,1}}- \frac{C_{K,2}}{C_{K,1}},
\end{equation}
where $\beta_K(\eta_{\textnormal{min}})\in\mathbb{R}$ is constant for a given $K$ and $\eta_{\textnormal{min}}$. W.r.t. the non-negative argument constraint of the $\ln(\cdot)$ function, (\ref{practical_limits_with_RFI_4}) and (\ref{practical_limits_with_RFI_5}) lead to the following condition:
\begin{condition}
\label{condition_kappa_eta_min_1}
$\kappa_K(\eta_{\textnormal{min}}) \in [0,1]$.
\end{condition}

From Appendix \ref{sec: proof_fund_limit_with_MU-interference} and (\ref{SNR_definition_with_MU-RFI_2})-(\ref{SNR_definition_with_MU-RFI_6}), it follows that
\begin{equation}
\label{gamma_upper_bound}
\gamma \leq  P_{\textnormal{max}}^s \big(\sum_{i=1}^2 X_i^2 \big) / \big[\tilde{P}_{\textnormal{min}}^i \big(\sum_{u=1}^{2U} Y_u^2 \big)\big] ,
\end{equation}
where $X_1 \eqdef \sqrt{2}\textnormal{Re}\{h\},  X_2 \eqdef \sqrt{2}\textnormal{Im}\{h\} \sim \mathcal{N}(0,1)$ and $Y_u \eqdef \sqrt{2}\textnormal{Re}\{g_u\},  Y_{u+U} \eqdef \sqrt{2}\textnormal{Im}\{g_u\} \sim \mathcal{N}(0,1)$, $\forall u\in[U]$, are all mutually independent standard normal RVs. Exploiting (\ref{gamma_upper_bound}) in the RHS of (\ref{practical_limits_with_RFI_7}) leads to
\begin{subequations}
\begin{align}
\label{practical_limits_with_RFI_8}
\hspace{-3mm}p(\eta_{\textnormal{min}})  \leq & \mathbb{P}\big( \sum_{i=1}^2 X_i^2/\big[\sum_{u=1}^{2U} Y_u^2 \big]  \geq \beta_K(\eta_{\textnormal{min}})\tilde{P}_{\textnormal{min}}^i/P_{\textnormal{max}}^s \big)   \\
\label{practical_limits_with_RFI_9}
=&\mathbb{P}\big(  Z_1/Z_2 \geq \beta_K(\eta_{\textnormal{min}}) \tilde{P}_{\textnormal{min}}^i/P_{\textnormal{max}}^s \big)     \\ 
\label{practical_limits_with_RFI_10}
=&\mathbb{P}\big(  (Z_1/2)/(Z_2/2U) \geq \beta_K(\eta_{\textnormal{min}}) U\tilde{P}_{\textnormal{min}}^i/P_{\textnormal{max}}^s \big)   \\ 
\label{practical_limits_with_RFI_11}
=&\mathbb{P}\big(  R \geq \beta_K(\eta_{\textnormal{min}}) U\tilde{P}_{\textnormal{min}}^i/P_{\textnormal{max}}^s \big),  
\end{align}
\end{subequations}
where $\big(Z_1, Z_2\big)\eqdef \big(\sum_{i=1}^2 X_i^2, \sum_{u=1}^{2U} Y_u^2\big)$ such that $Z_1 \sim \chi_2^2$ and  $Z_2 \sim \chi_{2U}^2$; $R \eqdef (Z_1/2)/(Z_2/2U)$ and hence $R \sim F_{2,2U}$ (see Appendix \ref{sec: proof_fund_limit_with_MU-interference}). If $\beta_K(\eta_{\textnormal{min}}) < 0$, the RHS of (\ref{practical_limits_with_RFI_11}) is always one, as $R$ is always a non-negative real number. Therefore, the feasible condition on $\beta_K(\eta_{\textnormal{min}})$ is $\beta_K(\eta_{\textnormal{min}}) \geq 0$. Deploying this condition in (\ref{beta_eta_min_defn}), the following condition ensues: 
\begin{condition}
\label{beta_eta_min_condition}
For $\alpha=e^{C_{K,2}}/(1+e^{C_{K,2}})$, $\kappa_K(\eta_{\textnormal{min}})\in [\alpha, \infty)$. 
\end{condition}
Under the satisfaction of Conditions \ref{condition_kappa_eta_min_1} and \ref{beta_eta_min_condition}, the RHS of (\ref{practical_limits_with_RFI_11}) is simplified using Markov’s inequality \cite[Proposition 1.2.4]{vershynin_2018} as $p(\eta_{\textnormal{min}})  \leq \mathbb{E}\{R\}/\tilde{c}$ for $\tilde{c} = \beta_K(\eta_{\textnormal{min}}) U\tilde{P}_{\textnormal{min}}^i/P_{\textnormal{max}}^s$:
\begin{subequations}
\begin{align}
\label{practical_limits_with_RFI_12}
p(\eta_{\textnormal{min}})   & \leq \mathbb{E}\{R\}P_{\textnormal{max}}^s/\beta_K(\eta_{\textnormal{min}}) U\tilde{P}_{\textnormal{min}}^i \\
\label{practical_limits_with_RFI_13}
&=2U/(2U-2) \times  P_{\textnormal{max}}^s/\beta_K(\eta_{\textnormal{min}}) U\tilde{P}_{\textnormal{min}}^i   \\
\label{practical_limits_with_RFI_14}
&= P_{\textnormal{max}}^s/[\beta_K(\eta_{\textnormal{min}}) (U-1)\tilde{P}_{\textnormal{min}}^i]. 
\end{align}
\end{subequations} 

For the RHS of (\ref{practical_limits_with_RFI_14}) to be a valid probability bound, $P_{\textnormal{max}}^s/[\beta_K(\eta_{\textnormal{min}}) (U-1)\tilde{P}_{\textnormal{min}}^i]$ must be less than or equal to one. This leads to the following condition:
\begin{condition}
\label{upper_bound_consstraint_eta_min_1}
$P_{\textnormal{max}}^s \leq \beta_K(\eta_{\textnormal{min}}) (U-1) \tilde{P}_{\textnormal{min}}^i$.
\end{condition}
Under the satisfaction of Conditions \ref{condition_kappa_eta_min_1}, \ref{beta_eta_min_condition}, and \ref{upper_bound_consstraint_eta_min_1}, it follows via (\ref{practical_limits_with_RFI_14}) that $p(\eta_{\textnormal{min}}) \leq P_{\textnormal{max}}^s/[\beta_K(\eta_{\textnormal{min}}) (U-1)\tilde{P}_{\textnormal{min}}^i]$. This ends the proof of Theorem \ref{thm: practical_limits_with_MU-interference}'s first claim. Under the same conditions, if we now apply the limit and its properties to (\ref{practical_limits_with_RFI_14}), it follows that   
\begin{subequations}
\begin{align}
\label{lim_Practical_limits_ii_1}
\lim_{P_{\textnormal{max}}^s \to 0} p(\eta_{\textnormal{min}}) & \leq 0, \hspace{1mm} \textnormal{for $U>1$ and $\tilde{P}_{\textnormal{min}}^i \in (0, \infty)$};    \\
\label{lim_Practical_limits_iii_1}
\lim_{\tilde{P}_{\textnormal{min}}^i \to \infty} p(\eta_{\textnormal{min}}) & \leq 0, \hspace{1mm} \textnormal{for $U>1$ and $ P_{\textnormal{max}}^s\in (0, \infty)$};  \\
\label{lim_Practical_limits_iv_1}
\lim_{U \to \infty} p(\eta_{\textnormal{min}}) & \leq 0, \hspace{1mm} \textnormal{for $P_{\textnormal{max}}^s, \tilde{P}_{\textnormal{min}}^i \in (0, \infty)$}. 
\end{align} 
\end{subequations}
From the axiomatic constraint of probability \cite{DPJN08}, it is true that $0 \leq  p(\eta_{\textnormal{min}}) \leq 1$, and hence 
\begin{subequations}
		\begin{align}
		\label{lim_Practical_limits_ii_2}
		0 \leq \lim_{P_{\textnormal{max}}^s \to 0} p(\eta_{\textnormal{min}}) & \leq 1;    \\
		\label{lim_Practical_limits_iii_2}
		0 \leq \lim_{\tilde{P}_{\textnormal{min}}^i \to \infty} p(\eta_{\textnormal{min}}) & \leq 1;  \\
		\label{lim_Practical_limits_iv_2}
		0 \leq \lim_{U \to \infty} p(\eta_{\textnormal{min}}) & \leq 1. 
	\end{align}
\end{subequations}
Intersecting (\ref{lim_Practical_limits_ii_1}) and (\ref{lim_Practical_limits_ii_2}),  (\ref{lim_Practical_limits_iii_1}) and (\ref{lim_Practical_limits_iii_2}), and (\ref{lim_Practical_limits_iv_1}) and (\ref{lim_Practical_limits_iv_2}) yield $\lim_{P_{\textnormal{max}}^s \to 0} p(\eta_{\textnormal{min}}) = 0$, $\lim_{\tilde{P}_{\textnormal{min}}^i \to \infty} p(\eta_{\textnormal{min}}) = 0$, and $\lim_{U \to \infty} p(\eta_{\textnormal{min}}) = 0$, respectively -- proving Theorem \ref{thm: practical_limits_with_MU-interference}'s\linebreak second, third, and fourth claim that are justified when Conditions \ref{condition_kappa_eta_min_1}, \ref{beta_eta_min_condition}, and \ref{upper_bound_consstraint_eta_min_1} are met. This ends Theorem \ref{thm: practical_limits_with_MU-interference}'s proof.   \QEDclosed

\section*{Acknowledgments}
The first author acknowledges Dr. Hamid Gharavi (\textit{Life Fellow, IEEE}) of NIST, MD, USA for funding and leadership support. The authors acknowledge the Editor's and anonymous reviewers' criticisms that have guided the improvement of their previously submitted manuscript, and the Digital Research Alliance of Canada for a computational support.

\section*{Disclaimer}
The identification of any commercial product or trade name does not imply endorsement or recommendation by NIST, nor is it intended to imply that the materials or equipment identified are necessarily the best available for the purpose.


\begin{IEEEbiography}[{\includegraphics[width=1.1in,height=1.25in,clip,keepaspectratio]{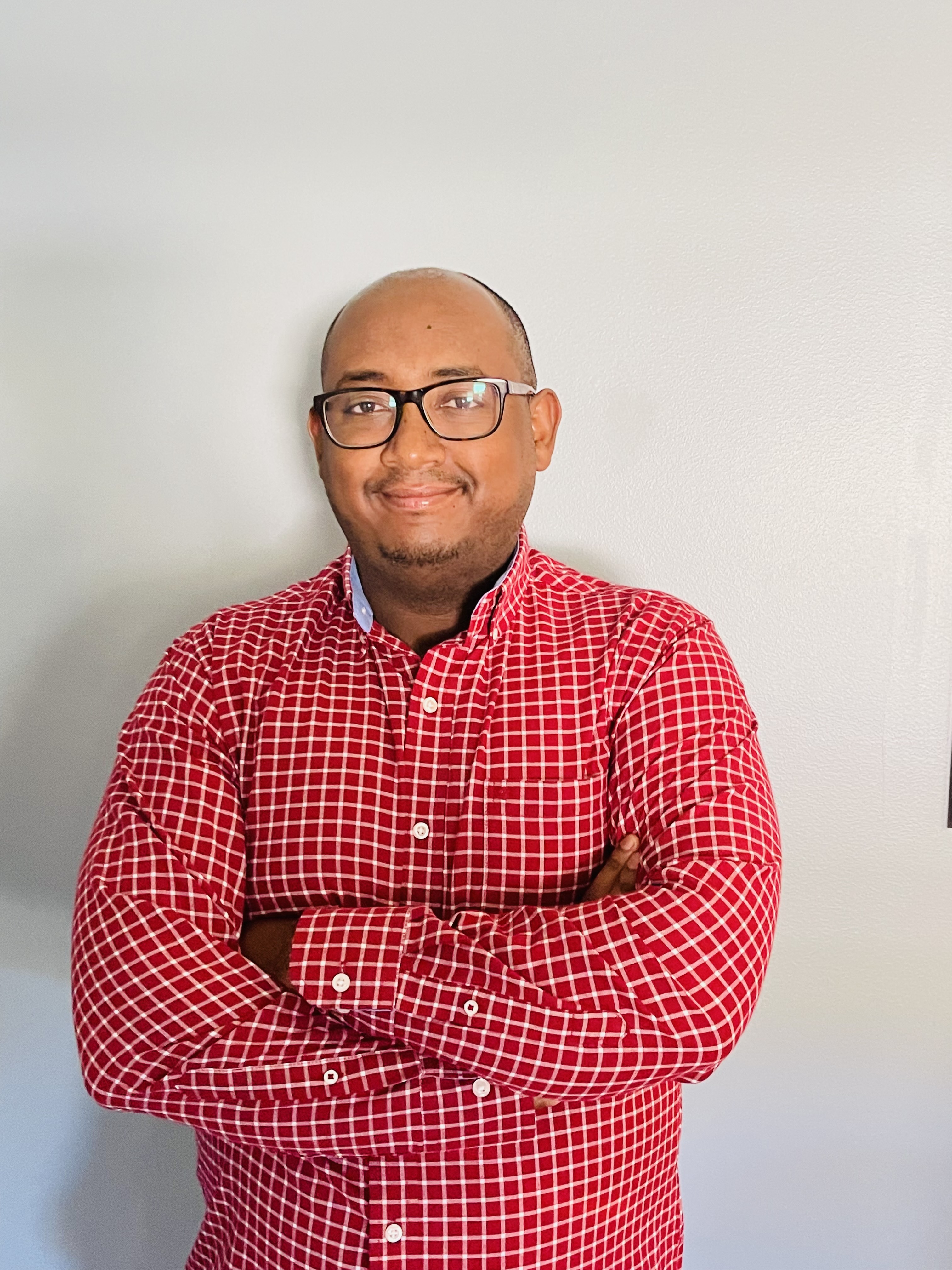}}]{\textbf{Tilahun M. Getu}} (M'19) earned the Ph.D. degree (with highest honor) in electrical engineering from the \'Ecole de Technologie Sup\'erieure (\'ETS), Montreal, QC, Canada in 2019. He is currently a Post-doctoral Fellow with the \'ETS. His transdisciplinary fundamental research interests span the numerous fields of classical and quantum \textbf{STEM} (\textbf{S}cience, \textbf{T}echnology, \textbf{E}ngineering, and \textbf{M}athematics) at the nexus of communications, signal processing, and networking (all types); intelligence (both artificial and natural); robotics; computing; security; optimization; high-dimensional statistics; and high-dimensional causal inference.   
	
Dr. Getu has received several awards, including the 2019 ÉTS Board of Director’s Doctoral Excellence Award in recognition of his Ph.D. dissertation selected as the 2019 \'ETS all-university best Ph.D. dissertation.
\end{IEEEbiography}

\begin{IEEEbiography}[{\includegraphics[width=1.1in,height=1.25in,clip,keepaspectratio]{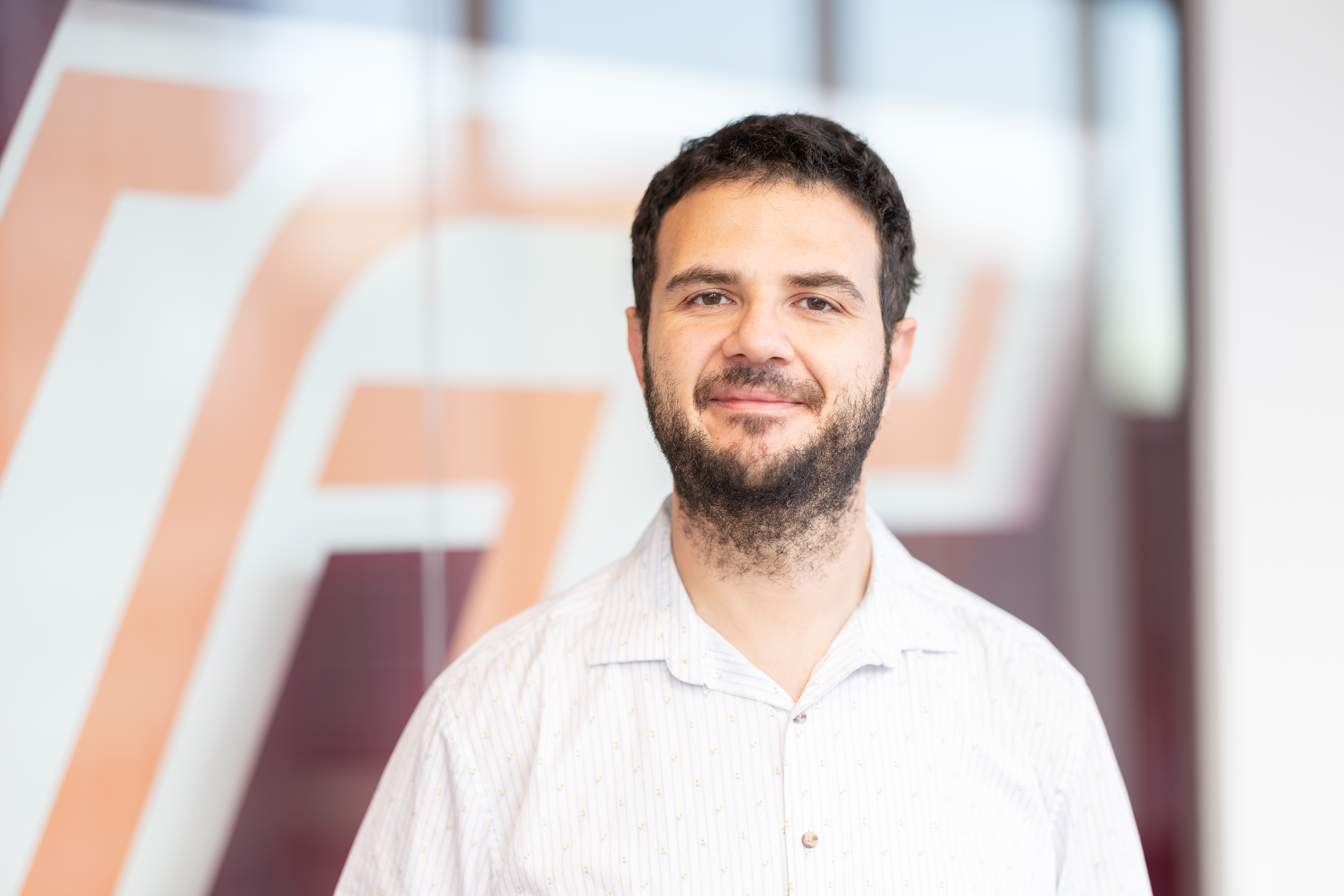}}]{\textbf{Walid Saad}} (S'07, M'10, SM'15, F'19) received his Ph.D degree from the University of Oslo, Norway in 2010. He is currently a Professor at the Department of Electrical and Computer Engineering at Virginia Tech, where he leads the \textbf{N}etwork sci\textbf{E}nce, \textbf{W}ireless, and \textbf{S}ecurity (\textbf{NEWS}) laboratory. His research interests include wireless networks (5G/6G/beyond), machine learning, game theory, security, UAVs, semantic communications, cyber-physical systems, and network science. Dr. Saad is a Fellow of the IEEE. He is also the recipient of the NSF CAREER award in 2013, the AFOSR summer faculty fellowship in 2014, and the Young Investigator Award from the Office of Naval Research (ONR) in 2015. He was the (co-)author of twelve conference best paper awards at IEEE WiOpt in 2009, ICIMP in 2010, IEEE WCNC in 2012, IEEE PIMRC in 2015, IEEE SmartGridComm in 2015, EuCNC in 2017, IEEE GLOBECOM (2018 and 2020), IFIP NTMS in 2019, IEEE ICC (2020 and 2022), and IEEE QCE in 2023. He is the recipient of the 2015 and 2022 Fred W. Ellersick Prize from the IEEE Communications Society, of the IEEE Communications Society Marconi Prize Award in 2023, and of the IEEE Communications Society Award for Advances in Communication in 2023. He was also a co-author of the papers that received the IEEE Communications Society Young Author Best Paper award in 2019, 2021, and 2023. Other recognitions include the 2017 IEEE ComSoc Best Young Professional in Academia award, the 2018 IEEE ComSoc Radio Communications Committee Early Achievement Award, and the 2019 IEEE ComSoc Communication Theory Technical Committee Early Achievement Award. From 2015-2017, Dr. Saad was named the Stephen O. Lane Junior Faculty Fellow at Virginia Tech and, in 2017, he was named College of Engineering Faculty Fellow. He received the Dean's award for Research Excellence from Virginia Tech in 2019. He was also an IEEE Distinguished Lecturer in 2019-2020. He has been annually listed in the Clarivate Web of Science Highly Cited Researcher List since 2019. He currently serves as an Area Editor for the IEEE Transactions on Communications. He is the Editor-in-Chief for the IEEE Transactions on Machine Learning in Communications and Networking. 
\end{IEEEbiography}


\begin{IEEEbiography}[{\includegraphics[width=1.1in,height=1.25in,clip,keepaspectratio]{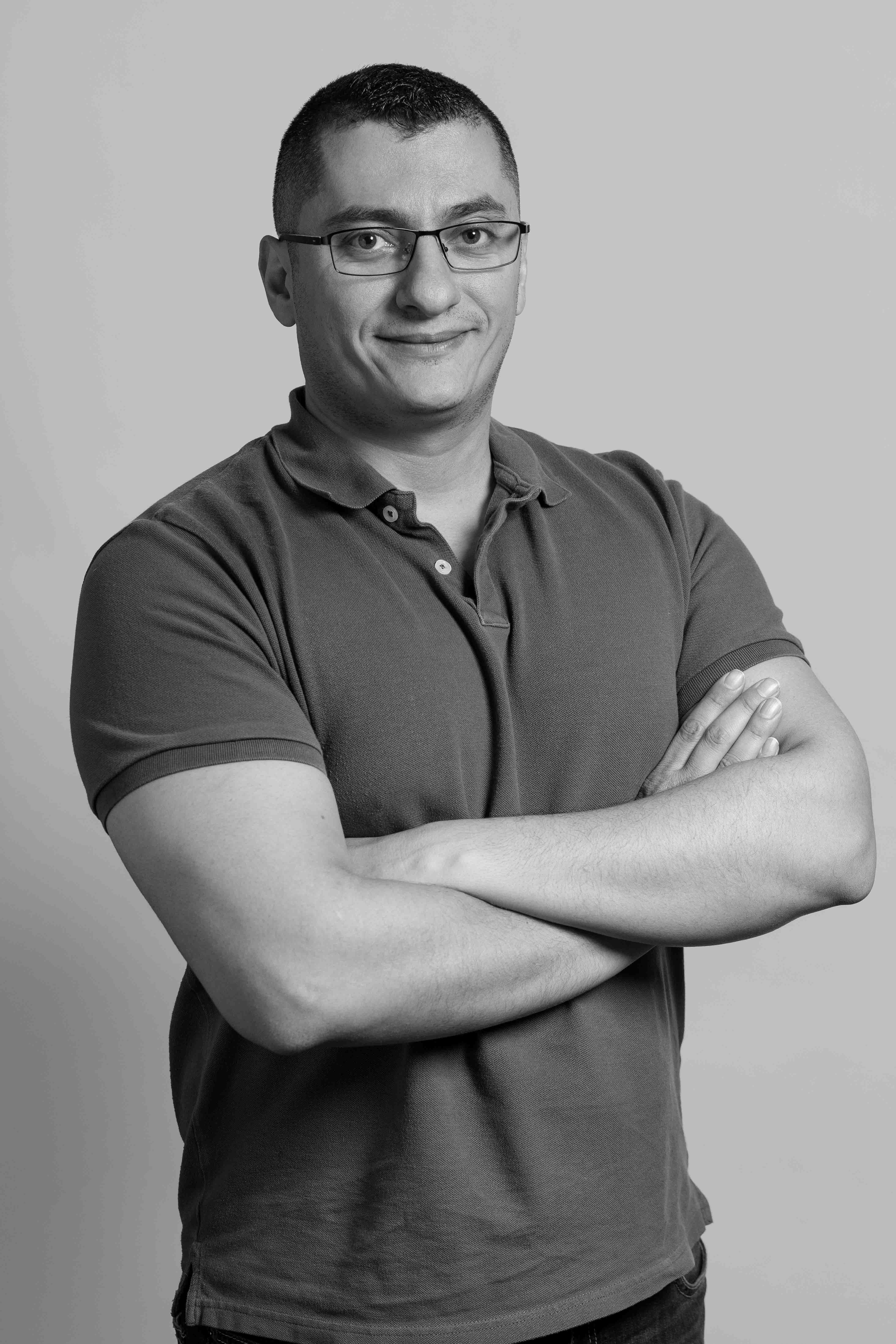}}]{\textbf{Georges Kaddoum}} (M'11--SM'20) is a professor and Tier 2 Canada Research Chair with the École de Technologie Supérieure (ÉTS), Université du Québec, Montréal, Canada. He is also a Faculty Fellow in the Cyber Security Systems and Applied AI Research Center at Lebanese American University. His recent research activities cover 5G/6G networks, tactical communications, resource allocations, and security.  Dr. Kaddoum has received many prestigious national and international awards in recognition of his outstanding research outcomes. Currently, Prof. Kaddoum serves as an Area Editor for the IEEE Transactions on Machine Learning in Communications and Networking and an Associate Editor for IEEE Transactions on Information Forensics and Security, and IEEE Transactions on Communications. 
\end{IEEEbiography}

\begin{IEEEbiography}[{\includegraphics[width=1.1in,height=1.25in,clip,keepaspectratio]{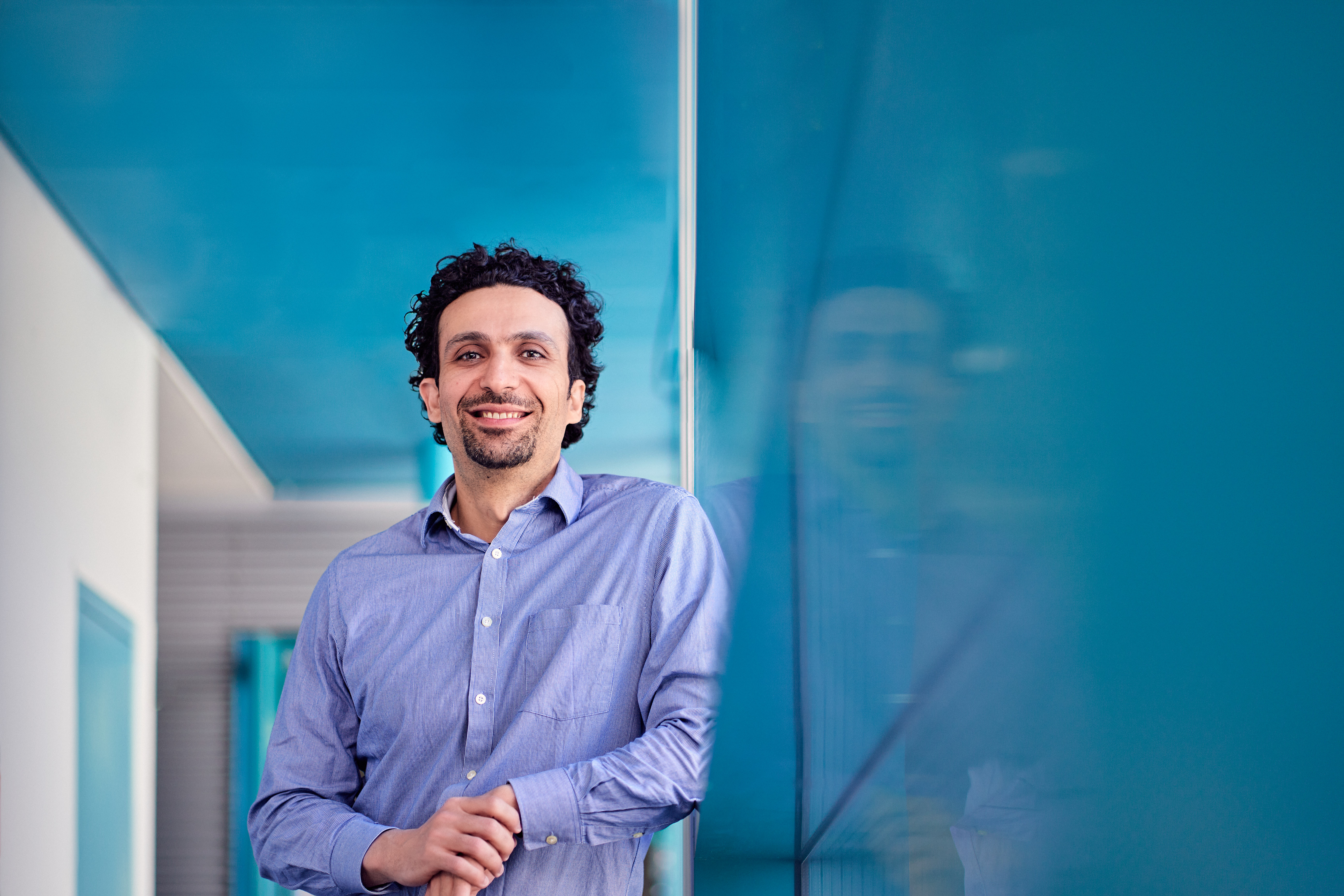}}]{\textbf{Mehdi Bennis}} (F’20) is a full tenured Professor at the Centre for Wireless Communications, University of Oulu, Finland and head of the \textbf{I}ntelligent \textbf{CO}nnectivity and \textbf{N}etworks/Systems Group (\textbf{ICON}). His main research interests are in radio resource management, game theory and distributed AI in 5G/6G networks. He has published more than 200 research papers in international conferences, journals and book chapters. He has been the recipient of several prestigious awards. Dr. Bennis is an editor of IEEE TCOM and Specialty Chief Editor for Data Science for Communications in the Frontiers in Communications and Networks journal.
\end{IEEEbiography}



\begin{thebibliography}{10}
	\providecommand{\url}[1]{#1}
	\csname url@samestyle\endcsname
	\providecommand{\newblock}{\relax}
	\providecommand{\bibinfo}[2]{#2}
	\providecommand{\BIBentrySTDinterwordspacing}{\spaceskip=0pt\relax}
	\providecommand{\BIBentryALTinterwordstretchfactor}{4}
	\providecommand{\BIBentryALTinterwordspacing}{\spaceskip=\fontdimen2\font plus
		\BIBentryALTinterwordstretchfactor\fontdimen3\font minus
		\fontdimen4\font\relax}
	\providecommand{\BIBforeignlanguage}[2]{{%
			\expandafter\ifx\csname l@#1\endcsname\relax
			\typeout{** WARNING: IEEEtran.bst: No hyphenation pattern has been}%
			\typeout{** loaded for the language `#1'. Using the pattern for}%
			\typeout{** the default language instead.}%
			\else
			\language=\csname l@#1\endcsname
			\fi
			#2}}
	\providecommand{\BIBdecl}{\relax}
	\BIBdecl
	
	\bibitem{Shannon_Weaver_Math_Theory_Commun'49}
	C.~E. Shannon and W.~Weaver, \emph{The Mathematical Theory of
		Communication}.\hskip 1em plus 0.5em minus 0.4em\relax Urbana, IL, USA: Univ.
	Illinois Press, 1949.
	
	\bibitem{Chaccour_Building_NG_SemCom_Networks'22}
	\BIBentryALTinterwordspacing
	C.~Chaccour, W.~Saad, M.~Debbah, Z.~Han, and H.~V. Poor, ``Less data, more
	knowledge: Building next generation semantic communication networks,'' 2022.
	[Online]. Available: \url{https://arxiv.org/pdf/2211.14343.pdf}
	\BIBentrySTDinterwordspacing
	
	\bibitem{Yang_SemCom_ComST’23}
	W.~Yang, H.~Du, Z.~Q. Liew, W.~Y.~B. Lim, Z.~Xiong, D.~Niyato, X.~Chi, X.~Shen,
	and C.~Miao, ``Semantic communications for future internet: Fundamentals,
	applications, and challenges,'' \emph{IEEE Commun. Surv. Tutor.}, vol.~25,
	no.~1, pp. 213--250, 2023.
	
	\bibitem{Tong_FL_ASC'21}
	H.~Tong, Z.~Yang, S.~Wang, Y.~Hu, O.~Semiari, W.~Saad, and C.~Yin, ``Federated
	learning for audio semantic communication,'' \emph{Front. Comms. Net.},
	vol.~2, 2021.
	
	\bibitem{Xie_DL-based_SemCom'21}
	H.~Xie, Z.~Qin, G.~Li, and B.-H. Juang, ``Deep learning enabled semantic
	communication systems,'' \emph{IEEE Trans. Signal Process.}, vol.~69, pp.
	2663--2675, Apr. 2021.
	
	\bibitem{Kountouris_Semantics_EmpoweredCF'21}
	M.~Kountouris and N.~Pappas, ``Semantics-empowered communication for networked
	intelligent systems,'' \emph{IEEE Commun. Mag.}, vol.~59, pp. 96--102, 2021.
	
	\bibitem{Luo_SemCom_Overview'22}
	X.~Luo, H.-H. Chen, and Q.~Guo, ``Semantic communications: Overview, open
	issues, and future research directions,'' \emph{IEEE Wirel. Commun.},
	vol.~29, no.~1, pp. 210--219, 2022.
	
	\bibitem{Shi_SemCom_ComMag’21}
	G.~Shi, Y.~Xiao, Y.~Li, and X.~Xie, ``From semantic communication to
	semantic-aware networking: Model, architecture, and open problems,''
	\emph{IEEE Commun. Mag.}, vol.~59, no.~8, pp. 44--50, 2021.
	
	\bibitem{Bao_Towards_Theory_SemCom'11}
	J.~Bao, P.~Basu, M.~Dean, C.~Partridge, A.~Swami, W.~Leland, and J.~A. Hendler,
	``Towards a theory of semantic communication,'' \emph{Proc. IEEE Network
		Science Workshop}, pp. 110--117, 2011.
	
	\bibitem{Saad_6G_Vision_20}
	W.~{Saad}, M.~{Bennis}, and M.~{Chen}, ``A vision of {6G} wireless systems:
	Applications, trends, technologies, and open research problems,'' \emph{IEEE
		Netw.}, vol.~34, no.~3, pp. 134--142, 2020.
	
	\bibitem{Letaief_Edge_AI_Vision'22}
	K.~B. Letaief, Y.~Shi, J.~Lu, and J.~Lu, ``Edge artificial intelligence for
	{6G}: Vision, enabling technologies, and applications,'' \emph{IEEE J. Sel.
		Areas Commun.}, vol.~40, no.~1, pp. 5--36, 2022.
	
	\bibitem{Alwis_Survey_GG_Networks'21}
	C.~D. Alwis, A.~Kalla, Q.-V. Pham, P.~Kumar, K.~Dev, W.-J. Hwang, and
	M.~Liyanage, ``Survey on {6G} frontiers: Trends, applications, requirements,
	technologies and future research,'' \emph{IEEE Open J. Commun. Soc.}, vol.~2,
	pp. 836--886, 2021.
	
	\bibitem{YYBGH_15}
	{Y. LeCun, Y. Bengio, and G. Hinton}, ``Deep learning,'' \emph{Nature}, vol.
	521, no. 436, pp. 436--444, 2015.
	
	\bibitem{Trands_in_DL-Based_NLP'18}
	T.~Young, D.~Hazarika, S.~Poria, and E.~Cambria, ``Recent trends in deep
	learning based natural language processing [review article],'' \emph{IEEE
		Comput. Intell. Mag.}, vol.~13, no.~3, pp. 55--75, 2018.
	
	\bibitem{Eirina_JSCC'19}
	E.~Bourtsoulatze, D.~Burth~Kurka, and D.~Gündüz, ``Deep joint source-channel
	coding for wireless image transmission,'' \emph{IEEE Trans. Cogn. Commun.},
	vol.~5, no.~3, pp. 567--579, 2019.
	
	\bibitem{Wang_JSAC_Semantic_Transmission’23}
	S.~Wang, J.~Dai, Z.~Liang, K.~Niu, Z.~Si, C.~Dong, X.~Qin, and P.~Zhang,
	``Wireless deep video semantic transmission,'' \emph{IEEE J. Sel. Areas
		Commun.}, vol.~41, no.~1, pp. 214--229, 2023.
	
	\bibitem{Xie_MU-SemCom’22}
	H.~Xie, Z.~Qin, and G.~Y. Li, ``Task-oriented multi-user semantic
	communications for {VQA},'' \emph{IEEE Wirel. Commun. Lett.}, vol.~11,
	no.~3, pp. 553--557, 2022.
	
	\bibitem{Hu_Robust_SemCom’23}
	Q.~Hu, G.~Zhang, Z.~Qin, Y.~Cai, G.~Yu, and G.~Y. Li, ``Robust semantic
	communications with masked {VQ-VAE} enabled codebook,'' \emph{IEEE Trans.
		Wirel. Commun.}, pp. 1--1, 2023.
	
	\bibitem{Shi_to_Semantic_Fidelity'21}
	\BIBentryALTinterwordspacing
	G.~Shi, D.~Gao, X.~Song, J.~Chai, M.~Yang, X.~Xie, L.~Li, and X.~Li, ``A new
	communication paradigm: from bit accuracy to semantic fidelity,'' 2021.
	[Online]. Available: \url{https://arxiv.org/pdf/2101.12649.pdf}
	\BIBentrySTDinterwordspacing
	
	\bibitem{Sagduyu_Is_SemCom_Secure'22}
	\BIBentryALTinterwordspacing
	Y.~E. Sagduyu, T.~Erpek, S.~Ulukus, and A.~Yener, ``Is semantic communications
	secure? a tale of multi-domain adversarial attacks,'' 2022. [Online].
	Available: \url{https://arxiv.org/pdf/2212.10438.pdf}
	\BIBentrySTDinterwordspacing
	
	\bibitem{TMWAR_TWC_18}
	{T. M. Getu, W. Ajib, and R. Jr. Landry}, ``Performance analysis of
	energy-based {RFI} detector,'' \emph{IEEE Trans. Wirel. Commun.}, vol.~17,
	no.~10, pp. 6601--6616, Oct. 2018.
	
	\bibitem{TWR_WCL_2018}
	T.~M. Getu, W.~Ajib, and {R. Jr. Landry}, ``Power-based broadband {RF}
	interference detector for wireless communication systems,'' \emph{IEEE Wirel.
		Commun. Lett.}, vol.~7, no.~6, pp. 1002--1005, Dec. 2018.
	
	\bibitem{Getu_dissertation_19}
	{T. M. Getu}, ``Advanced {RFI} detection, {RFI} excision, and spectrum sensing:
	Algorithms and performance analyses,'' {Ph.D.} dissertation, {\'Ecole de
		Technologie Sup\'erieure (\'ETS), Montr\'eal, QC, Canada}, 2019.
	
	\bibitem{Poggio_Theo_Issues_Dnets_2020}
	T.~Poggio, A.~Banburski, and Q.~Liao, ``Theoretical issues in deep networks,''
	\emph{Proc. Natl. Acad. Sci. U.S.A.}, Jun. 2020.
	
	\bibitem{arXiv_Getu_Fundamental_Limits'23_v1}
	\BIBentryALTinterwordspacing
	T.~M. Getu and G.~Kaddoum, ``Fundamental limits of deep learning-based binary
	classifiers trained with hinge loss,'' 2023. [Online]. Available:
	\url{https://arxiv.org/pdf/2309.06774.pdf}
	\BIBentrySTDinterwordspacing
	
	\bibitem{Tong_Zhu__6G'21}
	{W. Tong and P. Zhu (Eds.)}, \emph{{6G}: The Next Horizon: From Connected
		People and Things to Connected Intelligence}.\hskip 1em plus 0.5em minus
	0.4em\relax Cambridge, UK: Cambridge Univ. Press, 2021.
	
	\bibitem{Tong_Nine_Challenges’22}
	W.~Tong and G.~Y. Li, ``Nine challenges in artificial intelligence and wireless
	communications for {6G},'' \emph{IEEE Wirel. Commun.}, vol.~29, no.~4, pp.
	140--145, 2022.
	
	\bibitem{NJKB_Vol_I'94}
	{{N. L. Johnson \textit{et al.}}}, \emph{Continuous Univariate Distributions},
	2nd~ed., ser. Wiley series in probability and mathematical statistics.
	Applied probability and statistics, {N. L. Johnson, S. Kotz, and N.
		Balakrishnan}, Ed.\hskip 1em plus 0.5em minus 0.4em\relax Hoboken, NJ, USA:
	Wiley, 1994, vol.~1.
	
	\bibitem{NJKB_Vol_II'95}
	{N. L. Johnson \textit{et al.}}, \emph{Continuous Univariate Distributions},
	2nd~ed., ser. Wiley series in probability and mathematical statistics.
	Applied probability and statistics, {N. L. Johnson, S. Kotz, and N.
		Balakrishnan}, Ed.\hskip 1em plus 0.5em minus 0.4em\relax Hoboken, NJ, USA:
	Wiley, 1995, vol.~2.
	
	\bibitem{Mu_Heterogeneous_Commun_JSAC'23}
	X.~Mu, Y.~Liu, L.~Guo, and N.~Al-Dhahir, ``Heterogeneous semantic and bit
	communications: A semi-{NOMA} scheme,'' \emph{IEEE J. Sel. Areas Commun.},
	vol.~41, no.~1, pp. 155--169, Jan. 2023.
	
	\bibitem{Jiang_Reliable_SemCom'22}
	S.~Jiang, Y.~Liu, Y.~Zhang, P.~Luo, K.~Cao, J.~Xiong, H.~Zhao, and J.~Wei,
	``Reliable semantic communication system enabled by knowledge graph,''
	\emph{Entropy}, vol.~24, no.~6, 2022.
	
	\bibitem{Devlin’19_BERT}
	\BIBentryALTinterwordspacing
	J.~Devlin, M.-W. Chang, K.~Lee, and K.~Toutanova, ``{BERT}: Pre-training of
	deep bidirectional transformers for language understanding,'' 2019. [Online].
	Available: \url{https://arxiv.org/pdf/1810.04805.pdf}
	\BIBentrySTDinterwordspacing
	
	\bibitem{Getu_IEEE_Access'23}
	T.~M. Getu, G.~Kaddoum, and M.~Bennis, ``Making sense of meaning: A survey on
	metrics for semantic and goal-oriented communication,'' \emph{IEEE Access},
	vol.~11, pp. 45\,456--45\,492, 2023.
	
	\bibitem{Simon_Alouni_DC_over_Fading_Channels'05}
	M.~K. Simon and M.~S. Alouini, \emph{Digital Communication over Fading
		Channels}, 2nd~ed.\hskip 1em plus 0.5em minus 0.4em\relax Hoboken, NJ, USA:
	Wiley, 2005.
	
	\bibitem{Gollakota_dissertation_13}
	{S. Gollakota}, ``Embracing interference in wireless systems,'' {Ph.D.}
	dissertation, {MIT, Cambridge, MA, USA}, 2013.
	
	\bibitem{Tech_Report_NIST-1885_2015}
	{G. Koepke, W. Young, J. Ladburg, and J. Coder}, ``Complexities of testing
	interference and coexistence of wireless systems in critical
	infrastructure,'' RF Technology Division, Communications Technology
	Laboratory, NIST, Tech. Rep., Jul. 2015.
	
	\bibitem{Synthe_lect_DLLR_16}
	Z.~Chen and B.~Liu, ``Lifelong machine learning,'' \emph{Synth. Lect. Artif.
		Intell. Mach. Learn.}, vol.~10, no.~3, pp. 1--145, 2016.
	
	\bibitem{DeLange_CL_Survey'22}
	M.~D. Lange, R.~Aljundi, M.~Masana, S.~Parisot, X.~Jia, A.~Leonardis, G.~G.
	Slabaugh, and T.~Tuytelaars, ``A continual learning survey: Defying
	forgetting in classification tasks,'' \emph{IEEE Trans. Pattern Anal. Mach.
		Intell.}, vol.~44, pp. 3366--3385, 2022.
	
	\bibitem{Masana_Class_IL'20}
	\BIBentryALTinterwordspacing
	M.~Masana, X.~Liu, B.~Twardowski, M.~Menta, A.~D. Bagdanov, and J.~van~de
	Weijer, ``Class-incremental learning: survey and performance evaluation on
	image classification,'' 2020. [Online]. Available:
	\url{https://arxiv.org/pdf/2010.15277.pdf}
	\BIBentrySTDinterwordspacing
	
	\bibitem{arXiv_Getu_DeepSC_Performance_Limits'23_v2}
	\BIBentryALTinterwordspacing
	T.~M. Getu, W.~Saad, G.~Kaddoum, and M.~Bennis, ``Performance limits of a deep
	learning-enabled text semantic communication under interference,'' 2023.
	[Online]. Available: \url{https://arxiv.org/pdf/2302.14702v2.pdf}
	\BIBentrySTDinterwordspacing
	
	\bibitem{TMGWA17}
	{T. M. Getu, W. Ajib, and O. A. Yeste-Ojeda}, ``Tensor-based efficient
	multi-interferer {RFI} excision algorithms for {SIMO} systems,'' \emph{IEEE
		Trans. Commun.}, vol.~65, no.~7, pp. 3037--3052, Jul. 2017.
	
	\bibitem{GeAR_conf_16}
	T.~M. Getu, W.~Ajib, and {R. Jr. Landry}, ``Oversampling-based algorithm for
	efficient {RF} interference excision in {SIMO} systems,'' in \emph{Proc. IEEE
		Global Conf. on Signal and Inform. Process. (IEEE GlobalSIP)}, Washington DC,
	DC, USA, Dec. 2016, pp. 1423--1427.
	
	\bibitem{Marvin_K_SImon_Handbook'06}
	M.~K. Simon, \emph{Probability Distributions Involving {G}aussian Random
		Variables : A Handbook for Engineers and Scientists}.\hskip 1em plus 0.5em
	minus 0.4em\relax Springer, 2006.
	
	\bibitem{Getu_DSFC_Estimation'22}
	\BIBentryALTinterwordspacing
	T.~M. Getu, N.~T. Golmie, and D.~W. Griffith, ``Blind estimation of a doubly
	selective {OFDM} channel: A deep learning algorithm and theory,'' 2022.
	[Online]. Available: \url{https://arxiv.org/pdf/2206.07483.pdf}
	\BIBentrySTDinterwordspacing
	
	\bibitem{DPJN08}
	D.~P. Bertsekas and J.~N. Tsitsiklis, \emph{Introduction to Probability},
	2nd~ed.\hskip 1em plus 0.5em minus 0.4em\relax Belmont, MA, USA: Athena
	Scientific, 2008.
	
	\bibitem{Prob_Stat_Random_Process'14}
	H.~Pishro-Nik, \emph{Introduction to Probability, Statistics, and Random
		Processes}.\hskip 1em plus 0.5em minus 0.4em\relax Kappa Research, LLC, 2014.
	
	\bibitem{ISGI07}
	I.~S. Gradshteyn and I.~M. Ryzhik, \emph{Table of Integrals, Series, and
		Products}, 7th~ed.\hskip 1em plus 0.5em minus 0.4em\relax Academic Press,
	Burlington, MA, USA, 2007.
	
	\bibitem{vershynin_2018}
	R.~Vershynin, \emph{High-Dimensional Probability: An Introduction with
		Applications in Data Science}.\hskip 1em plus 0.5em minus 0.4em\relax
	Cambridge Univ. Press, 2018.
	
\end{thebibliography}
\end{document}